\documentclass[preprint,aps]{revtex4}
\usepackage{amssymb,amsmath}
\usepackage{graphics}
\usepackage{dcolumn}
\usepackage{bm}

\begin{document}
\preprint{}
\title{The internal color degrees of freedom 
for the weakly interacting quarks and gluons}
\author{Ismail Zakout}
\affiliation{ITP and FIAS, Frankfurt university, 
Frankfurt am main, Germany}
\date{\today} 
\begin{abstract}
When the phase transition from the hadronic matter 
to the deconfined quark-gluon plasma 
or quark-gluon liquid is reached, 
the color degrees of freedom become important 
and appear explicitly in the equation of state.
Under the extreme conditions, the color degrees of freedom 
can not decouple from the other degrees of freedom.
The conservation of color charges is maintained by 
introducing the color chemical potentials and their fugacities.
We demonstrate the explicit role of color degrees 
of freedom in the hot and dense matter 
of the weakly interacting quarks and gluons.
In order to illustrate our approach, we calculate 
the effective colored quark and gluon propagators 
as well as the hard thermal colored quark and gluon loops.
The calculations are preformed under the assumption
of the hard thermal loop approximation.
Finally, we present the decay rate for the hard thermal colored quark. 
The present calculation is relevant 
to the quarks and gluons which are not truly 
deconfined quark-gluon plasma and behave as a fluid in the RHIC energy. 
The model can be explored in the LHC energy.
\end{abstract}
\maketitle
\section{Introduction}

The phase transition diagram from the hadronic 
phase to the quark-gluon plasma is very rich and non-trivial.
It is thought that when the hadronic matter 
is compressed and cooled down at extreme baryonic densities it 
undergoes phase transition 
to new forms of strongly interacting matter 
such as the color superconducting quark matter 
and the color-flavor locked matter. 
On the other hand, when the hadronic matter is heated up to high
temperatures at low baryonic density, it undergoes smooth 
phase transition or multiple phase transitions
to the quark-gluon fluid 
at extreme temperature before explosive and true
deconfined quark-gluon plasma emerges in the system
~\cite{Zakout:2010ep,Zakout:2007nb,Zakout:2006zj}
\cite{Begun:2009an,Abir:2009sh,Ferroni:2008ej}.
Moreover, it is suggested that there 
is a possibility for the existence of a new class 
of matter, namely quarkyonic, 
with a specific internal structure when 
the system is heated up at intermediate baryonic density. 
The quarkyonic (i.e. Hagedorn) matter corresponds
to non-deconfined 
and weakly interacting quark and gluon 
blobs which maintain specific internal symmetries 
(with color-singlet states).
It is reasonable to imagine the existence of 
such intermediate hadronic phases 
in order to soften the equation of state 
below the borderline of 
the true deconfined quark-gluon plasma.
There is a strong argument that the deconfinement phase transition 
diagram has a critical point and underneath that point
the hadronic matter undergoes a higher order phase 
transition to the Hagedorn matter
at extreme temperature and it could be followed 
by multi phase transition precesses prior 
to the eventual deconfinement phase transition 
to quark-gluon plasma
~\cite{Zakout:2010ep,Zakout:2007nb,Zakout:2006zj,Begun:2009an,
Abir:2009sh,Ferroni:2008ej}. 
The role of the Hagedorn matter in RHIC has been considered 
in 
Refs.~\cite{NoronhaHostler:2009cf,NoronhaHostler:2007jf,
NoronhaHostler:2008ju,NoronhaHostler:2009tz}
 
The quark-gluon plasma is found 
a perfect fluid with a low viscosity. 
This discovery means that the quark-gluon plasma 
is a weakly interacting matter 
of quarks and gluons rather 
to be free mobile quark-gluon plasma.
Since at sufficiently high 
temperature the effective gauge running coupling constant
is small, the system of the weakly coupled quarks and gluons 
can be treated by perturbative methods 
due to the asymptotic freedom in QCD. 
The thermal calculations is preformed by a version of 
the thermal perturbation theory that has been 
developed and improved by several people. 
A nice review of the field theory at finite temperature 
and density is given 
by Landsman and van Weert
~\cite{Landsman:1986uw}. 
The procedure has been significantly improved 
and extended by Braaten and Pisarski 
in order to improve the convergence 
for the higher order loop calculations
~\cite{Braaten:1989mz,Braaten:1989kk,Frenkel:1989br,
Braaten:1990it,Braaten:1991gm,Braaten:1992gd}.
The physics of the thermal QCD is reviewed 
in Ref.~\cite{Smilga:1996cm}.
This theory is based on the hard thermal loop 
(HTL) re-summation program
instead of relying on the bare Green functions.
The HTL approximation is equivalent to the leading term 
of the high temperature expansion of the diagram 
under consideration. 
Excellent review for the collective dynamics 
and hard thermal loops is given 
by Blaizot and Iancu
~\cite{Blaizot:2001nr}.
%
Fortunately, the standard version 
of the HTL re-summation program is sufficient  
to calculate the dynamical quantities such 
as the quasi-particle properties 
and the production and loss rates.
However, the standard re-summation 
program in the thermal perturbation
theory may not work at certain order of 
the running coupling constant because
of the inherently non-perturbative effects 
of non-Abelian gauge theories. 
For instance, there are some calculations, 
such as the production rate of real photons 
in the plasma, 
are associated with the non-perturbative effect.
In this case, the convergence 
of the re-summation procedure 
must be considered carefully
~\cite{Blaizot:1999ap,Blaizot:2003tw}.
However, the non-perturbative effect 
does not affect much on 
the physical quantities which 
are considered in the present work. 
Consequently, the standard HTL re-summation program
is sufficient to illustrate our approach how to 
include to color degrees of freedom explicitly 
in the calculation. This can be generalized 
to the non-perturbative calculation.
The effective field at finite 
density~\cite{Vija:1994is,Schafer:2003jn}  
and thermal field with a specific group symmetry
in the lattice have been considered~\cite{Levai:1997yx}. 

The explicit color degrees of freedom 
and the underlying internal color symmetry group 
of the quarks and gluons
are supposed to be crucial 
in the deconfinement phase transition diagram
in particular in the intermediate phases 
such as the Hagedorn, quarkyonic or quark-gluon fluid.
Those intermediate phases are supposed 
to be located above the low-lying hadronic matter and 
just below the explosive deconfined quark-gluon plasma.
Furthermore, the effect of weakly interacting 
colored quarks and gluons becomes important above 
the deconfinement phase transition borderline
to quark-gluon plasma in particular when the medium 
becomes ultra-extreme hot and/or dense.
The color degrees of freedom are the non-Abelian 
charges in QCD. 
They are parameterized by the eigenvalues 
of the $SU(N_{c})$ symmetry group 
and their eigenvectors are given by diagonal generators. 
The quarks carry $N_{c}$ 
fundamental color charges
while the gluons carry $N^{2}_{c}-1$ adjoint color charges. 
The conservation of the fundamental and adjoint color charges 
is maintained by the fundamental and adjoint 
imaginary chemical potentials, namely, 
$i\,\theta_{i}$ and $i\,\phi^{a}$ 
for quarks and gluons, respectively.
The fundamental and adjoint color indexes 
$i$ and $a$, respectively, run over 
$i=1,\cdots, N_c$ and $a=1,\cdots,N^2_c-1$, 
respectively.
The adjoint color chemical potentials are related 
to the fundamental ones in the following way,
\begin{eqnarray}
\phi^{a}&=&\phi^{\underbrace{(AB)}},
\nonumber\\
&=&
\left(\theta_{A}-\theta_{B}\right),
\end{eqnarray}
with the adjoint representation, namely, 
$a=\underbrace{(AB)}$ 
where $A,B=1,\cdots, N_c$ are fundamental-like indexes 
and they are given in order to maintain 
the conservation of the color charges. 
Furthermore, the imaginary fundamental and adjoint 
color chemical potentials can be transformed 
to real ones by applying the Wick rotation 
procedure as follows
\begin{eqnarray} 
i\,\frac{1}{\beta}\,\theta_{i}&=&{\mu_{C}}_{i},
\end{eqnarray}
and
\begin{eqnarray}
i\,\frac{1}{\beta}\,\phi^{a}&=&{\mu_{C}}^{a},
\nonumber\\
i\,\frac{1}{\beta}\,\phi^{a}&\rightarrow& 
i\,\frac{1}{\beta}\,\phi^{\underbrace{(AB)}},
\nonumber\\
&=&
i\,\frac{1}{\beta}\,\left(\theta_{A}-\theta_{B}\right),
\nonumber\\
&=& 
\left({\mu_{C}}_{A}-{\mu_{C}}_{B}\right). 
\end{eqnarray}
The fugacities of the fundamental and adjoint color charges 
are given by the fundamental and adjoint 
real (i.e. not imaginary) color chemical potentials. 
The real color chemical potentials
${\mu_{C}}_{i}$ where $i=1,\cdots, N_{c}$
are essential to control the color charge densities 
in the equilibrium reaction.

It is relevant to note that the color chemical 
potentials are related to the Polyakov loop 
$\Phi$ 
in the following way,
\begin{eqnarray}
\Phi&=&\frac{1}{N_c}\,\mbox{tr}\, L(x).
\end{eqnarray}
The Polyakov loop matrix is given by
\begin{eqnarray}
L(x)&=&{\cal P}\, 
\exp\left[
-\int^{\beta}_{0}d\tau {\bf t}^{a} A^{a}_{0}(x,\tau)   
\right],
\end{eqnarray}
where ${\cal P}$ is the path ordering.
The Polyakov gauge, $L\equiv L(x)$
in the mean field approximation 
is written as follows
\begin{eqnarray}
L&=&
\mbox{diag}
\left(\{e^{i\theta_{i}}\}\right)
\rightarrow
\mbox{diag}
\left(\{e^{\beta\,{\mu_{C}}_{i}}\}\right).
\end{eqnarray}
Hence, the Polyakov loop is reduced to
\begin{eqnarray}
\Phi&=&\frac{1}{N_c}\,\mbox{tr}\, L
\rightarrow
\frac{1}{N_c}\,\sum^{N_c}_{i=1} e^{i\theta_{i}}
\rightarrow
\frac{1}{N_c}\,\sum^{N_c}_{i=1} e^{\beta\,{\mu_{C}}_{i}}.
\end{eqnarray}
For the symmetric group $SU(3)$, there 
is two independent variables namely the
Polyakov loop $\Phi$ and its complex conjugate 
$\Phi^{*}$. Those two Polyakov variables correspond to
the two independent chemical potentials in the $SU(3)$,
namely, ${\mu_{C}}_{1}$ and ${\mu_{C}}_{2}$.
The Polyakov loop is the parameter of
the spontaneous $Z_{3}$ symmetry breaking. 
This is associated with the deconfinement 
phase transition.
However, the Polyakov loop parametrization 
becomes questionable above 
the deconfinement phase transition 
in particular 
when the color and flavor degrees 
of freedom are coupled with each other
in a way that the color degrees 
of freedom can not be separated from 
the rest of the other degrees of freedom. 
In the deconfinement phase transition, 
it is naively 
to assume that the quarks and gluons carry  
explicit color degrees of freedom with specific 
color chemical potentials in order to control 
the color charges in the equilibrium.

Recently, the explicit color degrees of freedom have been considered 
to study the semi quark-gluon-plasma 
using the double line basis at finite $N_c$ of $SU(N_c)$ symmetry group
and $Z(N_c)$ interface~\cite{Hidaka:2008dr,Hidaka:2009hs,Hidaka:2009ma}.
The approach considered in the present work resembles to some extent that one
has been considered in 
Ref.~\cite{Hidaka:2008dr,Hidaka:2009hs,Hidaka:2009ma}, 
though they apparently look different.

The outline of the present paper is as follows:
In Section \ref{section1}, the Cartan sub-algebra
is reviewed for the fundamental and adjoint $SU(N_c)$
symmetry group. 
The fundamental and adjoint constructions 
are essential to define the color degrees of freedom
for quarks and gluons, respectively, in the QCD.
In section \ref{section2}, the thermal field theory 
in the imaginary-time formalism is extended 
to include the color degrees of freedom explicitly.
The thermal propagators for the colored quarks and gluons 
are introduced in the context of the mixed-time representation
of the imaginary-time formulation.
In section \ref{section3}, 
the contraction and reduction 
of the fundamental and adjoint color indexes
of the $SU(N_c)$ symmetry group are given 
for the quark's self-energy and 
the gluon's polarization tensor as illustrative examples 
of the present procedure. 
Furthermore, a special attention is given 
for the color contraction
for the colored quark screening frequency 
and the colored gluon Debye mass.
In section \ref{section4}, the HTL self-energy
for the soft colored quark is calculated.
The quark's external momentum is assumed 
soft in comparison with the hard momentum 
of the internal loop. 
In this approximation, when the soft external
momentum is assumed to be of order $g\,T$, 
the momentum of internal loop becomes of order $T$.
The ratio between the soft momentum 
over the hard momentum is of order $g$ (i.e. $p/k\sim g$).
In the calculation course of the quark self-energy,
the color degrees of freedom 
are considered explicitly for the fundamental quarks 
and adjoint gluons.
The calculations of the colored plasma frequency 
and the effective propagator
for the soft colored quark are given in detail.
In section \ref{section5}, 
the HTL polarization tensor 
for the soft colored gluon 
is calculated. 
The gluon's external momentum is taken soft 
in comparison with the hard internal loop momentum.
The external momentum is considered soft 
and of order $p\sim\, g\,T$ while the hard internal 
momentum is taken of order $T$.
The Debye mass and the effective propagator 
for the soft colored gluon are presented.
In section \ref{section6}, 
the effective quark-quark-gluon vertex 
and 
the effective two-quarks and two-gluons vertex 
up to the relevant corrections of order 
$g^{3}$ and $g^{4}$, respectively, are introduced.
In section \ref{section7}, 
the effective self-energy 
and the decay rate for the colored (hard) quark 
are presented to illustrate 
the procedure of the HTL approximation 
when the color degrees of freedom is considered 
explicitly in the calculation 
and they can not be separated 
from the other degrees of freedom.
Finally, the discussions and conclusions
are given in section \ref{section8}.
%
%
%

\section{\label{section1}
Fundamental and Adjoint $SU(N_c)$ Cartan sub-algebra}

A semi-simple Lie algebra can be written 
as a direct sum of simple Lie algebras.
The necessary and sufficient condition to be semi-simple
is that group matrix, namely, $g$ is nonsingular, 
i.e.$\det(g)\neq 0$.
The Cartan metric can tell us directly whether or not
a Lie algebra is semi-simple.
We have the freedom of taking a new basis vectors.
The set of new basis vectors can be written
as linear combinations of the old ones.
The overt form of the commutation relation
is changed when the the basis vectors are transformed
from one set to another.
The commutation relations look rather different,
but of course they are completely equivalent
to the first form.
The Cartan way of presenting the commutation relation
is a generalization of lowering and raising operators.
In the fundamental representation,
the maximal number of the commuting fundamental generators
of the $SU(N_c)$ symmetry group is $N_c-1$ while for $U(N_c)$ is $N_c$.
The maximal set of commuting generators
$\left[{\bf t}_{i},{\bf t}_{j}\right]=0$
form a bases set for the Cartan
sub algebra and the number of such generators
is the rank of group.
The simultaneous eigenvalues of the ${\bf t}_{i}$ will
be used to label
the states of any representation
$\left[{\bf t}_{i},{\bf t}_{j}\right]=0$, where the lower index $i$ refers 
to the fundamental index $i=1,\cdots,N_c-1$ 
for the diagonal fundamental generators.
Analogous to the lowering and raising operator
in the quantum mechanics,
we take linear combination ${{\bf t}^{*}}^{\alpha}$
of the remaining $N^2_c-N_c$ 
generators so that they have
the property of step operators with respect to all of the
${\bf t}_{i}$, namely
\begin{eqnarray}
[{\bf t}^{*}_i,{{\bf t}^{*}}^{\alpha}]\propto {{\bf t}^{*}}^{\alpha}.
\end{eqnarray}
Our objective of the Cartan sub-algebra
is to write the commutation relations
of the fundamental generators in order to achieve
the diagonalizing of the adjoint action
of ${\bf t}_{i}$. The goal is to reach a class of good
quantum numbers requires that the matrix elements 
for $\phi^{a}\,{\bf T}^{a}$
is diagonal in the adjoint representation 
where ${\bf T}^{a}$ 
are the adjoint generators and the upper index $a$ refers to the adjoint index
$a=1,\cdots,N^2_c-1$.
 
\subsection{The group generators in the fundamental representation}

The discussion in the present section is applicable to the theoretic 
symmetry group $SU(N_c)$ or $U(N_c)$ for any $N_c$. 
The focus on the theoretic symmetry group $SU(3)$ is for the sake 
of simplicity and to illustrate the method strategy in a visible way.

The maximal set of the commuting generators for the theoretic 
symmetry group $SU(3)$ with respect to fundamental generators 
$\left[{\lambda}^{3},{\lambda}^{8}\right]=0$, 
where $\left\{{\lambda}^{a}\right\}$
are the Gell-mann matrices, forms a bases set for 
the Cartan sub-algebra. 
The generators 
$\left\{{\lambda}^{a}\right\}$ 
are the common generators 
which are usually adopted in QCD.
The fundamental matrices ${\lambda}^{3}$ and
${\lambda}^{8}$ are real traceless diagonal matrices.
On the other hand, the rest of the fundamental generators 
do not commute with each others  
$\left[{\lambda}^{a},{\lambda}^{b}\right]\neq 0$ 
for $a\neq b$ and 
$a,b=1,2,4,5,6,7$.
The symmetry group $SU(N_c)$ has rank $N_c-1$. 
The group rank is the number of independent 
real traceless diagonal fundamental matrices. 
It is possible to introduce a new set of fundamental 
generators 
${\bf t}_i$, where $i=1,\cdots,N_c-1$. 
In the case of symmetry group $SU(3)$, we have 
$\{{\bf t}_1, {\bf t}_2\}\equiv \{{\lambda}^{3},{\lambda}^{8}\}$.
The non-diagonal fundamental matrices are assigned by 
${{\bf t}^{*}}^{{\cal J}}$ where ${\cal J}=1,\cdots,N^2_c-N_c$.
In the symmetry group SU(3), the new generators ${{\bf t}^{*}}^{\cal J}$ 
are written as linear combinations of the old ones, 
namely, 
${{\bf t}^{*}}^{\cal J}\equiv 
\{{\lambda}^{1},{\lambda}^{2},{\lambda}^{4},
{\lambda}^{5},{\lambda}^{6},{\lambda}^{7}\}$.
It is possible to introduce $N^2_c-N_c$ 
fundamental matrices for $SU(N_c)$.
These new fundamental matrices are defined in the following way
\begin{eqnarray}
{{\bf t}^{*}}^{\cal J}\rightarrow {{\bf t}^{*}}^{\underbrace{(AB)}},  
~~(\mbox{where} A\neq B), 
\end{eqnarray}
whose entries are just a 1 in the $\underbrace{(AB)}$
place and zero elsewhere, i.e.
\begin{eqnarray}
\left({{\bf t}^{*}}^{\underbrace{(AB)}}\right)_{kl}
=\delta_{Ak}\delta_{Bl}.
\end{eqnarray}
The both fundamental-like indexes $A$ and $B$ run over 
$1, \cdots, N_c$.
In the case of $SU(3)$ symmetry group, 
the fundamental generators of this transformation 
are related to the Gell-mann matrices, namely, 
$\{\lambda^{i}\}$ as follows
\begin{eqnarray}
{{\bf t}^{*}}^{\underbrace{(12)}}&=&
({\bf{\lambda}}^{1} + i {\bf{\lambda}}^{2}),
~{{\bf t}^{*}}^{\underbrace{(13)}} =
({\bf{\lambda}}^{4} + i {\bf{\lambda}}^{5}),
~{{\bf t}^{*}}^{\underbrace{(23)}} =
({\bf{\lambda}}^{6} + i {\bf{\lambda}}^{7}),
\nonumber\\
{{\bf t}^{*}}^{\underbrace{(21)}}&=&
({\bf{\lambda}}^{1} - i {\bf{\lambda}}^{2}),
~{{\bf t}^{*}}^{\underbrace{(31)}} =({\bf{\lambda}}^{4} - i {\bf{\lambda}}^{5}),
~{{\bf t}^{*}}^{\underbrace{(32)}} =({\bf{\lambda}}^{6} - i {\bf{\lambda}}^{7}).
\end{eqnarray} 
For the definiteness,
the traceless diagonal matrices ${\bf t}_i$ which are
the bases for the Cartan sub-algebra 
with the index $i$ that runs over $(i=1,\,\cdots,\,N_c-1)$
are defined as follows
\begin{eqnarray}
\left({\bf t}_{i}\right)_{k l}
&=&\delta_{i k}\,\delta_{i l}\,
-\,\frac{1}{N_c}\,\delta_{l k}.
\end{eqnarray}
It is worth to note that for the $U(N_c)$ symmetry group, 
we have the representation 
$\left({\bf t}^{\underbrace{(AB)}}\right)_{ij}=
\delta_{Ai}\,\delta_{Bj}$ 
where $A,B=1,\,\cdots,\,N_c$. 
Furthermore, in the $SU(N_c)$ symmetry group, 
we have the unimodular constraint 
over the $U(N_c)$ symmetry group.
The complete set of fundamental generators 
for the $SU(N_c)$ symmetry group are reduced to following set
\begin{eqnarray}
\{{\bf t}^{a}\}&\rightarrow& 
\left(
\left\{ {{\bf t}^{*}}^{\cal J} \right\}, {\cal J}=1,\cdots\,N^2_c-N_c, 
~\mbox{and}~ 
\left\{{\bf t}_{i}\right\}, i=1,\cdots,N_c-1
\right),
\nonumber\\
&\rightarrow& 
\left(
\begin{array}{l}
\left.
\left\{ {{\bf t}^{*}}^{\underbrace{(AB)}}
\right\}
\right|_{A\neq B}, 
\underbrace{(A B)}=1,\cdots,N^2_c-N_c
\\
\left\{{\bf t}_{i}\right\}, i=1,\cdots,N_c-1
\end{array}
\right),
\end{eqnarray}
where the adjoint index runs over 
$a=1, \cdots, N^2_c-1$.
Nonetheless, 
it is more comfortable in the realistic calculations 
to define new fundamental generators 
that are transformed as follows 
\begin{eqnarray}
x_i {\bf t}_{i}&=& 
\mbox{diag}\left(\theta_1,\theta_2,\cdots,\theta_{N_c}\right),
\end{eqnarray}
where the constraint $\sum^{N_c}_{i=1}\theta_i=0$ or
$\theta_{N_c}=-\sum^{N_c-1}_{i=1}\theta_i$ is imposed.
The new set is related to the old one as follows
\begin{eqnarray}
x_i {\bf t}_{i}&=&\sum^{N_c}_{i=1}\theta_i  \tilde{\bf t}_i,
\nonumber\\
&=&\sum^{N_c-1}_{i=1}\theta_i  \tilde{\bf t}^{'}_i.
\end{eqnarray}
where $\left(\tilde{\bf t}_i\right)_{nm}=\delta_{in}\,\delta_{im}$
or 
$\left(\tilde{\bf t}^{'}_i\right)_{nm}
=\left(\delta_{in}\,-\,\delta_{nN_c}\right)\,\delta_{nm}$.
For the sake of simplification, we define
\begin{eqnarray} 
{\bf t}^{\underbrace{(AA)}}=\tilde{\bf t}_A,
\end{eqnarray}
where the index $A=i$ acts as 
a fundamental-like index and it runs over 
$(1,\,\cdots,\,N_c)$.
Hence, the new generators are related to the
Gell-mann generators as follows
\begin{eqnarray}
\tilde{\bf t}_{1}&=&\frac{\sqrt{3}}{6}
{\bf{\lambda}}^{8}
+\frac{1}{2}{\bf{\lambda}}^{3}+\frac{1}{3}I, 
\nonumber\\
\tilde{\bf t}_{2}&=&\frac{\sqrt{3}}{6}{\bf{\lambda}}^{8}
-\frac{1}{2}{\bf{\lambda}}^{3}+\frac{1}{3}I, 
\nonumber\\
\tilde{\bf t}_{3}&=&\frac{1}{3}I-\frac{\sqrt{3}}{3}
{\bf{\lambda}}^{8},
\end{eqnarray}
where the following constraint is imposed 
\begin{eqnarray}
\sum^{N_c=3}_{i=1}\theta_i&=&0, 
~~\mbox{and}~~ (\theta_1, \theta_2, \theta_3).
\end{eqnarray}
Furthermore, it is possible to define another 
new set of fundamental generators in the following way
\begin{eqnarray}
\tilde{\bf t}{'}_{1}&=&\frac{1}{2}
\left(\sqrt{3}{\bf{\lambda}}^{8}+{\bf{\lambda}}^{3}\right), 
\nonumber\\
\tilde{\bf t}{'}_{2}&=&
\frac{1}{2}
\left(\sqrt{3}{\bf{\lambda}}^{8}-{\bf{\lambda}}^{3}\right), 
\nonumber\\
I,&~& ~\mbox{and}~~ \left(\theta_1, \theta_2\right).
\end{eqnarray}
This means that it is always possible to find a linear combination 
for the diagonal matrices from one form to another.
The diagonal representation of any new set must commute 
with each other
\begin{eqnarray}
\left[{\bf t}_i,{\bf t}_j\right]\equiv
\left[\tilde{\bf t}_i,\tilde{\bf t}_j\right]\equiv
\left[\tilde{\bf t}{'}_i,\tilde{\bf t}{'}_j\right]=0.
\end{eqnarray}

\subsection{Adjoint representation and 
the diagonalizing of the 
$\left(\theta_{a}\,{\bf T}^{a}\right)$}

The simultaneous eigenvalues of the fundamental generators 
${\bf t}_i$ will be used to label the states of any representation.
The commutation relations of the Cartan sub-algebra bases 
can be cast as 
\begin{eqnarray}
\left[{\bf t}_i, {\bf t}_j\right]&=& 0\,{\bf t}_j, 
~~(i, j=1, \cdots, N_c-1).
\end{eqnarray}
The diagonal representation  $x_{i}\,{\bf t}_i$ commutes 
with any linear combination 
of the diagonal set $\{{\bf{t}}_i\}$ as follows
\begin{eqnarray}
[x_i{\bf t}_i,c_j{\bf t}_j]=0.
\end{eqnarray}
The Cartan sub-algebra for the fundamental diagonal 
generators gives zero eigenvalues.
The remaining non-diagonal generators of the fundamental group
can be transformed to another new set that is represented by 
\begin{eqnarray}
{{\bf t}^{*}}^{\cal J}\rightarrow 
{{\bf t}^{*}}^{\underbrace{(AB)}}, 
~~ (A\neq B), (A,B = 1, \cdots, N_c),
\end{eqnarray} 
where ${\cal J}=1,\cdots N^2_c-N_c$ 
and ${\cal J}\equiv \underbrace{(AB)}$. 
These bases have the property of step operator
with respect to all of the ${\bf t}_i$ and satisfy 
the following commutation relation 
\begin{eqnarray}
\left[x_i {\bf t}_i, {{\bf t}^{*}}^{{\cal J}}\right]
\propto {{\bf t}^{*}}^{{\cal J}}
\rightarrow
\left[x_i {\bf t}_i, 
{{\bf t}^{*}}^{\underbrace{(AB)}}\right]
\propto {{\bf t}^{*}}^{\underbrace{(AB)}}.
\end{eqnarray}
Furthermore, it is possible to generalize the commutation relation
to include the both diagonal and non-diagonal 
bases as follows
\begin{eqnarray}
\left[x_j{\bf t}_j,{{\bf t}^{*}}^{a}\right]&\propto& 
{{\bf t}^{*}}^{\alpha},
\nonumber\\
&=& -\lambda {{\bf t}^{*}}^{a}, (a = 1, \cdots, N^2_c-1),
\end{eqnarray}
where the constant $\lambda$ is the proportionality constant 
that is inverted 
to be the eigenvalue of the operator (it should not to be confused 
with the Gell-mann generators ${\lambda}^{a}$). 
The set of fundamental generators $\{{\bf{t}}^{a}\}$ 
with the adjoint index $a$ runs 
over $1, \cdots, N_c^2-1$ 
consists of the non-diagonal fundamental generators
$\{{\bf{t}}^{\cal J}\}$ 
where ${\cal J}$ runs over $1, \cdots, N_c^2-N_c$ 
and the diagonal fundamental generators $\{{\bf{t}}_i\}$
with the fundamental index $i$ runs over $1, \cdots, N_c-1$.
These commutation relations have $N_c-1$ zero eigenvalues.
The fundamental theorem of Cartan is that 
the nonzero eigenvalues are non degenerate.

In order to achieve the diagonalizing of 
the adjoint action of  $x\cdot {\bf T}$, 
it is possible to write commutation relations 
as follows
\begin{eqnarray}
\left[x_j{\bf t}_j,{{\bf t}^{*}}^{a}\right]&=&
x_j\, \left[i\, f_{j a b}\, {{\bf t}^{*}}^{b}\right],\nonumber\\
&=&
-x_j\, \left({\bf T}^{j}\right)_{a b} {{\bf t}^{*}}^{b},
\nonumber\\
&=&
-\left(x\cdot {\bf T}\right)_{a b}\, {{\bf t}^{*}}^{b}.
\end{eqnarray}
The last line is derived due to the fact that
the commutation constants of the fundamental generators
are the adjoint generators of the group,
\begin{eqnarray}
\left({\bf T}^{c}\right)_{a b}=-i\, f_{c a b}.
\end{eqnarray}
When a second order commutation relation is applied, 
the adjoint operation emerges
\begin{eqnarray}
\left[x_i {\bf t}_i,\,
\left[x_j{\bf t}_j,\,{{\bf t}^{*}}^{a}\right]\right]
&=& -\lambda \left[x_j {\bf t}_j\,,\, {{\bf t}^{*}}^{a}\right],
\nonumber\\
&=& -\lambda\,\delta_{a b}\, 
\left[x_j {\bf t}_j\,,\, {{\bf t}^{*}}^{b}\right],\nonumber\\
&=&
-\left(x\cdot {\bf T}\right)_{a b}\,
\left[x_j {\bf t}_j\,,\,{{\bf t}^{*}}^{b}\right].
\end{eqnarray}
The diagonalizing of the adjoint 
generator is done in the following way
\begin{eqnarray}
\left[\left(x\cdot {\bf T}\right)_{a b}\,-\,
\lambda\delta_{a b}\right]\,X=0,
\end{eqnarray}
with the eigenvector
\begin{eqnarray}
X=\left[x_j {\bf t}_j\,,\,{{\bf t}^{*}}^{b}\right].
\end{eqnarray}
This means that we are looking for the eigenvalues 
of the following equation
\begin{eqnarray}
\det\left(
\left(x\cdot {\bf T}\right)_{a b}-
\lambda\delta_{a b}
\right)=0.
\end{eqnarray}
It is noted that $x_i\, f_{i a b}$ 
is real when $x_i$ is assumed to be real.
Hence, the operator
$x_i\,\left({\bf T}^{i}\right)_{a b}$ 
is pure imaginary and antisymmetric 
and therefore it is Hermitian matrix with real eigenvalues. 
The eigenvalues of the adjoint matrix are identified by $N_c-1$ 
zero eigenvalues
and $N^2_c-N_c$ nonzero eigenvalues.
In order to determine the eigenvalues, 
we assume the following algebra
\begin{eqnarray}
\{ {{\bf t}^{*}}^{\cal J}\}&\equiv& 
\{ {{\bf t}^{*}}^{\underbrace{(AB)}} \},
~~(\mbox{with}~~ 
A\neq B, ~\mbox{and}~ {\cal J} = 1,\cdots, N^2-N_c),
\nonumber\\
\left[x_i {\bf t}_i,
{{\bf t}^{*}}^{\underbrace{(AB)}} \right]
&=&
\left[x_c {\bf t}^{c},
{{\bf t}^{*}}^{\underbrace{(AB)}} \right],
~~(\mbox{with}~~ i=1,\cdots, N_c-1).
\end{eqnarray}
The commutation relation is reduced to
\begin{eqnarray}
\left[x_i {\bf t}_i,
{{\bf t}^{*}}^{\underbrace{(AB)}} \right]
&=&
\sum^{N_c}_{C=1} \theta_C \hat{e}_C
\cdot
\left(\hat{e}_A -\hat{e}_B\right)
{{\bf t}^{*}}^{\underbrace{(AB)}},
\nonumber\\
&=&
-\left(\theta_{B}-\theta_{A}\right)
{{\bf t}^{*}}^{\underbrace{(AB)}}.
\end{eqnarray}
The affine transformation bases 
$\hat{e}_{A}$ and $\hat{e}_{B}$ 
are unit $N_c$-dimensional vectors pointing 
in the $A$ and $B$ directions respectively.
The representation ${{\bf t}^{*}}^{\underbrace{(AB)}}$ 
is the root generator
(i.e. vector).
The summary of the commutation relations with $N_c-1$ zero 
eigenvalues and
$N_c^2-N_c$ nonzero eigenvalues is given by
\begin{eqnarray}
\left[x\cdot {\bf t}, {\bf t}^{a} \right]\rightarrow
\left(\begin{array}{c}
\left[x\cdot {\bf t}, 
{{\bf t}^{*}}^{\underbrace{(AB)}} \right]=
-(\theta_B-\theta_A) 
{{\bf t}^{*}}^{\underbrace{(AB)}} \\ 
\left[x\cdot {\bf t}, {\bf t}_{i} \right]= 0 {\bf t}_{i}
\end{array}\right).
\end{eqnarray}
The resulting eigenvalues read
\begin{eqnarray}
\lambda\equiv\{ \lambda^{a} \}=
\mbox{diag}
\left(\left\{
\left(\theta_B-\theta_A\right)_{A\neq B}
\right\}_{1,\cdots,\,N^2_{c}-N_{c}}, \{0\}_{1,\cdots,N_c-1}\right), 
\end{eqnarray}
or any permutation of the diagonal elements, 
where $a=1\,\cdots\, N^2_c-1$. 
In the realistic calculation
such as $SU(3)$, 
the zero eigenvalues are for the generators 
$\lambda^{3}$ and 
$\lambda^{8}$ and the other eigenvalues are nonzero 
for the generators $\lambda^{1}$, $\lambda^{2}$,
$\lambda^{4}$, $\lambda^{5}$, $\lambda^{6}$ and $\lambda^{7}$
and according
to the fundamental theorem of Cartan: the nonzero eigenvalues 
are non degenerate.
In case of $U(3)$ and $SU(3)$, 
the fundamental and adjoint representations read
\begin{eqnarray}
x\cdot {\bf t}&=&
\left(\begin{array}{ccc}
\theta_1&0&0\\
0&\theta_2&0\\
0&0&\theta_3
\end{array}\right),
\end{eqnarray}
and
\begin{eqnarray}
x\cdot{\bf T}&=&
\left(\begin{array}{cccccccc}
(\theta_1-\theta_2)&0&0&0&0&0&0&0\\
0&(\theta_2-\theta_1)&0&0&0&0&0&0\\
0&0&0&0&0&0&0&0\\
0&0&0&(\theta_1-\theta_3)&0&0&0&0\\
0&0&0&0&(\theta_3-\theta_1)&0&0&0\\
0&0&0&0&0&(\theta_2-\theta_3)&0&0\\
0&0&0&0&0&0&(\theta_3-\theta_2)&0\\
0&0&0&0&0&0&0&0
\end{array}
\right),
\end{eqnarray}
where $\theta_3=-(\theta_1+\theta_2)$ for $SU(3)$.
It should be noted that the operators 
$x\cdot {\bf t}$ and $x\cdot {\bf T}$ 
can be transformed from one form to another 
and even the diagonal elements can be permuted
in different ways
but a thorough consideration must be followed in order 
to preserve the conservation of the color charges.
%
%
%
%
%
%
%
%
%
%
\section{\label{section2}
The thermal propagators in the mixed-time representation}

The Lagrangian for the non-Abelian fundamental 
and adjoint particles in the theoretic $SU(N_c)$ 
symmetry group reads,
\begin{eqnarray}
{\cal L}=
-\frac{1}{4} {F^{\mu\nu}}^{a} {F_{\mu\nu}}^{a}
+\sum_{f}\overline{\psi}_{f}\left(
{i}\gamma^{\mu}\,{D_{\mu}}-m\right)\psi_{f}.
\end{eqnarray}
The fundamental operator acting of the fundamental field 
is given by
\begin{eqnarray}
D_{\mu}=\partial_{\mu} - {i}\, g\, {A_{\mu}}^{a}\, {\bf t}^{a},
\end{eqnarray}
while the adjoint field is given by
\begin{eqnarray}
{F_{\mu\nu}}^{a}&=&[D_{\mu},D_{\nu}]/{i} g\nonumber\\
&=&\partial_{\mu} {A_{\nu}}^{a}-
\partial_{\nu}{A_{\mu}}^{a}
+g\, f^{abc}\, {A_{\mu}}^{b}\, {A_{\nu}}^{c},
\end{eqnarray}
where the adjoint index $a$ runs 
over ($1,\,\cdots,\,N^2_c-1$).
%
%
%
\subsection{Thermal Quark propagator}
The propagator, for free fundamental particle in the limit 
$g\rightarrow 0$, reads
\begin{eqnarray}
~~~ ~~~ {{\cal S}_0}_{Q}(k)&=&
i\frac{1}{\left(\gamma_0 k_0 -\vec{\gamma}\cdot \vec{k}-m_{Q}\right)},
\nonumber\\
\mbox{or}~~
i\,{{\cal S}_0}_{Q}(k)&=&
\frac{-1}{\left(\gamma_0 k_0 -\vec{\gamma}\cdot \vec{k}-m_{Q}\right)}.
\end{eqnarray}
When the fundamental particles are embedded 
in the $SU(N_c)$ color symmetry group representation,
it is transformed 
as follows 
\begin{eqnarray}
k_0\rightarrow k_0+ i \frac{1}{\beta} \theta\cdot {\bf t}.
\end{eqnarray}
Under the flavor charge conservation 
$U(1)_{\mbox{B}}$, it is transformed to
\begin{eqnarray}
k_0&\rightarrow& k_0 + \mu_{Q} + i\frac{1}{\beta} 
\theta\cdot {\bf t},   
\end{eqnarray}
where $\mu_{Q}=\mu_{\mbox{B}}+\mu_{\mbox{S}}+\cdots$ 
is the flavor chemical potential that is satisfying the (flavor-) charge
conservation
such as the baryonic and strangeness charges.
In the context of the diagonal fundamental representation 
transformation 
with fundamental operators that commute with
the Hamiltonian of fundamental particles, 
the fundamental representation is reduced to 
\begin{eqnarray}
\theta\cdot {\bf t}=
\left(
\begin{array}{ccc}
\theta_1 & 0 & 0 \\
\vdots & \ddots & \vdots \\
0 & 0 &\theta_{N_c}
\end{array}
\right), ~~\mbox{with constraint} 
~\sum^{N_c}_{i=1} \theta_i= 0.
\end{eqnarray}
This leaves $N_c-1$ conservative color charges. 
It is more convenience for the practical calculation 
to factorize the propagator of the fundamental 
particle into positive and negative energy 
components as follows
\begin{eqnarray} 
{\cal S}_{Q}(k_0,\vec{k})
&=&~ 
{{\cal S}_0}_{Q}\left(
k_0+\mu_{Q}+\frac{1}{\beta}i\theta\cdot {\bf t},\vec{k}
\right),
\nonumber\\
&=&~
i\frac{1}{2}
\left[\gamma_0-
\frac{1}{\epsilon_{Q}(\vec{k})}
\left(\vec{k}\cdot\vec{\gamma}-m_{Q}\right)\right]
\frac{1}
{k_0-\left[\epsilon_{Q}(\vec{k})
-\left(\mu_{Q}+i\frac{1}{\beta}\theta\cdot {\bf t}
\right)\right]}
\nonumber\\
&~&+
i\frac{1}{2}
\left[\gamma_0+
\frac{1}{\epsilon_{Q}(\vec{k})}
\left(\vec{k}\cdot\vec{\gamma}-m_{Q}\right)\right]
\frac{1}
{k_0+\left[\epsilon_{Q}(\vec{k})
+\left(\mu_{Q}+i\frac{1}{\beta}\theta\cdot {\bf t}
\right)\right]}.
\end{eqnarray}
It can be written in terms of the
Foldy-Wouthuysen decomposition:
\begin{eqnarray} 
{\cal S}_{Q}(k_0,\vec{k})
&=&
i\left[
\frac{{\Lambda^{(+)}_{Q}}(\vec{k})\,\gamma_0}
{k_0-\left[\epsilon_{Q}(\vec{k})
-\left(\mu_{Q}+i\frac{1}{\beta}\theta\cdot {\bf t}
\right)\right]}
+
\frac{{\Lambda^{(-)}_{Q}}(\vec{k})\,\gamma_0}
{k_0+\left[\epsilon_{Q}(\vec{k})
+\left(\mu_{Q}+i\frac{1}{\beta}\theta\cdot {\bf t}
\right)\right]}
\right].
\end{eqnarray}
The projections ${\Lambda^{(\pm)}_{Q}}(\vec{k})$ 
are the Foldy-Wouthuysen 
positive and negative energy projections 
and they are given, 
respectively, by
\begin{eqnarray}
{\Lambda^{(+)}_{Q}}(\vec{k})&=&
\frac{1}{2\epsilon_{Q}(\vec{k})}
\left[
\epsilon_{Q}(\vec{k})
+\gamma_0\,\left(
\vec{\gamma}\cdot\vec{k}+m_{Q}
\right)\right],
\nonumber\\
{\Lambda^{(-)}_{Q}}(\vec{k})&=&
\frac{1}{2\epsilon_{Q}(\vec{k})}
\left[
\epsilon_{Q}(\vec{k})
-\gamma_0\,\left(
\vec{\gamma}\cdot\vec{k}+m_{Q}
\right)\right].
\end{eqnarray}
Furthermore, under the diagonal transformation of the $SU(N_c)$ 
with color charges commuting with the particle energy, 
the fundamental particle propagator is 
transformed as follows 
\begin{eqnarray} 
{ {{\cal S}_{Q}}_{ij} }(k_0,\vec{k})
&=&i\,\delta_{ij}\,
\left[
\frac{ {\Lambda^{(+)}_{Q}}(\vec{k})\,\gamma_0 }
{k_0-\left[\epsilon_{Q}(\vec{k})
-\left(\mu_{Q}+i\frac{\theta_{j}}{\beta}\right)\right]}
+
\frac{ {\Lambda^{(-)}_{Q}}(\vec{k})\,\gamma_0 }
{k_0+\left[\epsilon_{Q}(\vec{k})
+\left(\mu_{Q}+i\frac{\theta_{j}}{\beta}\right)\right]}
\right],
\end{eqnarray}
where $i$ and $j$ are the fundamental indexes of the colored particles.
In order to proceed to the thermal field framework, 
we adopt the imaginary-time formalism. 
In the context of the imaginary-time formalism, 
the Matsubara frequency for the fermion 
is introduced 
\begin{eqnarray}
\omega_n=(2n+1)\pi/\beta.
\end{eqnarray}
The momentum's time-component is transformed 
to
\begin{eqnarray}
k_0\rightarrow 
i\,\omega_n + 
\mu_{Q} + i\,\frac{1}{\beta}\theta\cdot {\bf t}.
\end{eqnarray}
It is more convenience to work in the mixed-time representation 
in the context of the imaginary-time formalism.
The quark propagator is reduced to
\begin{eqnarray}
{{\cal S}}_{Q}(\tau,\vec{k})&=&
\frac{1}{\beta}
\sum_n e^{-i\omega_n \tau}\,
{{\cal S}_0}_{Q}(i\omega_n+\mu_{Q}
+i\,\frac{1}{\beta}\theta\cdot{\bf t},\vec{k}),
\nonumber\\
&=& -\oint \frac{d\omega}{2 i\pi} 
\frac{e^{-\omega\tau}}{e^{-\beta\omega}+1}\, 
{{\cal S}_0}_{Q}(\omega+\mu_{Q}+i\frac{1}{\beta}\theta\cdot{\bf t},\vec{k}).
\end{eqnarray}
The above equation is equivalent to the following form
\begin{eqnarray}
{{\cal S}}_{Q}(\tau,\vec{k})&=& 
-\oint \frac{d\omega}{2 i\pi} 
\frac{e^{\tau\left(\mu_{Q}+i\frac{1}{\beta}\theta\cdot{\bf t}\right)}
\,e^{-\omega\tau}}
{e^{\beta\left(\mu_{Q}+i\frac{1}{\beta}\theta\cdot{\bf t}\right)}
\,e^{-\beta\omega}+1}
\,
{{\cal S}_0}_{Q}(\omega,\vec{k}).
\end{eqnarray}
The quark propagator is decomposed into the $SU(N_c)$ 
fundamental color components
as follows 
\begin{eqnarray}
{ {{\cal S}_{Q}}_{ij} }(\tau,\vec{k})
&=& -\oint \frac{d\omega}{2i\pi} 
\left(\frac{
e^{\tau\left(\mu_{Q}+i\frac{1}{\beta}\theta_i\right)}
e^{-\omega\tau}
}
{
e^{\beta\left(\mu_{Q}+i\frac{1}{\beta}\theta_i\right)}
e^{-\beta\omega}+1
}
\,{{\cal S}_0}_{Q}(\omega,\vec{k})\right)\,  \delta_{ij},
\nonumber\\
&~&~~~~~~~~~~\mbox{with the constraint} 
~\sum^{N_c}_{i=1}\theta_i=0.
\end{eqnarray}
The integral is evaluated using the method 
of the residues in the complex calculus.
The two poles are found at 
\begin{eqnarray}
k_{0}&=&k^{(+)}_{0}=\left(
\epsilon_{Q}(\vec{k})-\mu_{Q}-i\frac{1}{\beta}\theta_{i}
\right) ~\mbox{and}~~~ 
k_{0}=k^{(-)}_{0}=-\left(
\epsilon_{Q}(\vec{k})+\mu_{Q}+i\frac{1}{\beta}\theta_{i}
\right).
\end{eqnarray}
The quark propagator in the mixed-time representation 
is evaluated and projected into positive 
and negative energy components 
as follows
\begin{eqnarray}
i\,{ {{\cal S}_{Q}}_{ij} }(\tau,\vec{k})
&=&
\delta_{ij}\,\left\{
{\Lambda^{(+)}_{Q}}(\vec{k})\,\gamma_{0}\,
\left[
1-n_{F}\left(
\epsilon_{Q}(\vec{k})-\mu_{Q}-i\frac{1}{\beta}\theta_i
\right)\right]
\right.
\nonumber\\
&~&
~~~ ~~~\times\exp\left[-\left(
\epsilon_{Q}(\vec{k})-\mu_{Q}-i\frac{1}{\beta}\theta_i
\right)\tau
\right]
\nonumber\\
&~&
+
{\Lambda^{(-)}_{Q}}(\vec{k})\,\gamma_0\, 
\left[n_{F}\left(\epsilon_{Q}(\vec{k})
+\mu_{Q}+i\frac{1}{\beta}\theta_i\right)\right]\nonumber\\
&~&
\left.
~~~ ~~~\times\exp\left[+\left(
\epsilon_{Q}(\vec{k})+\mu_{Q}+i\frac{1}{\beta}\theta_i
\right)\tau
\right]\right\}.
\end{eqnarray}

%
%
%
\subsection{Thermal gluon propagator}

The gluon propagator can be introduced in several ways 
depending on the type of the gauge. 
The straightforward way is to write it as follows
\begin{eqnarray}
{{\cal G}_{\mu\nu}}(k_0,\vec{k})
&=& {{\cal G}}(k_0,\vec{k}) 
\widehat{g}_{\mu\nu}\left(k_0,\vec{k}\right).
\end{eqnarray}
The scalar propagator ${\cal G}(k_0,\vec{k})$ reads
\begin{eqnarray}
{{\cal G}}(k_0,\vec{k})&=& 
\frac{-1}{k^2_0-\vec{k}^2-m^2_G},\nonumber\\
&=&\frac{-1}{2\epsilon_G(\vec{k})}
\left[\frac{1}{k_0-\epsilon_{G}(\vec{k})}
-\frac{1}{k_0+\epsilon_{G}(\vec{k})}\right],
\end{eqnarray}
where ${\epsilon_{G}}(\vec{k})=\sqrt{\vec{k}^2+m_G^2}$ 
and $m_{G}=0$ for the gluon in the vacuum.
The gluon propagator numerator 
$\widehat{g}_{\mu\nu}\left(k_0,\vec{k}\right)$ 
is gauge dependent. It can be written 
in the covariant gauge with a gauge-fixing 
$(\partial^{\mu} A^a_{\mu})^2/2\xi$ as 
$\left[-g_{\mu\nu}+\xi k_{\mu}k_{\nu}/k^2\right]$. 
In the Feynman gauge: we have
$\widehat{g}_{\mu\nu}\left(k_0,\vec{k}\right)=-g_{\mu\nu}$.
Nonetheless, it has been shown the quark and gluon self-energies
are gauge independent. 
In the sake of the calculation simplicity, 
we shall the present calculations in the Feynman gauge.
The same procedure is also valid for the covariant gauge.
Fortunately, the final results is independent on the gauge
in the HTL approximation.
The gluons are adjoint particles of the $SU(N_c)$ 
symmetry group. 
The adjoint representation for the gluon 
is given by the following transformation 
\begin{eqnarray}
k_0&\rightarrow& k_0+\frac{1}{\beta}\theta_i f_{iab},
\nonumber\\
&\rightarrow& 
k_0 + i\frac{1}{\beta}\left(\theta \cdot {\bf T}\right)_{ab}. 
\end{eqnarray} 
In the imaginary-time formalism, the Matsubara frequency 
for gluon is introduced as follows
\begin{eqnarray}
k_0&\rightarrow& 
i \omega_n + i\frac{1}{\beta}
\left(\theta \cdot {\bf T}\right)_{ab}, 
\end{eqnarray}
where $\omega_n=2 n\pi\beta$ is the even Matsubara frequency.
Furthermore, in the context of the mixed-time representation,
the gluon propagator is transformed in the following way
\begin{eqnarray}
{\cal G}_{\mu\nu}(\tau,\vec{k})&=&\frac{1}{\beta}
\sum_n e^{-i \omega_n \tau}
{\cal G}_{\mu\nu}
\left(
i\omega_n+i\frac{1}{\beta}\theta\cdot {\bf T},\vec{k}
\right),
\nonumber\\
&=&
\frac{1}{(2\pi i)}\oint d\omega \frac{e^{-\omega\tau}}
{e^{-\beta\omega}-1}
{\cal G}_{\mu\nu}\left(
\omega+i\frac{1}{\beta}\theta\cdot{\bf T},\vec{k}
\right),
\nonumber\\
&=&
\frac{1}{(2\pi i)}\oint d\omega \frac{
e^{-\left(\omega-i\frac{1}{\beta}\theta\cdot{\bf T}\right)\tau}
}
{
e^{-\beta\left(
\omega-i\frac{1}{\beta}\theta\cdot{\bf T}
\right)}
-1}
{\cal G}_{\mu\nu}\left(
\omega,\vec{k}
\right).
\end{eqnarray}
The adjoint representation in the theoretic $SU(N_c)$ 
symmetry group can be transformed into a diagonal matrix. 
The advantage of the diagonal adjoint representation 
is that it commutes with the diagonal fundamental representation
beside its commutation with the adjoint particle's energy state. 
Furthermore, it preserves the $SU(N_c)$ fundamental color charges 
$\theta_i$ 
with the unimodular constraint 
$\sum^{N_c}_{i=1}\theta_{i}=0$. 
To this end, we get the following diagonal transformation
\begin{eqnarray}
\left(\theta\cdot{\bf T}\right)_{ab}\rightarrow \phi^{a}\, \delta^{ab}.
\end{eqnarray}
For the definiteness, the adjoint representation can be diagonalized 
and written in terms of the conserved fundamental color 
charges $\theta_i$ as follows
\begin{eqnarray}
\mbox{tr} \left[e^{i\theta\cdot{\bf T}}\right]&=&
\sum^{N^2_c-1}_{a=1} e^{i\phi^{a}},
\nonumber\\
&=&
\sum^{N_c}_{ij} e^{i(\theta_i-\theta_j)}-1.
\end{eqnarray}
The diagonal adjoint representation elements 
are good quantum numbers.
The conserved adjoint charges depend basically 
on the conserved fundamental
charges $\{\phi^{a}\}=\{(\theta_i-\theta_j)\}$ where 
$i, j=1, \cdots, N_c$ are fundamental indexes 
while $a=\underbrace{(ij)}=1, \cdots, (N^2_c-1)$
is the adjoint index.
The gluon propagator in the mixed-time representation
can be transformed to a diagonal adjoint 
color matrix as follows
\begin{eqnarray}
{ {{\cal G}_{\mu\nu}}^{ab} }(\tau,\vec{k})
&=&
\frac{1}{(2\pi i)}\oint d\omega \frac{
e^{-\left(\omega-i\frac{1}{\beta}\phi^{a}\right)\tau}
}
{
e^{-\beta\left(
\omega-i\frac{1}{\beta}\phi^{a}
\right)}
-1}
\,{\cal G}_{\mu\nu}\left(
\omega,\vec{k}
\right)\,\delta^{ab}.
\end{eqnarray}
It is noted that the gluon propagator is gauge dependent 
and can be written as follows
\begin{eqnarray}
{ {{\cal G}_{\mu\nu}}^{ab} }(\tau,\vec{k})
&=&
\frac{1}{(2\pi i)}\oint d\omega \frac{
e^{-\left(\omega-i\frac{1}{\beta}\phi^{a}\right)\tau}
}
{
e^{-\beta\left(
\omega-i\frac{1}{\beta}\phi^{a}
\right)}
-1}
\,{\cal G}\left(
\omega,\vec{k}
\right)
\, 
\widehat{g}_{\mu\nu}(\omega,\vec{k})\,\delta^{ab}.
\end{eqnarray}
In the Feynman gauge, it is reduced to
\begin{eqnarray}
{ {{\cal G}_{\mu\nu}}^{ab} }(\tau,\vec{k})
&=&
{ {\cal G}^{ab} }(\tau,\vec{k}) g_{\mu\nu}
\nonumber\\
&=&
\frac{1}{(2\pi i)}\oint d\omega \frac{
e^{-\left(\omega-i\frac{1}{\beta}\phi^{a}\right)\tau}
}
{
e^{-\beta\left(
\omega-i\frac{1}{\beta}\phi^{a}
\right)}
-1}
\,
{\cal G}\left(
\omega,\vec{k}
\right)
\, \delta^{ab}\, g_{\mu\nu}.
\end{eqnarray} 
Hence, the scalar part of the gluon propagator is given by
\begin{eqnarray}
{\cal G}^{ab}(\tau,\vec{k})
&=&
\frac{1}{(2\pi i)}\oint d\omega \frac{
e^{-\left(\omega-i\frac{1}{\beta}\phi^{a}\right)\tau}
}
{
e^{-\beta\left(
\omega-i\frac{1}{\beta}\phi^{a}
\right)}
-1}
\,
{\cal G}\left(
\omega,\vec{k}
\right)\, \delta^{ab}.
\end{eqnarray} 
The gluon propagator in the mixed-time representation 
is evaluated trivially by using the complex calculus of residues. 
The poles are found in the following locations 
\begin{eqnarray}
k_0&=&k^{(+)}_{0}=\epsilon_{G}(\vec{k})-i\frac{1}{\beta}\phi^{a},
~\mbox{and}~
k_0=k^{(-)}_{0}=
-
\left(
\epsilon_{G}(\vec{k})+i\frac{1}{\beta}\phi^{a}
\right).
\end{eqnarray}
After evaluating the integral, 
the scalar part of the gluon propagator becomes
\begin{eqnarray}
{\bf {\cal G}}^{ab}(\tau,\vec{k})
&=&
\,\delta^{ab}\,
\frac{1}{2\epsilon_{G}(\vec{k})}
\left[
1+N_G\left(\epsilon_{G}(\vec{k})-i\frac{1}{\beta}\phi^{a}
\right)\right]
e^{-\left(\epsilon_{G}(\vec{k})-i\frac{1}{\beta}\phi^{a}\right)\tau}
\nonumber\\
&~&
+
\,\delta^{ab}\,
\frac{1}{2\epsilon_{G}(\vec{k})}
\left[
N_G\left(\epsilon_{G}(\vec{k})+i\frac{1}{\beta}\phi^{a}
\right)
\right]
e^{+\left(\epsilon_{G}(\vec{k})+i\frac{1}{\beta}\phi^{a}\right)\tau}.
\end{eqnarray}
It is interesting to note that it is possible to generalize 
the standard procedure used in the Feynman gauge to other 
gauge whereas the gauge does not affect the propagator singularity. 
The generalization of the gluon propagator is written as follows
\begin{eqnarray}
{ {{\cal G}_{\mu\nu}}^{a\,b} }(\tau,\vec{k})&=&
\delta^{a\,b}\,{ {{\cal G}_{\mu\nu}}^{a}}(\tau,\vec{k}),
\nonumber\\
{ {{\cal G}_{\mu\nu}}^{a} }(\tau,\vec{k})
&=&\frac{1}{2\epsilon_{G}(\vec{k})}
\left[
1+N_G\left(\epsilon_{G}(\vec{k})-i\frac{1}{\beta}\phi^{a}
\right)\right]
\exp\left[
-\left(\epsilon_{G}(\vec{k})-i\frac{1}{\beta}\phi^{a}\right)\tau
\right]
\,
\widehat{g}_{\mu\nu}\left(k^{(+)}_{0}\right)
\nonumber\\
&~&
+\frac{1}{2\epsilon_{G}(\vec{k})}
\left[
N_G\left(\epsilon_{G}(\vec{k})+i\frac{1}{\beta}\phi^{a}
\right)
\right]
\exp\left[
\left(\epsilon_{G}(\vec{k})+i\frac{1}{\beta}\phi^{a}\right)\tau
\right]
\,
\widehat{g}_{\mu\nu}\left(k^{(-)}_{0}\right).
\end{eqnarray}
In the mixed-time representation of the imaginary-time formalism,
it is useful to introduce some notations 
and to write the following results
\begin{eqnarray}
\left[k^2{\cal G}\right]^{ab}\left(\tau,\vec{k}\right)&=&
\left(\frac{1}{2\pi i}\oint d\omega 
\frac{
e^{i \frac{\tau}{\beta}\theta\cdot{\bf T}}
e^{-\omega\tau} 
}
{
e^{i\theta\cdot{\bf T}} 
e^{-\beta\omega}-1}
k^2 {{\cal G}_{0}}^{ab}(\omega,\vec{k})\right),
\nonumber\\
&=&-\,\delta(\tau)\,\delta^{a b},
\end{eqnarray}
and
\begin{eqnarray}
\left[k^2_0{\cal G}\right]^{ab}(\tau,\vec{k})
&=&
\frac{1}{2\pi i}
\left(
\oint d\omega 
\frac{
e^{i \frac{\tau}{\beta}\theta\cdot{\bf T}}
e^{-\omega\tau} 
}
{
e^{i\theta\cdot{\bf T}} 
e^{-\beta\omega}-1}
\omega^{2}
{ {{\cal G}_{0}}^{ab} }(\omega,\vec{k})\right),
\nonumber\\
&=&
\,\frac{\epsilon_G(k)}{2}
\left[
1+N_G\left(\epsilon_G(k)-i\frac{1}{\beta}\phi^{a}
\right)\right]\exp\left[
-\left(\epsilon_{G}(k)-i\frac{1}{\beta}\phi^{a}
\right)\tau\right]
\,\delta^{a b}
\nonumber\\
&~&
+\frac{\epsilon_{G}(k)}{2}
\left[
N_G\left(\epsilon_{G}(k)+i\frac{1}{\beta}\phi^{a}
\right)
\right]\exp\left[+\left(
\epsilon_{G}(k)+i\frac{1}{\beta}\phi^{a}
\right)\tau\right]
\,\delta^{a b},
\end{eqnarray}
and finally,
\begin{eqnarray}
\left[k_0{\cal G}\right]^{ab}(\tau,\vec{k})
&=&
\left(\frac{1}{2\pi i}\oint d\omega 
\frac{
e^{i \frac{\tau}{\beta}\theta\cdot{\bf T}}
e^{-\omega\tau} 
}
{
e^{i\theta\cdot{\bf T}} 
e^{-\beta\omega}-1}
\omega { {{\cal G}_0}^{ab} }(\omega,\vec{k})\right),
\nonumber\\
&=&
~~\frac{1}{2}
\left[
1+N_G\left(\epsilon_{G}(k)-i\frac{1}{\beta}\phi^{a}
\right)\right]
\exp\left[-\left(
\epsilon_{G}(k)-i\frac{1}{\beta}\phi^{a}
\right)\tau\right]
\,\delta^{a b}
\nonumber\\
&~&
-\frac{1}{2}
\left[
N_G\left(\epsilon_{G}(k)+i\frac{1}{\beta}\phi^{a}
\right)
\right]
\exp\left[+\left(
\epsilon_{G}(k)+i\frac{1}{\beta}\phi^{a}
\right)\tau\right]
\,\delta^{a b}.
\end{eqnarray}
\section{\label{section3}
Color contraction}
\subsection{Quark self-energy}
In the calculations of the quark-self energy and the quark plasma frequency
(i.e. Landau frequency),
the color structure for the plasma Landau frequency is displayed 
as follows
\begin{eqnarray}
\left(
\omega^{2}_{0\,Q}
\right)_{ij}&=&
{\bf t}^{a}_{in}\, {\bf t}^{b}_{mj}\,
{\left(
\omega^{2}_{0\,Q}
\right)}^{ab}_{nm},
\nonumber\\
&=&
{\bf t}^{a}_{in}\, {\bf t}^{b}_{mj}\, 
\delta^{a b}\, \delta_{n m}\,
{\left(
\omega^{2}_{0\,Q}
\right)}^{a}_{n},
\nonumber\\
&=&
{\bf t}^{a}_{in}\, ({\bf t}^{b\dagger})_{mj}\, 
\delta^{a b}\,\delta_{n m}\,
{\left(
\omega^{2}_{0\,Q}
\right)}^{a}_{n},
\nonumber\\
&=&
{\bf t}^{a}_{in}\, {\bf t}^{b}_{mj}\, 
\left[
{\left(
\omega^{2}_{0\,Q (Q)}
\right)}_n
+
{\left(
\omega^{2}_{0\,Q (G)}
\right)}^{a}
\right]\, 
\delta^{a b}\,\delta_{nm}.
\end{eqnarray}
The contraction of the fundamental indexes reduces
the quark plasma frequency to
\begin{eqnarray}
{\left(
\omega^{2}_{0\,Q}
\right)}^{a b}_{nm}
&=&
{\left(\omega^{2}_{0\,Q}\right)}^{a}_{n}\,
\delta^{a b}\,
\delta_{n m},
\end{eqnarray}
where
\begin{eqnarray}
{\left(
\omega^{2}_{0\,Q}
\right)}^{a}_{n}&=&
\left[
{\left(\omega^{2}_{0\,Q (G)}\right)}^{a}
+
{\left(\omega^{2}_{0\,Q (Q)}\right)}_{n}
\right].
\end{eqnarray}
The fundamental generators are defined in the following way
\begin{eqnarray}
{\bf t}^{a}&=&\frac{1}{\sqrt{2}}\tau^{a},\nonumber\\
     &=& \frac{1}{\sqrt{2}} \tau^{\underbrace{(AB)}}, 
~\mbox{where}~ a\equiv \underbrace{(AB)}, 
\end{eqnarray}
where the matrix elements are expressed as follows
\begin{eqnarray}
{\bf t}^{\underbrace{(AB)}}_{ij} &=& 
\frac{1}{\sqrt{2}}\left[
\delta_{A i}\,\delta_{B j}
-\frac{1}{N_c}\,\delta_{A B}\,\delta_{i j}
\right].
\end{eqnarray}
The Hermitian fundamental matrices are represented by
\begin{eqnarray}
({\bf t}^{\underbrace{(A B)}\dagger})_{i j}&=& 
{\bf t}^{\underbrace{(B A)}}_{i j},
\nonumber\\
&=&\frac{1}{\sqrt{2}}\left[
\delta_{Bi}\,\delta_{Aj}
-\frac{1}{N_c}\,\delta_{A B}\,\delta_{i j}
\right]. 
\end{eqnarray}
The color contraction of the quark segment of the quark self-energy 
is reduced to 
\begin{eqnarray}
{\bf t}^{a}_{i n}\, {\bf t}^{b}_{m j}\,
{\left(\omega^{2}_{0\,Q (Q)}\right)}_n
\,\delta_{n m}\,\delta^{a b}
&\equiv& 
\frac{1}{2} 
\tau^{a}_{in}\, (\tau^{a\dagger})_{nj}\, 
{\left(\omega^{2}_{0\,Q (Q)}\right)}_{n},
\nonumber\\
&=&
\frac{1}{2} 
\tau^{\underbrace{(AB)}}_{in}\, \tau^{\underbrace{(BA)}}_{nj}\, 
{\left(\omega^{2}_{0\,Q (Q)}\right)}_n.
\end{eqnarray}
It becomes
\begin{eqnarray}
{\bf t}^{a}_{in}\, {\bf t}^{b}_{mj}\, 
{\left(\omega^{2}_{0\,Q (Q)}\right)}_n\, 
\delta_{nm}\,\delta^{ab}
&=&
\frac{N^2_c-1}{2N_c}\,\delta_{ij}\, 
{\left({\omega}^{*\,2}_{0\,Q (Q)}\right)}_{i},
\end{eqnarray}
where
\begin{eqnarray}
{\left({\omega}^{*\,2}_{0\,Q (Q)}\right)}_{i}
&=&\frac{1}{N^2_c-1}\,
\left[
N_c\,\sum^{N_c}_{n=1} 
{\left({\omega}^{2}_{0\,Q (Q)}\right)}_{n}
-
{\left({\omega}^{2}_{0\,Q (Q)}\right)}_{i}
\right],
\nonumber\\
&=&
\frac{N_c}{N^2_c-1}\,
\left[
\sum^{N_c}_{n=1}
\left(
1-\frac{\delta_{in}}{N_c}
\right)\,
{\left({\omega}^{2}_{0\,Q (Q)}\right)}_{n}
\right],
\nonumber\\
&=&
\frac{N_c}{N^2_c-1}\,
\sum^{N_c}_{n=1}
\sum^{N_c}_{m=1}
\,\left[
\left(
1-\frac{\delta_{nm}}{N_c}
\right)\,\delta_{mi}\,
{\left({\omega}^{2}_{0\,Q (Q)}\right)}_{n}
\right].
\end{eqnarray}
Furthermore, the color contraction of the gluon segment 
of the quark self-energy is reduced to
\begin{eqnarray}
{\bf t}^{a}_{in}\, {\bf t}^{b}_{mj}\,
{\left({\omega}^{2}_{0\,Q (G)}\right)}^{a}\, 
\delta_{nm}\,\delta^{ab}
&\equiv& 
\frac{1}{2} \tau^{a}_{in}\, 
\left(\tau^{a\dagger}\right)_{nj}\, 
{\left({\omega}^{2}_{0\,Q (G)}\right)}^{a},
\nonumber\\
&\equiv& 
\frac{1}{2}\, \tau^{\underbrace{(AB)}}_{in}\, 
\tau^{\underbrace{(BA)}}_{nj}\,
{\left(
{\omega}^{2}_{0\,Q (G)}
\right)}^{\underbrace{(AB)}},
\nonumber\\
&=&
\frac{1}{2 N_c}\,\delta_{ij}\,
\left[
N_c \sum^{N_c}_{n=1}
{\left(
{\omega}^{2}_{0\,Q (G)}
\right)}^{\underbrace{(i n)}}
-
{\left(
{\omega}^{2}_{0\,Q (G)}
\right)}^{\underbrace{(ii)}}
\right].
\end{eqnarray}
It is reduced to
\begin{eqnarray}
{\bf t}^{a}_{in}\, {\bf t}^{b}_{mj}\,
{\left({\omega}^{2}_{0\,Q (G)}\right)}^{a}
\,\delta_{nm}\,\delta^{ab}
&=&
\frac{N^2_c-1}{2N_c}\,\delta_{ij}\,
{\left({\omega}^{*\,2}_{0\,Q (G)}\right)}_{i},
\end{eqnarray}
where
\begin{eqnarray}
{\left({\omega}^{*\,2}_{0\,Q (G)}\right)}_{i}
&=&
\frac{1}{N^2_c-1}
\,\left[
N_c\sum^{N_c}_{n=1}
{\left(
{\omega}^{2}_{0\,Q (G)}
\right)}^{\underbrace{(in)}}
-
{\left(
{\omega}^{2}_{0\,Q (G)}
\right)}^{\underbrace{(ii)}}
\right],
\nonumber\\
&=&
\frac{N_c}{N^2_c-1}\,
\sum^{N_c}_{n=1}\,\sum^{N_c}_{m=1}
\left[
\left(
1-
\frac{\delta_{nm}}{N_c}
\right)\,\delta_{mi}
{\left({\omega}^{2}_{0\,Q (G)}
\right)}^{\underbrace{(mn)}}
\right].
\end{eqnarray}
The adjoint color indexes are represented 
in terms of the fundamental color indexes
as follows
\begin{eqnarray}
a\rightarrow \underbrace{(AB)} \rightarrow \underbrace{(ij)}.
\end{eqnarray}
The quark Landau frequency is coupled 
with two fundamental generators as follows,
\begin{eqnarray}
{\bf t}^{a}_{in}\, {\bf t}^{a}_{nj}\,
{\left(
{\omega}^{2}_{0\,Q}
\right)}^a_n
&=&
{\bf t}^{a}_{in}\, {\bf t}^{a}_{nj}\, 
\left[
{\left(
{\omega}^{2}_{0\,Q (Q)}
\right)}_{n}
+
{\left(
{\omega}^{2}_{0\,Q (G)}
\right)}^{a}
\right],
\nonumber\\
&=&
\frac{N^2_c-1}{2N_c}\,\delta_{ij}\,
{\left(
{\omega}^{* 2}_{0\,Q}
\right)}_{i},
\end{eqnarray}
where
\begin{eqnarray}
{\left(
{\omega}^{* 2}_{0\,Q}
\right)}_{i}
&=& \left[
{\left({\omega}^{* 2}_{0\,Q (Q)}\right)}_{i} 
+ {\left({\omega}^{* 2}_{0\,Q (G)}\right)}_{i}
\right],
\nonumber\\
&=&
\frac{N_c}{N^2_c-1}\,
\sum^{N_c}_{n=1}\,\sum^{N_c}_{m=1}
\left(
1-
\frac{\delta_{nm}}{N_c}
\right)\,\delta_{mi}\,
\left[
{\left(
{\omega}^{2}_{0\,Q (Q)}
\right)}_{n}
+
{\left(
{\omega}^{2}_{0\,Q (G)}
\right)}^{\underbrace{(mn)}}
\right].
\end{eqnarray}
In the case that the quark Landau frequency is decoupled from the color 
degree of freedom, the plasma frequency components 
for the quark and gluon segments (i.e. lines) 
of the quark self-energy are reduced to
\begin{eqnarray}
{\left(
{\omega}^{2}_{0\,Q (Q)}
\right)}_{n}
&\rightarrow&
\overline{{\omega}}^{2}_{0\,Q (Q)},
\nonumber\\
{\left(
{\omega}^{2}_{0\,Q (G)}
\right)}^{\underbrace{(mn)}}
&\rightarrow&
\overline{
{\omega}}^{2}_{0\,Q (G)},
\end{eqnarray}
respectively.
Hence, the quark Landau frequency is reduced to
\begin{eqnarray}
{\left(
{\omega}^{*\,2}_{0\,Q}
\right)}_{i}
&=&
{\overline{\omega}}^{2}_{0\,Q (Q)}
+
{\overline{\omega}}^{2}_{0\,Q (G)}.
\end{eqnarray}
The coupling of the internal Landau frequency, namely, 
${\left({\omega}^{2}_{0\,Q}\right)}^{a}_{n}$  
with the fundamental generators contracts 
the color indexes in the following way
\begin{eqnarray}
{\bf t}^{a}_{in}\, {\bf t}^{a}_{nj}\, 
{\left({\omega}^{2}_{0\,Q}
\right)}^{a}_{n}
&=&
\frac{N^2_c-1}{2N_c}\,\delta_{ij}\,
\left[
{\overline{\omega}}^{2}_{0\,Q (Q)}
+
{\overline{\omega}}^{2}_{0\,Q (G)}
\right].
\end{eqnarray}
%
%
%
%

\subsection{Gluon Polarization tensor}
\subsubsection{Gluon Polarization tensor:
the color indexes for  
gluon and ghost loops and tadpole}

The multiplication of two adjoint matrices appears 
in the Feynman diagrams for the gluon self-energy. 
It is given by
\begin{eqnarray}
({\bf T}^{a'})_{c'b'} ({\bf T}^a)_{b c}
\equiv
({\bf T}^{a'})_{c'b'} ({\bf T}^{a\dagger})_{b c}.
\end{eqnarray}
The adjoint matrix can be represented in terms 
of the fundamental representations as follows,
\begin{eqnarray}
\left({\bf T}^{a}\right)_{bc}&=&
-i f^{abc},\nonumber\\
&=& -2 \mbox{tr}
\left(\left[
{\bf t}^{a},{\bf t}^{b}
\right]{\bf t}^{c}\right).
\end{eqnarray}
The adjoint conjugate matrix becomes
\begin{eqnarray}
\left({\bf T}^{a\dagger}\right)_{bc}
&=&
- 2 \mbox{tr}
\left([{\bf t}^{a\dagger},{\bf t}^{b\dagger}]
{\bf t}^{c\dagger}\right).
\end{eqnarray}
The multiplication of two adjoint matrices is 
written in terms of fundamental matrices as follows,
\begin{eqnarray}
({\bf T}^{a'})_{c'b'} ({\bf T}^{a\dagger})_{bc}&=&
4 \mbox{tr}\left([{\bf t}^{a'},{\bf t}^{c'}]
{\bf t}^{b'}\right)\times
\mbox{tr}
\left([{\bf t}^{a\dagger},{\bf t}^{b\dagger}]
{\bf t}^{c\dagger}\right),
\nonumber\\
&=&
-4 \mbox{tr}
\left(
[{\bf t}^{a'},{\bf t}^{b'}]{\bf t}^{c'}
\right)\times
\mbox{tr}\left([{\bf t}^{a\dagger},{\bf t}^{b\dagger}]
{\bf t}^{c\dagger}\right),
\nonumber\\
&=&
-4 \mbox{tr}
\left(
[{\bf t}^{a'},{\bf t}^{c'}]{\bf t}^{b'}
\right)\times
\mbox{tr}\left([{\bf t}^{a\dagger},{\bf t}^{c\dagger}]
{\bf t}^{b \dagger}\right).
\end{eqnarray}
The  $\delta^{c'\,c}$ index contraction 
is performed in the following way,
\begin{eqnarray}
({\bf T}^{a'})_{c' b'}\, 
({\bf T}^{a\dagger})_{b c}\,\delta^{c' c}
&=&
-2[{\bf t}^{a'},{\bf t}^{b'}]_{in}\,
[{\bf t}^{a\dagger},{\bf t}^{b\dagger}]_{jm}\,
\tau^{\underbrace{(AB)}}_{ni}\,
\tau^{\underbrace{(BA)}}_{mj},
\nonumber\\
&=&
-2\left(
[{\bf t}^{a'},{\bf t}^{b'}]_{ij}\,
[{\bf t}^{a\dagger},{\bf t}^{b\dagger}]_{ji}
-\frac{1}{N_c}\,
[{\bf t}^{a'},{\bf t}^{b'}]_{ii}\,
[{\bf t}^{a\dagger},{\bf t}^{b\dagger}]_{jj}
\right),
\nonumber\\
&=&
-2\left(
[{\bf t}^{a'},{\bf t}^{b'}]_{ij} \,
[{\bf t}^{a\dagger},{\bf t}^{b\dagger}]_{ji}
\right).
\label{weighted1}
\end{eqnarray}
In a similar manner, the contraction over 
$\delta^{b'\,b}$ leads to the following result
\begin{eqnarray}
({\bf T}^{a'})_{c' b'}\, 
({\bf T}^{a\dagger})_{b c}\,\delta^{b' b}
&=&
-2\left(
[{\bf t}^{a'},{\bf t}^{c'}]_{ij} \,
[{\bf t}^{a\dagger},{\bf t}^{c\dagger}]_{ji}
\right).
\label{weighted2}
\end{eqnarray}
Therefore, the contraction over either 
$\delta^{b'\,b}\,
\left({\bf m}^{2}_{D\,(G)}\right)^{b}$ 
for Eq.~(\ref{weighted1})
or
$\delta^{c'\,c}\,
\left({\bf m}^{2}_{D\,(G)}\right)^{c}$ 
for Eq.~(\ref{weighted2}) leads, respectively,
to
\begin{eqnarray}
({\bf T}^{a'})_{c' b'}\, 
({\bf T}^{a\dagger})_{b c}\,\delta^{c' c}
\,\delta^{b' b}
\left({\bf m}^{2}_{D\,(G)}\right)^{b}
&=&
-2\left(
[{\bf t}^{a'},{\bf t}^{b}]_{ij} \,
[{\bf t}^{a\dagger},{\bf t}^{b\dagger}]_{ji}
\right)\,
\left({\bf m}^{2}_{D\,(G)}\right)^{b},
\label{contract-adj-gluon1}
\end{eqnarray}
or 
\begin{eqnarray}
({\bf T}^{a'})_{c' b'}\, 
({\bf T}^{a\dagger})_{b c}\,\delta^{b' b}
\,\delta^{c' c}
\left({\bf m}^{2}_{D\,(G)}\right)^{c}
&=&
-2\left(
[{\bf t}^{a'},{\bf t}^{c}]_{ij} \,
[{\bf t}^{a\dagger},{\bf t}^{c\dagger}]_{ji}
\right)\,
\left({\bf m}^{2}_{D\,(G)}\right)^{c}.
\label{contract-adj-gluon2}
\end{eqnarray}
The results of Eqns.~(\ref{contract-adj-gluon1})
and (\ref{contract-adj-gluon2}) leads 
to the folowing symmetry relation 
\begin{eqnarray}
({\bf T}^{a'})_{c' b'}\, 
({\bf T}^{a\dagger})_{b c}\,\delta^{c' c}
\,\delta^{b' b}
\left({\bf m}^{2}_{D\,(G)}\right)^{b}
&=&({\bf T}^{a'})_{c' b'}\, 
({\bf T}^{a\dagger})_{b c}\,\delta^{b' b}
\,\delta^{c' c}
\left({\bf m}^{2}_{D\,(G)}\right)^{c},
\nonumber\\
&=&
({\bf T}^{a'})_{c b}\, 
({\bf T}^{a\dagger})_{b c}
\,\left({\bf m}^{2}_{D\,(G)}\right)^{b},
\nonumber\\
&=&
({\bf T}^{a'})_{c b}\, 
({\bf T}^{a\dagger})_{b c}
\,\left({\bf m}^{2}_{D\,(G)}\right)^{c},
\nonumber\\
&~&
~~~(\mbox{sum over the repeated adjoint color indexes}),
\label{adjoint-symmetry-b-and-c}
\end{eqnarray}
when the summation over the internal adjoint color 
indexes $b$ and $c$
is taken for the Debye mass 
that  is weighted by the Casimir-like operator,
namely, $({\bf T}^{a'})_{c b}\, ({\bf T}^{a\dagger})_{b c}$.
Moreover, the term 
$\left(\Delta{\bf m}^{2}_{D\,(G)}\right)^{b}$
which generates the quadratic temperature
has the following symmetry relations with respect 
to the interchange of adjoint color indexes 
$b$ and $c$:
\begin{eqnarray}
({\bf T}^{a'})_{c b}\, 
({\bf T}^{a\dagger})_{b c}
\,\left(\Delta{\bf m}^{2}_{D\,(G)}\right)^{b}
&=&
({\bf T}^{a'})_{c b}\, 
({\bf T}^{a\dagger})_{b c}
\,\left(\Delta{\bf m}^{2}_{D\,(G)}\right)^{c}, 
\nonumber\\
&~&
~~~(\mbox{sum over the repeated adjoint color indexes}).
\label{adjoint-quadratic-symmetry}
\end{eqnarray}
Evidently, the symmetry in Eq.~(\ref{adjoint-quadratic-symmetry}) 
eliminates the quadratic temperature terms
from gluon polarization tensor with the external line legs those 
are labeled by the external adjoint color indexes, namely, $a'\,a$.
These quadratic temperature terms usually appear 
in the internal loop segments 
which are associated with the three gluon vertexes 
and they are usually labelled by the internal adjoint color indexes.

The adjoint color index $c$ is replaced 
with the fundamental-like $A$ and $B$ 
indexes using the following relation, 
\begin{eqnarray}
c\equiv \underbrace{(AB)} ~\mbox{or}~ \underbrace{(ij)} 
~\mbox{or}~ \underbrace{(nm)}.
\end{eqnarray}
The $\delta^{c' c}$ and $\delta^{b' b}$ contractions
with the coefficient 
$\left({\bf m}^2_{D\,(G)}\right)^{b}$  
reduce the multiplication of the two adjoint matrices 
to
\begin{eqnarray}
({\bf T}^{a'})_{c' b'}\, ({\bf T}^{a\dagger})_{b c}\,
\delta^{c' c}\,\delta^{b' b}\,
\left({\bf m}^2_{D\,(G)}\right)^{b}
&=& 
-2\left(
{\bf t}^{a'}_{in}\, {\bf t}^{b}_{nj}\,
{\bf t}^{a\dagger}_{jm}\, {\bf t}^{b\dagger}_{mi}\,
-\,
{\bf t}^{a'}_{in}\, {\bf t}^{b}_{nj}\, 
{\bf t}^{b\dagger}_{jm}\, {\bf t}^{a\dagger}_{mi}
\right.
\nonumber\\
&~&
\left.~
-
{\bf t}^{b}_{in}\, {\bf t}^{a'}_{nj}\, 
{\bf t}^{a\dagger}_{jm}\, {\bf t}^{b\dagger}_{mi}\,
+\,
{\bf t}^{b}_{in}\, {\bf t}^{a'}_{nj}\, 
{\bf t}^{b\dagger}_{jm}\, {\bf t}^{a\dagger}_{mi}
\right)\,
\left({\bf m}^2_{D\,(G)}\right)^{b},
\nonumber\\
&=& 
4 \left(
{\bf t}^{a'}_{in}\, {\bf t}^{b}_{nj}\, 
{\bf t}^{b\dagger}_{jm}\, {\bf t}^{a\dagger}_{mi}\,
-\,
{\bf t}^{a'}_{in}\, {\bf t}^{b}_{nj}\, 
{\bf t}^{a\dagger}_{jm}\, {\bf t}^{b\dagger}_{mi}
\right)\,
\left({\bf m}^2_{D\,(G)}\right)^{b}.
\label{adjoint_two_coeff1}
\end{eqnarray}
It is useful to derive the contraction relation 
over the adjoint index $b$ for the following 
fundamental matrices operation
\begin{eqnarray}
{\bf t}^{b}_{nj}\, {\bf t}^{b\dagger}_{jm}\, 
\left({\bf m}^2_{D\,(G)}\right)^{b}
&=&
\frac{1}{2} \tau^{\underbrace{(AB)}}_{nj}\, 
\tau^{\underbrace{(BA)}}_{jm}\,
\left({\bf m}^2_{D\,(G)}\right)^{\underbrace{(AB)}},
\nonumber\\
&=&
\frac{1}{2}\left[
\delta_{nm}\,\sum^{N_c}_{l=1}
\left({\bf m}^2_{D\,(G)}\right)^{\underbrace{(nl)}}
-
\frac{1}{N_{c}}
\delta_{nm}\,
\left({\bf m}^2_{D\,(G)}\right)^{\underbrace{(00)}}
\right].
\label{adjoint_rel_1}
\end{eqnarray}
Furthermore, we have the following contraction relation,
\begin{eqnarray}
{\bf t}^{b}_{nj}\, {\bf t}^{b\dagger}_{mi}\, 
\left({\bf m}^2_{D\,(G)}\right)^{b}
&=&
\frac{1}{2}\, \tau^{\underbrace{(AB)}}_{nj}\, 
\tau^{\underbrace{(BA)}}_{mi}\,
\left({\bf m}^2_{D\,(G)}
\right)^{\underbrace{(AB)}}, 
~~\left(\mbox{where}\, b\,\equiv\,
\underbrace{(AB)}\right),
\nonumber\\
&=&
\frac{1}{2}
\left[
\delta_{in}\,\delta_{jm}
\left(
{\bf m}^2_{D\,(G)}
\right)^{\underbrace{(ij)}}
-
\frac{1}{N_c}\delta_{mi}\,\delta_{nj}\,
\left(
{\bf m}^2_{D\,(G)}
\right)^{\underbrace{(00)}}
\right].
\label{adjoint_rel_2}
\end{eqnarray}
Using Eq.~(\ref{adjoint_rel_1}), the first term in 
Eq.~(\ref{adjoint_two_coeff1}) is reduced to
\begin{eqnarray}
{\bf t}^{a'}_{in}\, {\bf t}^{a\dagger}_{mi}\,
{\bf t}^{b}_{nj}\, {\bf t}^{b\dagger}_{jm}\, 
\left({\bf m}^2_{D\,(G)}\right)^{b}
&=&
\frac{1}{2}
\left[
{\bf t}^{a'}_{in}\, {\bf t}^{a\dagger}_{ni}\,
\sum^{N_c}_{l=1}
\left({\bf m}^2_{D\,(G)}\right)^{\underbrace{(nl)}}
-
\frac{1}{N_c}
{\bf t}^{a'}_{in}\, {\bf t}^{a\dagger}_{ni}\,
\left({\bf m}^2_{D\,(G)}\right)^{\underbrace{(00)}}
\right].
\label{adj_contr1}
\end{eqnarray}
While using Eq.~(\ref{adjoint_rel_2}) reduces 
the second term 
in Eq.~(\ref{adjoint_two_coeff1}) to 
\begin{eqnarray}
{\bf t}^{a'}_{in}\, {\bf t}^{a\dagger}_{jm}\,
{\bf t}^{b}_{nj}\, {\bf t}^{b\dagger}_{mi}\, 
\left({\bf m}^2_{D\,(G)}\right)^{b}
&=&
\frac{1}{2}\left[
{\bf t}^{a'}_{ii}\, {\bf t}^{a\dagger}_{jj}\,
\left({\bf m}^2_{D\,(G)}\right)^{\underbrace{(ij)}}
-
\frac{1}{N_c}
{\bf t}^{a'}_{ij}\, {\bf t}^{a\dagger}_{ji}\,
\left({\bf m}^2_{D\,(G)}\right)^{\underbrace{(00)}}
\right].
\label{adj_contr2}
\end{eqnarray}
Furthermore, the contraction relations 
of the fundamental matrices which are given by 
Eqns.~(\ref{adj_contr1}) and (\ref{adj_contr2})
reduce Eq.~(\ref{adjoint_two_coeff1}) to
\begin{eqnarray}
({\bf T}^{a'})_{c b}\,
({\bf T}^{a})_{b c}^{\dagger}\,
\left({\bf m}^2_{D\,(G)}\right)^{b}
&=& 2 \left(
{\bf t}^{a'}_{ij}\, {\bf t}^{a\dagger}_{ji}\,
\sum^{N_c}_{l=1}
\left({\bf m}^2_{D\,(G)}\right)^{\underbrace{(jl)}}
-
{\bf t}^{a'}_{ii}\, {\bf t}^{a\dagger}_{jj}\,
\left({\bf m}^2_{D\,(G)}\right)^{\underbrace{(ij)}}
\right).
\label{adjoint-mass1}
\end{eqnarray}
The first term on the right hand side of 
Eq.~(\ref{adjoint-mass1}) is reduced to
\begin{eqnarray}
2\,{\bf t}^{a'}_{in}\, {\bf t}^{a\dagger}_{ni}\,
\sum^{N_c}_{l=1}
\left({\bf m}^2_{D\,(G)}\right)^{\underbrace{(nl)}}
&=& 
2\,{\bf t}^{\underbrace{(A'B')}}_{in}\, 
{\bf t}^{\underbrace{(BA)}}_{ni}\,
\sum^{N_c}_{l=1}
\left({\bf m}^2_{D\,(G)}\right)^{\underbrace{(nl)}},
\nonumber\\
&=&
\delta_{A'A}\delta_{B'B}\,\sum^{N_c}_{l=1}
\left({\bf m}^2_{D\,(G)}\right)^{\underbrace{(Bl)}}
\nonumber\\
&~&
-\frac{1}{N_c}\,\delta_{A' B'}\delta_{A B}\,
\left[
\sum^{N_c}_{l=1}
\left({\bf m}^2_{D\,(G)}\right)^{\underbrace{(B'l)}}
+
\sum^{N_c}_{l=1}
\left({\bf m}^2_{D\,(G)}\right)^{\underbrace{(Bl)}}
\right.
\nonumber\\
&~&
\left.
-\frac{1}{N_{c}}
\sum^{N_c}_{k=1}\sum^{N_c}_{l=1}
\left({\bf m}^2_{D\,(G)}\right)^{\underbrace{(kl)}}
\right].
\label{first-term-adj-adj}
\end{eqnarray}
The second term in Eq.~(\ref{adjoint-mass1}) 
reads
\begin{eqnarray}
2\,{\bf t}^{a'}_{ii}\, {\bf t}^{a\dagger}_{jj}\,
\left({\bf m}^2_{D\,(G)}\right)^{\underbrace{(ij)}}
&=& 
2\,\sum^{N_c}_{ij}\,
{\bf t}^{\underbrace{(A'B')}}_{ii}\, 
{\bf t}^{\underbrace{(BA)}}_{jj}\,
\left({\bf m}^2_{D\,(G)}\right)^{\underbrace{(ij)}},
\nonumber\\
&=&
\delta_{A' B'}\delta_{A B}\,
\left({\bf m}^2_{D\,(G)}\right)^{\underbrace{(A'A)}}
\nonumber\\
&~&
-\frac{1}{N_c}\,\delta_{A' B'}\delta_{A B}\,
\left[
\sum^{N_c}_{l=1}
\left({\bf m}^2_{D\,(G)}\right)^{\underbrace{(A' l)}}
+
\sum^{N_c}_{l=1}
\left({\bf m}^2_{D\,(G)}\right)^{\underbrace{(l A)}}
\right.
\nonumber\\
&~&
\left.
-\frac{1}{N_{c}}
\sum^{N_c}_{k=1}\sum^{N_c}_{l=1}
\left({\bf m}^2_{D\,(G)}\right)^{\underbrace{(k l)}}
\right].
\label{second-term-adj-adj}
\end{eqnarray}
By considering the results of 
Eq.~(\ref{first-term-adj-adj}) and 
(\ref{second-term-adj-adj}), 
Eq.~(\ref{adjoint-mass1}) becomes
\begin{eqnarray}
({\bf T}^{a'})_{c\,b}\,({\bf T}^{a})_{b\,c}^{\dagger}\,
\left({\bf m}^2_{D\,(G)}\right)^{b}
&=&
\delta_{A'A}\delta_{B'B}\,\sum^{N_c}_{l=1}
\left({\bf m}^2_{D\,(G)}\right)^{\underbrace{(Bl)}}
-\delta_{A' B'}\delta_{A B}\,
\left({\bf m}^2_{D\,(G)}\right)^{\underbrace{(B'B)}}
\nonumber\\
&-&
\frac{1}{N_{c}}\delta_{A' B'}\delta_{A B}\,
\left[
\sum^{N_c}_{l=1}\left(
\left({\bf m}^2_{D\,(G)}\right)^{\underbrace{(B l)}}
-
\left({\bf m}^2_{D\,(G)}\right)^{\underbrace{(l B)}}
\right)
\right].
\label{debye-adjoint-reduction1}
\end{eqnarray}
The term 
$\left({\bf m}^2_{D\,(G)}\right)^{\underbrace{(ij)}}$
is symmetric over the fundamental-like indexes 
$i$ and $j$. 
Furthermore, the second term on the right hand side
of Eq.~(\ref{debye-adjoint-reduction1})
can be averaged as follows
\begin{eqnarray}
\left(
{\bf m}^2_{D\,(G)}
\right)^{\underbrace{(B' B)}}
\, \delta_{A' B'}\delta_{A B}
&\equiv&
\frac{1}{N_{c}}
\sum^{N_{c}}_{l=1}
\left({\bf m}^2_{D\,(G)}
\right)^{\underbrace{(B l)}}
\, \delta_{A' B'}\delta_{A B}.
\end{eqnarray}
Therefore, Eq.~(\ref{debye-adjoint-reduction1}) 
is reduced to
\begin{eqnarray}
({\bf T}^{a'})_{c\,b}\,({\bf T}^{a})_{b\,c}^{\dagger}\,
\left({\bf m}^2_{D\,(G)}\right)^{b}
&=&
\left(\delta_{A'A}\delta_{B'B}
-\frac{1}{N_c}\delta_{A'B'}\delta_{AB}\right)
\sum^{N_c}_{l=1}
\left({\bf m}^2_{D\,(G)}\right)^{\underbrace{(Bl)}},
\nonumber\\
&=&
N_c\,\delta^{a' a}\,
\left({\bf m}^2_{D\,(G)}\right)^{a},
\label{adjoint_cont_f1}
\end{eqnarray}
where 
\begin{eqnarray}
\left({\bf m}^2_{D\,(G)}\right)^{a}
&\equiv&
\frac{1}{N_c}
\sum^{N_c}_{l=1}
\left({\bf m}^2_{D\,(G)}\right)^{\underbrace{(Bl)}}, 
~~~(\mbox{note that}~ a\equiv \underbrace{(AB)} ).
\label{m-D-G-a}
\end{eqnarray}
Eq.~(\ref{m-D-G-a}) is re-written as follows
\begin{eqnarray}
\left({\bf m}^2_{D\,(G)}\right)^{a}
&=&
\frac{1}{N_c}\,
\left[
\sqrt{2} 
\sum^{N_c}_{ij}\sum^{N_c}_{n}
({\bf t}^{a})_{in} \left({\bf m}^2_{D\,(G)}
\right)^{\underbrace{(nj)}}
+
\frac{1}{N_c}\delta_{AB}
\sum^{N_c}_{ij}\left({\bf m}^2_{D\,(G)}
\right)^{\underbrace{(ij)}}
\right].
\end{eqnarray}
The last term with the Kronecker delta $\delta_{AB}$ corresponds the 
diagonal fundamental generators with the fundamental index 
$i=1,\cdots,N_c-1$ which commute 
with the entire fundamental generator set
$[{\bf t}_{i},{\bf t}^{a}]=0$ where the adjoint index 
$a=1,\cdots,N^{2}_c-1$
and they commute with the energy eigenstates as well.
Furthermore,  the Kronecker delta $\delta_{AB}$ can be replaced 
by $\delta_{a i}$.
Hence, the above equation becomes
 \begin{eqnarray}
\left({\bf m}^2_{D\,(G)}\right)^{a}
&\equiv&
\frac{1}{N_c}\,
\left[
\sqrt{2} 
\sum^{N_c}_{ij}\sum^{N_c}_{n}
({\bf t}^{a})_{in} \left({\bf m}^2_{D\,(G)}
\right)^{\underbrace{(nj)}}
+
\frac{1}{N_c}\delta_{a i}
\sum^{N_c}_{ij}\left({\bf m}^2_{D\,(G)}
\right)^{\underbrace{(ij)}}
\right],
\end{eqnarray} 
where $i$ is the fundamental index that represents 
the diagonal fundamental generators 
where its off-diagonal elements are identical to zero.

\subsubsection{Gluon Polarization tensor: quark-loop color indexes}

The multiplication of the two fundamental color matrices which appears 
in the quark loop is usually given by
\begin{eqnarray}
{\bf t}^{a'}_{i' j'}\, {\bf t}^{a\dagger}_{j i}\,
\delta_{j' j}\,\delta_{i' i}\,
\left({\bf m}^2_{D\,(Q)}\right)_{i}
&=&
{\bf t}^{a'}_{i j}\, ({\bf t}^{a\dagger})_{j i}\,
\left({\bf m}^2_{D\,(Q)}\right)_{i}.
\end{eqnarray}
It can be written as follows
\begin{eqnarray}
{\bf t}^{a'}_{i j}\, ({\bf t}^{a\dagger})_{ji}\,
\left({\bf m}^2_{D\,(Q)}\right)_{i}
&=&
{\bf t}^{\underbrace{(A'B')}}_{i j}\, 
{\bf t}^{\underbrace{(BA)}}_{ji}\,
\left({\bf m}^2_{D\,(Q)}\right)_{i},
\nonumber\\
&=&
\frac{1}{2}\left(
\delta_{A'A}\,\delta_{B'B}
-\frac{1}{N_c}\,\delta_{A'B'}\,\delta_{AB}
\right)\,\left({\bf m}^2_{D\,(Q)}\right)_{A},
\nonumber\\
&=&
\frac{1}{2}
\,\delta^{a'a}\,
\left({\bf m}^2_{D\,(Q)}\right)_{A},
\nonumber\\
&=&
\frac{1}{2}\delta^{a' a}\, 
\left({\bf m}^2_{D\,(Q)}\right)^{a}.
\label{quark-loop1}
\end{eqnarray}
where
\begin{eqnarray}
\left({\bf m}^2_{D\,(Q)}\right)^{a}
&=&
\left({\bf m}^2_{D\,(Q)}\right)_{A}, 
~~~(\mbox{note that}~ a\equiv \underbrace{(AB)}.
\end{eqnarray}
The quark-loop part of the Debye mass 
with an external adjoint color index $a$
is written as follows
\begin{eqnarray}
\left({\bf m}^2_{D\,(Q)}\right)^{a}
&=&
\left[
\sqrt{2}\sum^{N_c}_{ij} ({\bf t}^{a})_{ij}\, 
\left({\bf m}^2_{D\,(Q)}\right)_{i}
+
\frac{1}{N_c}\delta_{AB}\,\left({\bf m}^2_{D\,(Q)}\right)_{i}
\right].
\label{m-D-Q-a}
\end{eqnarray}
The second term comes from the diagonal elements of the representation
that represents the $N_c-1$ fundamental generators which commute with the
energy states. The fundamental generators are given by ${\bf t}_i$ 
where $\left[{\bf t}_i,{\bf t}^{a}\right]=0$. 
The Kronecker delta $\delta_{AB}$ is replaced by $\delta_{a i}$ where 
$a=1,\cdots,N^2_c-1$ is the adjoint color index for the entire set 
of generators 
while $i=1,\cdots, N_c-1$ is the fundamental color index 
for fundamental generators which  their off-diagonal elements 
are identical to zero. They commute with energy eigenstates 
as well as the rest of fundamental color generators. 
Hence, Eq.~(\ref{m-D-Q-a}) becomes
\begin{eqnarray}
\left({\bf m}^2_{D\,(Q)}\right)^{a}
&=&
\left[
\sqrt{2}\sum^{N_c}_{ij} ({\bf t}^{a})_{ij}\, 
\left({\bf m}^2_{D\,(Q)}\right)_{i}
+
\frac{1}{N_c}\delta_{a i}\,\left({\bf m}^2_{D\,(Q)}\right)_{i}
\right].
\end{eqnarray}
It is worth to remind the reader to the following dual,
\begin{eqnarray}
{\bf t}^{a}_{i'\,j'}\, {\bf t}^{b}_{j\,i}\rightarrow 
{\bf t}^{a}_{i'\,j'}\, ({\bf t}^{b\dagger})_{j\,i},
\end{eqnarray} 
\subsubsection{Gluon self-energy: quark, gluon, ghost loops 
and tadpole}
The colored gluon Debye mass is given by adding the contributions
of quark loop, gluon loop, ghost loop
and tadpole Feynman diagrams. 
The colored  Debye mass is reduced to
\begin{eqnarray}
\delta^{a' a}
\left({\bf m}^2_{D}\right)^{a}&=&
\left[
({\bf T}^{a'})_{cb}\,({\bf T}^{a})_{bc}^{\dagger}\,
\left({\bf m}^2_{D\,(G)}\right)^{b} 
+
{\bf t}^{a'}_{ij}\, ({\bf t}^{a\dagger})_{ji}\,
\left({\bf m}^2_{D\,(Q)}\right)_{i}
\right],
\end{eqnarray}
where
\begin{eqnarray}
\left({\bf m}^2_{D}\right)^{a}
&=&
N_c\,\left({\bf m}^2_{D\,(G)}\right)^{a}
+
\frac{1}{2}\,
\left({\bf m}^2_{D\,(Q)}\right)^{a}.
\end{eqnarray}

\section{\label{section4} Quark Self-energy}

\begin{figure} 
\includegraphics{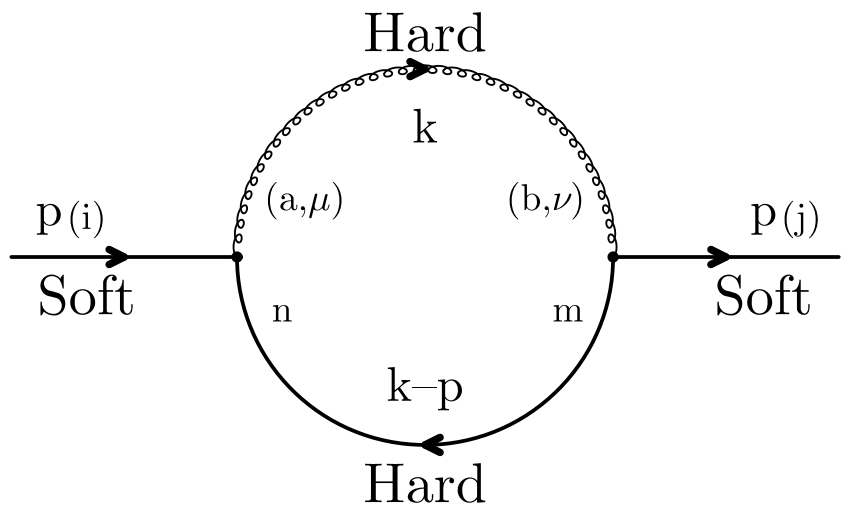}
\caption{\label{fig:quarkeff}
The soft quark self-energy correction.} 
\end{figure}

Fig.~(\ref{fig:quarkeff}) depicts the Feynman diagrams 
for the quark self-energy.  
The fundamental indexes for the quark segment 
and adjoint indexes for the gluon segment in the quark-gluon loop 
are shown explicitly.
The quark self-energy is calculated as follows,
\begin{eqnarray}
{\left.{\Sigma_{Q}}\right.}_{ij}(p)
&=&
{\left.{\Sigma_{Q}}\right.}_{ij}(p_0,\vec{p}),
\nonumber\\
&=&\int \frac{d k_0}{(2\pi)}\,
\int \frac{d^{3}\vec{k}}{(2\pi)^{3}}\,
{ {\left.{\Sigma_{Q}}\right.}_{i j} }(p,k),
\nonumber\\
&=&\int \frac{d k_0}{(2\pi)}\,
\int \frac{d^{3}\vec{k}}{(2\pi)^{3}}\,
{ {\left.{\Sigma_{Q}}\right.}_{ij} }
\left(p_0,\vec{p},k_{0},\vec{k}\right),
\end{eqnarray}
where $p$ and $k$ are the external and internal momentum, respectively. 
The fundamental indexes $i\,j$ refer 
to the quark line (i.e. two external legs). 
The conservation of color charges leads to color contraction $\delta_{ij}$
for the quark line and make the quark line to have 
a conserved color charge with fundamental index $i$. 
The quark self-energy kernel part for 
the quark-gluon loop is furnished as follows,
\begin{eqnarray}
{ {\Sigma_{Q}}_{ij} } (p,k)
&=&
\left[-g\,\gamma^{\mu} ({\bf t}^{a})_{i n}\right]
\,
i\, 
{ {{\cal S}_{Q}}_{n m} } \left(p_0-k_0,\vec{p}-\vec{k}\right)
\,
\left[-g\,\gamma^{\nu} ({\bf t}^{b})_{mj}\right]
\nonumber\\
&~&
\times
{ {{\cal G}_{\mu\nu}}^{a b} }\left(-k_0,-\vec{k}\right),
\nonumber\\
&=& 
\left({\bf t}^{a}\right)_{in}\,
\left({\bf t}^{b}\right)_{mj}\,
{ {\Sigma_{Q}}^{a b}_{n m} }\left(p_0,\vec{p},k_{0},\vec{k}\right),
\nonumber\\
&=& 
\left({\bf t}^{a}\right)_{in}\,
\left({\bf t}^{b}\right)_{mj}\,
{ {\left.{\Sigma_{Q}}\right.}^{a b}_{n m} }(p,k),
\end{eqnarray}
where $p$ and $k$ are the external and internal momenta, respectively.
The internal quark-gluon loop kernel, where the quark segment carries 
the fundamental color indexes $n$ and $m$ 
and the gluon segment carries the adjoint color indexes
$a$ and $b$, is given by 
\begin{eqnarray}
{ {\left.{\Sigma_{Q}}\right.}_{nm}^{ab} }(p_{0},\vec{p},k_{0},\vec{k})
&=&
g^2\,\left[
\gamma^{\mu}\,
i\,
{ {{\cal S}_{Q}}_{nm} }\left(p_0-k_0,\vec{p}-\vec{k}\right)
\,\gamma^{\nu}\,
{ {{\cal G}_{\mu\nu}}^{ab} }\left(-k_0,-\vec{k}\right)
\right].
\label{eq-quark-self_fey}
\end{eqnarray}
It is naturally to anticipate that the conservation of color charge contracts 
the gluon segment to $\delta^{ab}$ while the quark segment to $\delta_{nm}$ 
and subsequently the internal quark and gluon segments have 
the fundamental color index $n$  and the adjoint color index $a$, 
respectively. 
The quark self-energy can be projected with respect to the quark-quark-gluon
vertexes $({\bf t}^{a})_{in}$ and $({\bf t}^{b})_{mj}$ as follows
\begin{eqnarray}
{ {\left.{\Sigma_{Q}}\right.}_{ij} }(p_0,\vec{p})
&=&
\left({\bf t}^{a}\right)_{in}\,
\left({\bf t}^{b}\right)_{mj}\,
{ {\left.{\Sigma_{Q}}\right.}^{ab}_{nm} }(p_0,\vec{p}),
\end{eqnarray}
where the integration 
over the internal momentum $k$ 
(i.e. $k_{0}$ and $\vec{k}$) 
is given by
\begin{eqnarray}
{ {\left.{\Sigma_{Q}}\right.}^{ab}_{nm} }(p_0,\vec{p})
&=&
\int\frac{d^4 k}{(2\pi)^4}\,
{ {\left.{\Sigma_{Q}}\right.}^{ab}_{nm} }(p_0,\vec{p},k_{0},\vec{k}).
\end{eqnarray}
The quark self-energy kernel which is given by Eq.~(\ref{eq-quark-self_fey})
is written in the mixed-time representation of the imaginary-time formalism 
as follows,
\begin{eqnarray}
{ {\left.{\Sigma_{Q}}\right.}^{a b}_{n m} }(p_{0},\vec{p},k_{0},\vec{k})
&=&
\int^{\beta}_{0}{d\tau}\int^{\beta}_{0}{d\tau'} 
\exp\left(\left[
(p_0-k_0)-\mu_Q-i\frac{1}{\beta}\theta_n
\right]\tau'\right)
\,
\exp\left(
\left[ k_0-i\frac{1}{\beta}\phi^a \right]
\tau\right)
\nonumber\\
&~&
\times\,
{ \left.{ {\widetilde{\Sigma}}_Q }\right. }_{nm}^{ab} 
(\tau,\tau',\vec{p},\vec{k}).
\label{eq-quark-self_fey2-a}
\end{eqnarray}
It is re-written as follows
\begin{eqnarray}
{ {\left.{\Sigma_{Q}}\right.}_{ij} }(p_{0},\vec{p},k_{0},\vec{k})
&=& 
\left({\bf t}^{a}\right)_{in}\,
\left({\bf t}^{b}\right)_{mj}\,
\int^{\beta}_{0}{d\tau}\,\int^{\beta}_{0}{d\tau'}\, 
\exp\left(
\left[k_0-i\frac{1}{\beta}\phi^{a}\right]
\left[\tau-\tau'\right]
\right) 
\nonumber\\
&~&
~~~\times\exp\left(
\left[
p_0-\mu_Q-
i\frac{1}{\beta}\theta_n-i\frac{1}{\beta}\phi^a
\right]\tau'
\right)\,
{ \left.{{\widetilde{\Sigma}}_{Q}}\right. }^{ab}_{nm} 
(\tau,\tau',\vec{p},\vec{k}).
\label{eq-quark-self_fey2}
\end{eqnarray}
The last term under the integral 
in Eqns.~(\ref{eq-quark-self_fey2-a}) 
and (\ref{eq-quark-self_fey2}) 
is given by
\begin{eqnarray}
{ {\left.{\widetilde{\Sigma}_{Q}}\right.}_{nm}^{ab} }
(\tau,\tau',\vec{p},\vec{k})
&=&
g^2\, \gamma^{\mu} \,
i\,
{ {{\cal S}_{Q}}_{nm} }\left(\tau',\vec{p}-\vec{k}\right)
\,\gamma^{\nu}\,
{ {{\cal G}_{\mu\nu}}^{ab} }\left(\tau,\vec{k}\right).
\label{eq-quark-mix1}
\end{eqnarray}
After evaluating the integral over the internal time-momentum $k_{0}$, 
Eq.~(\ref{eq-quark-self_fey2}) is reduced to
\begin{eqnarray}
{ {\left.{\Sigma_{Q}}\right.}_{ij} }(p_{0},\vec{p},\vec{k})
&=&
\int \frac{d k_0}{(2\pi)}
\,
{ {\left.{\Sigma_{Q}}\right.}_{ij} }
(p_{0},\vec{p},k_{0},\vec{k}),
\nonumber\\
&=&
\left({\bf t}^{a}\right)_{in}
\,
\left({\bf t}^{b}\right)_{mj}
\,
\int^{\beta}_{0}{d\tau}
\exp\left(
\left[
p_0-\mu_Q
-i\frac{1}{\beta}\theta_n-i\frac{1}{\beta}\phi^{a}
\right]\tau\right)
\nonumber\\
&~&\times
{ {\left.{\widetilde{\Sigma}_{Q}}\right.}_{nm}^{ab} }
(\tau,\tau,\vec{p},\vec{k}).
\end{eqnarray}
In order to evaluate Eq.~(\ref{eq-quark-mix1}), the quark and gluon
propagators are decomposed to the energy plane components. 
The quark propagator part that is sandwiched by gamma matrices,
namely,
$\gamma^{\mu}\otimes\gamma^{\nu}$ 
in Eq.~(\ref{eq-quark-mix1}) is decomposed in the following way,
\begin{eqnarray}
{ {S_{Q}^{\mu\nu}}_{nm} }\left(\tau,\vec{q}\right)
&=&
\gamma^{\mu}\,
i\,
{ {{\cal S}_{Q}}_{nm} }(\tau,\vec{q})\,
\gamma^{\nu},
\nonumber\\
&=&
\left[
1\,-\,n_{F}\left(
\epsilon_{Q}(\vec{q})-\mu_Q-\frac{1}{\beta}i\theta_{n}
\right)\right]
\exp\left(-\left(\epsilon_{Q}(\vec{q})-\mu_Q
-\frac{1}{\beta}i\theta_n\right)\tau\right)
\nonumber\\
&~&
~~~ ~~~\times\,
\left[\gamma^{\mu}\,
{\Lambda_{Q}^{(+)}}(\vec{q})
\,\gamma_0\,\gamma^{\nu}\right]
\,\delta_{nm}
\nonumber\\
&+&
\left[n_{F}\left(\epsilon_{Q}(\vec{q})
+\mu_Q+\frac{1}{\beta}i\theta_{n}\right)\right]
\exp\left(\left(\epsilon_{Q}(\vec{q})+\mu_Q
+\frac{1}{\beta}i\theta_n\right)\tau\right)
\nonumber\\
&~&
~~~ ~~~\times\,
\left[
\gamma^{\mu}\,{\Lambda_{Q}^{(-)}}(\vec{q})
\,\gamma_0\,\gamma^{\nu}\right]
\,\delta_{nm},
\end{eqnarray}
where $n_{F}(x)=\frac{1}{e^{x}+1}$ 
is the Fermi-Dirac partition function, 
$\vec{q}=\vec{p}-\vec{k}$ 
and 
$\epsilon_{Q}=\sqrt{\vec{p}^2+m^2_{Q}}\approx |\vec{p}|$.
The Foldy-Wouthuysen positive 
and negative energy projectors read
\begin{eqnarray}
{\Lambda_{Q}^{(\pm)}}(\vec{q})&=&
\frac{1}{2\epsilon_{Q}(\vec{q})}
\left[
\epsilon_{Q}(\vec{q})
\pm\gamma_0\,\left(
\vec{\gamma}\cdot\vec{q}+m_{Q}
\right)\right].
\end{eqnarray}
In the case  that color degree of freedom is decoupled,
the quark screening mass (i.e.  the plasma Landau frequency)
is shown to be gauge independent in the HTL approximation.
The result is also gauge independent in the HTL approximation
for the case that the color degree of freedom is not decoupled. 
Hence, it is sufficient to work in the Feynman gauge 
to derive the gauge independent result.
The gluon propagator for the gluon segment in
Eq.~(\ref{eq-quark-mix1}) is given by
\begin{eqnarray}
{{\cal G}_{\mu\nu}}^{ab}(\tau,\vec{k})
&=& 
g_{\mu\nu}\,{\cal G}^{a}\left(\tau,\vec{k}\right)
\,\delta^{a b}.
\end{eqnarray}
The color contracted gluon propagator 
with an adjoint index $a$ is reduced to
\begin{eqnarray}
{\cal G}^{a}\left(\tau,\vec{k}\right)
&=&
\frac{1}{2\epsilon_{G}(\vec{k})}\,
\left[
1+N_{G}\left(
\epsilon_{G}(\vec{k})-i\frac{1}{\beta}\phi^{a}
\right)\right]\,
\exp\left[
-\left(\epsilon_{G}(\vec{k})
-i\frac{1}{\beta}\phi^{a}\right)\tau\right]
\nonumber\\
&~&
+\frac{1}
{2\epsilon_{G}(\vec{k})}\,
\left[
N_G\left(\epsilon_{G}(\vec{k})
+i\frac{1}{\beta}\phi^{a}
\right)
\right]
\exp\left[+\left(\epsilon_{G}(\vec{k})
+i\frac{1}{\beta}\phi^{a}\right)\tau\right],
\end{eqnarray}
where $N_{G}(x)=\frac{1}{e^{x}-1}$ 
is Bose-Einstein partition function and 
$\epsilon_{G}(\vec{k})=\sqrt{\vec{k}^2}=|\vec{k}|$.
The quark self-energy is calculated 
by integrating out the internal momentum 
as follows
\begin{eqnarray}
{ {\left.{\Sigma}_{Q}\right.}_{ij} }(p_0,\vec{p})
&=&
\left({\bf t}^{a}\right)_{in}\,
\left({\bf t}^{b}\right)_{mj}\,
{ {\left.{\Sigma}_{Q}\right.}_{nm}^{ab} }(p_0,\vec{p}),
\end{eqnarray}
where
\begin{eqnarray}
{ {\left.{\Sigma}_{Q}\right.}_{nm}^{ab} }(p_0,\vec{p})
&=&
\int\frac{d^{3}\vec{k}}{(2\pi)^3}
\int^{\beta}_{0} d\tau
\exp\left[
\left(
p_0-\mu_{Q}
-i\frac{1}{\beta}\theta_{n}-i\frac{1}{\beta}\phi^{a}
\right)\tau\right]
\nonumber\\
&~&~~~\times
\,{ {\left.{\widetilde{\Sigma}}_{Q}\right.}_{nm}^{ab} }
(\tau,\tau,\vec{p},\vec{k}).
\label{quark-self_p1}
\end{eqnarray}
The term ${ {\left.{\Sigma_{Q}}\right.}_{nm}^{ab} }(p_0,\vec{p})$
that is appeared in Eq.~(\ref{quark-self_p1}) 
can be projected using the Foldy-Wouthuysen 
transformation to positive and negative energy components. 
It is reduced to the following superposition function,
\begin{eqnarray}
{ {\left.{\Sigma_{Q}}\right.}_{nm}^{ab} }\left(p_0,\vec{p}\right)
&=& 
\sum_{r=\pm}^{\pm} \sum_{s=\pm}^{\pm} 
{ {\left.{\Sigma^{(r s)}_{Q}}\right.}^{ab}_{nm} }(p_0,\vec{p}),
\nonumber\\
&=& 
\sum_{r=\pm}^{\pm} \sum_{s=\pm}^{\pm} 
{ {\left.{\Sigma^{(r s)}_{Q}}\right.}^{a}_{n} }(p_0,\vec{p})
\,\delta_{nm}\,\delta^{ab}.
\label{quark-self_p2}
\end{eqnarray}
It is evident that Eq.~(\ref{quark-self_p2})
is diagonalized with respect to 
the adjoint and fundamental color indexes alike.
This result is a natural one since the fundamental color 
index $n$ represents the internal quark segment 
with the fundamental color charge conservation $\delta_{n\,m}$ 
while the adjoint color index $a$ represents 
the internal gluon segment with the adjoint color charge 
conservation $\delta^{a\,b}$.
The Foldy-Wouthuysen energy components of 
Eq.~(\ref{quark-self_p2}) is calculated as follows
\begin{eqnarray}
{ {\left.{\Sigma_{Q}^{(r s)}}\right.}^{a}_{n} }(p_0,\vec{p})
&=&
g^2\,\left(\int\frac{d^{3}\vec{k}}{(2\pi)^3}\,
g_{\mu\nu}\,
{ {\bf S}^{(r)}_{Q} }^{\mu\nu}(\vec{p}-\vec{k})
{ {\bf D}_{Q}^{(r s)} }^{a}_{n}(p,\vec{k})
\right), 
\label{quark-self_p3}
\end{eqnarray}
where $r=\pm$, $s=\pm$ are 
the positive and negative energy components.
The sandwiched Foldy-Wouthuysen projectors by
the gamma matrices 
$\gamma^{\mu}\otimes\gamma^{\nu}$
are given by,
\begin{eqnarray}
{{\bf S}^{(r)}_{Q}}^{\mu\nu}(\vec{p}-\vec{k})
&=&
\left[
\gamma^{\mu}\,
{\Lambda^{(r)}_{Q}}(\vec{p}-\vec{k})
\gamma_0\,\gamma^{\nu}
\right].
\end{eqnarray}
The explicit expressions for 
${ {\bf D}_{Q}^{(rs)} }^{a}_{n}(p,\vec{k})$
are reduced to
\begin{eqnarray}
\left(
{ {\bf D}_{Q}^{(++)} }^{a}_{n}(p,\vec{k})
\right)
&=&
-\frac{\left[1-
n_{F}\left(
\epsilon_{Q}(\vec{p}-\vec{k})-\mu_{Q}-i\frac{1}{\beta}\theta_{n}
\right)
+N_G\left(\epsilon_{G}(\vec{k})-i\frac{1}{\beta}\phi^{a}
\right)
\right]}
{
2\epsilon_{G}(\vec{k})
\left[
p_0-
\left(
\epsilon_{G}(\vec{k})+\epsilon_{Q}(\vec{p}-\vec{k})
\right)\right]
},
\end{eqnarray}
\begin{eqnarray}
\left(
{ {\bf D}_{Q}^{(+-)} }^{a}_{n}(p,\vec{k})
\right)
&=&
-\frac{\left[
n_{F}\left(
\epsilon_{Q}(\vec{p}-\vec{k})-\mu_Q-i\frac{1}{\beta}\theta_{n}
\right)
+
N_{G}\left(\epsilon_{G}(\vec{k})+i\frac{1}{\beta}\phi^{a}
\right)
\right]}
{
2\epsilon_{G}(\vec{k})
\left[
p_0
+\left(
\epsilon_{G}(\vec{k})-\epsilon_{Q}(\vec{p}-\vec{k})\right)
\right]
},
\end{eqnarray}
\begin{eqnarray}
\left(
{ {\bf D}_{Q}^{(-+)} }^{a}_{n}(p,\vec{k})
\right)
&=&
-\frac{\left[
n_{F}\left(
\epsilon_{Q}(\vec{p}-\vec{k})+\mu_{Q}+i\frac{1}{\beta}\theta_{n}
\right)
+
N_{G}\left(\epsilon_{G}(\vec{k})-i\frac{1}{\beta}\phi^{a}
\right)
\right]}
{
2\epsilon_{G}(\vec{k})
\left[
p_0
-\left(\epsilon_{G}(\vec{k})
-\epsilon_{Q}(\vec{p}-\vec{k})\right)
\right]},
\end{eqnarray}
and
\begin{eqnarray}
\left(
{{\bf D}_{Q}^{(--)}}^{a}_{n}(p,\vec{k})
\right)
&=&
-\frac{\left[
1-
n_{F}\left(
\epsilon_{Q}(\vec{p}-\vec{k})+\mu_Q+i\frac{1}{\beta}\theta_{n}
\right)
+
N_G\left(\epsilon_{G}(\vec{k})+i\frac{1}{\beta}\phi^{a}
\right)
\right]}
{
2\epsilon_{G}(\vec{k})
\left[
p_0+
\left(\epsilon_{G}(\vec{k})
+\epsilon_{Q}(\vec{p}-\vec{k})\right)
\right]},
\end{eqnarray}
for the positive-positive, positive-negative, negative-positive
and negative-negative energy components, respectively.
In the derivation of Eq.~(\ref{quark-self_p1}), 
we have considered the following relations 
for the Matsubara frequency for the momentum  
time-component in the imaginary-time formalism,
\begin{eqnarray}
\begin{array}{l}
p_0-\mu_{Q}-i\frac{1}{\beta}\theta_{n}
-i\frac{1}{\beta}\phi^{a}
=i\frac{(2m+1)\pi}{\beta},
\\
\exp\left(
p_0-\mu_{Q}-i\frac{1}{\beta}\theta_{n}
-i\frac{1}{\beta}\phi^{a}
\right)=-1,
\end{array}
\end{eqnarray}
where the Fermion Matsubara frequency is given by
$\omega=(2m+1)\pi/\beta$ and $m=1,2,\cdots\,$. 
The HTL are one-loop diagrams 
for which only the contribution 
from hard loop momenta of the order $T$
or larger is considered. 
In this approximation, 
the external momentum is soft 
$p\sim g\,T$, while the internal momentum is hard $k\sim T$. 
Hence, the internal momentum $k$ is assumed relatively large 
in comparison with the external one (i.e. $p/k\sim g$).
In order to simplify the calculations, 
it is useful to adopt the following approximations
\begin{eqnarray}
g_{\mu\nu}\,
{{\bf S}_{Q}^{(\pm)\,\mu\nu}}(\vec{p}-\vec{k})
&=&
g_{\mu\nu}\,\left[
\gamma^{\mu}
\,{\Lambda^{(\pm)}_{Q}}(\vec{p}-\vec{k})
\gamma_0\gamma^{\nu}\right],
\nonumber\\
&\approx&
-\left[\gamma_0\pm\vec{\gamma}\cdot\hat{k}\right].
\end{eqnarray}
It is also useful to introduce the following approximation 
for the composites with the hard internal momentum 
$k\sim\, T$ and the soft external momentum $p\sim\,g\,T$,
\begin{eqnarray}
\epsilon_{G}(\vec{k})+\epsilon_{Q}(\vec{p}-\vec{k})
&\approx&\, 2\, |\vec{k}|\sim\, 2\, T,
\end{eqnarray}
and 
\begin{eqnarray}
\epsilon_{G}(\vec{k})-\epsilon_{Q}(\vec{p}-\vec{k})
&\approx& \hat{k}\cdot\vec{p} = |\vec{p}|\cos\theta,
\nonumber\\
&\sim& g T \cos\theta.
\end{eqnarray}
Therefore, under the assumption of the HTL approximation, 
the positive-positive energy component 
of  the quark-gluon loop part of 
the quark self-energy is reduced to 
\begin{eqnarray}
{ {\left.{\Sigma_{Q}^{(++)}}\right.}^{a}_{n} }(p_0,\vec{p})
&=&
g^2\,\int\frac{d^3\vec{k}}{(2\pi)^3}\,
\frac{1}{2\epsilon_{G}(\vec{k})}
\frac{\left(\gamma_0+\vec{\gamma}\cdot\hat{k}\right)}
{\left(
p_0-\epsilon_{Q}(\vec{p}-\vec{k})-\epsilon_{G}(\vec{p})
\right)}
\nonumber\\
&~&
~~\times
\left[
-n_{F}\left(
\epsilon_{Q}(\vec{p}-\vec{k})-\mu_{Q}-i\frac{1}{\beta}\theta_{n}
\right)
+N_G\left(\epsilon_{G}(\vec{k})-i\frac{1}{\beta}\phi^{a}
\right)
\right],
\nonumber\\
&\approx&
g^2\int\frac{d^3\vec{k}}{(2\pi)^3}
\frac{1}{4|\vec{k}|^2}
\left(\gamma_0-\vec{\gamma}\cdot\hat{k}\right)
\nonumber\\
&~&
~~\times
\left[
n_{F}\left(
\epsilon_{Q}(\vec{k})-\mu_{Q}-i\frac{1}{\beta}\theta_{n}
\right)
-N_G\left(\epsilon_{G}(\vec{k})-i\frac{1}{\beta}\phi^{a}
\right)
\right].
\end{eqnarray}
The sum of the positive-positive and negative-negative 
energy components is approximated to
\begin{eqnarray}
\left[
{ {\left.{\Sigma_{Q}^{(++)}}\right.}^{a}_{n} }\left(p_0,\vec{p}\right)
+
{ {\left.{\Sigma_{Q}^{(--)}}\right.}^{a}_{n} }\left(p_0,\vec{p}\right)
\right]\approx 0.
\end{eqnarray}
The results for the positive-negative 
and negative-positive energy components
are finite in the HTL approximation.
The positive-negative energy component reads 
\begin{eqnarray}
{ {\left.{\Sigma_{Q}^{(+-)}}\right.}^{a}_{n} }\left(p_0,\vec{p}\right)
&=&
g^2\,\int\frac{d^3\vec{k}}{(2\pi)^3}\,
\frac{1}{2\epsilon_{G}(\vec{k})}\,
\frac{\left(\gamma_0+\vec{\gamma}\cdot\hat{k}\right)}
{\left(
p_0-\epsilon_{Q}(\vec{p}-\vec{k})+\epsilon_{G}(\vec{p})
\right)}
\nonumber\\
&~&
~~~
\times\left[
n_{F}\left(
\epsilon_{Q}(\vec{p}-\vec{k})-\mu_{Q}-i\frac{1}{\beta}\theta_{n}
\right)
+
N_G\left(\epsilon_{G}(\vec{k})+i\frac{1}{\beta}\phi^{a}
\right)
\right].
\label{quark-self-pn-a}
\end{eqnarray}
The $(+-)$-component with the assumptions of 
the soft external momentum $p$  and 
the hard internal momentum $k$ 
is simplified to
\begin{eqnarray}
{ {\left.{\Sigma_{Q}}\right.}^{a}_{n} }^{(+-)}\left(p_0,\vec{p}\right)
&\approx&
g^2\,\int\frac{d^3\vec{k}}{(2\pi)^3}\,
\frac{1}{2|\vec{k}|}
\left[\frac{\gamma_0-\vec{\gamma}\cdot\hat{k}}
{p_0-\hat{k}\cdot\vec{p}}\right]
\nonumber\\
&~&
~~~
\times
\left[
n_{F}\left(
\epsilon_{Q}(\vec{k})-\mu_{Q}-i\frac{1}{\beta}\theta_n
\right)
+
N_G\left(\epsilon_{G}(\vec{k})+i\frac{1}{\beta}\phi^{a}\right)
\right].
\end{eqnarray}
Furthermore, the symmetry over the polar integration 
for the negative-positive and positive-negative energy 
components leads to
\begin{eqnarray}
\int \frac{d\Omega}{4\pi}
\left[
\frac{\gamma_0+\vec{\gamma}\cdot\hat{k}}
{p_0+\vec{p}\cdot\hat{k}}
\right]&=&
\int \frac{d\Omega}{4\pi}
\left[
\frac{\gamma_0-\vec{\gamma}\cdot\hat{k}}
{p_0-\vec{p}\cdot\hat{k}}
\right].
\end{eqnarray}
The sum of the positive-negative and negative-positive energy components 
is found finite while the sum of positive-positive and negative-negative 
components is approximated to zero. 
Therefore, the quark self-energy component, namely, 
${ {\left.{\Sigma_{Q}}\right.}^{a}_{n} }\left(p_0,\vec{p}\right)$
is reduced to
\begin{eqnarray}
{ {\left.{\Sigma_{Q}}\right.}^{a}_{n} }\left(p_0,\vec{p}\right)
&=&
{ {\left.{\Sigma_{Q}^{(+-)}}\right.}^{a}_{n} }\left(p_0,\vec{p}\right)
+
{ {\left.{\Sigma_{Q}^{(-+)}}\right.}^{a}_{n} }\left(p_0,\vec{p}\right),
\nonumber\\
&=&
\left({\omega}^2_{0\,Q}\right)^{a}_{n}
\int \frac{d\Omega}{4\pi}\,
\left[
\frac{\gamma_0-\vec{\gamma}\cdot\hat{k}}
{p_0-\hat{k}\cdot\vec{p}}
\right].
\end{eqnarray}
The quark plasma frequency that is labeled 
by the internal quark and gluon color indexes namely 
$n$ and $a$, respectively, is defined by
\begin{eqnarray}
\left({\omega}^2_{0\,Q}\right)^{a}_{n}&=&
\left[
\left({\omega}^2_{0\,Q (Q)}\right)_{n}
+
\left({\omega}^2_{0\,Q (G)}\right)^{a}
\right].
\end{eqnarray}
The contributions of quark and gluon segments 
of the quark-gluon loop to the fermion Landau frequency 
are given by,
\begin{eqnarray}
\left({\omega}^2_{0\,Q (Q)}\right)_n
&=&
\frac{1}{2}g^2\int \frac{d|\vec{k}| |\vec{k}|}{2\pi^2 }
\left[
n_{F}\left(
\epsilon_{Q}(\vec{k})-\mu_{Q}-i\frac{1}{\beta}\theta_{n}
\right)
+
n_{F}\left(
\epsilon_{Q}(\vec{k})+\mu_{Q}+i\frac{1}{\beta}\theta_{n}
\right)\right],
\nonumber\\
&=&
\frac{g^2}{4\pi^2}\,
\left[
\frac{\pi^2}{6\,\beta^2}
+\frac{1}{2}
\left(
\mu_{Q}+i\,\frac{1}{\beta}\theta_{n}
\right)^2
\right],
\label{landau-freq-quark}
\end{eqnarray}
and
\begin{eqnarray}
\left({\omega}^2_{0\,Q (G)}\right)^{a}
&=&
\frac{1}{2}g^2\int \frac{d|\vec{k}| |\vec{k}|}{2\pi^2 }
\left[
N_G\left(\epsilon_{G}(\vec{k})-i\frac{1}{\beta}\phi^{a}\right)
+
N_G\left(\epsilon_{G}(\vec{k})+i\frac{1}{\beta}\phi^{a}\right)
\right],
\nonumber\\
&=&
\frac{g^2}{4\pi^2}\,
\left[
\frac{\pi^2}{3\,\beta^2}
-\frac{1}{2}\left(i\,\frac{1}{\beta}\phi^{a}\right)^2
+i\,\frac{\pi}{\beta}\left(i\,\frac{1}{\beta}\phi^{a}\right)
\right],
\nonumber\\
&=&
\frac{g^2}{4\pi^2\,\beta^{2}}\,
\left[
\frac{1}{2}\left(\phi^{a}\right)^2
-\pi\left(\phi^{a}\right)
+\frac{\pi^2}{3}
\right],
\label{landau-freq-gluon}
\end{eqnarray}
respectively.
In the case of real color chemical potentials, the fundamental 
and adjoint color chemical potentials are reduced to
$i\,\frac{1}{\beta}\theta_{n}\rightarrow {\mu_{C}}_{n}$
and
$i\,\frac{1}{\beta}\phi^{a}\equiv i\,\frac{1}{\beta}\phi^{\underbrace{(AB)}}
\rightarrow {\mu_{C}}^{a}\equiv\left({\mu_{C}}_{A}-{\mu_{C}}_{B}\right)$, 
respectively.
Hence, the quark and gluon contributions to the Landau fermion frequency
which are given by Eqns.~(\ref{landau-freq-quark}) and (\ref{landau-freq-gluon}), 
are reduced to
\begin{eqnarray}
\left({\omega}^2_{0\,Q (Q)}\right)_n
&=&
\frac{g^2}{4\pi^2}\,
\left[
\frac{\pi^2}{6\,\beta^2}
+\frac{1}{2}
\left(
\mu_{Q}+{\mu_{C}}_{n}
\right)^2
\right],
\end{eqnarray}
and
\begin{eqnarray}
\left({\omega}^2_{0\,Q (G)}\right)^{a}
&=&
\frac{g^2}{4\pi^2}\,
\left[
\frac{\pi^2}{3\,\beta^2}
-\frac{1}{2}\left(
{\mu_{C}}^{a}
\right)^2
+i\,\frac{\pi}{\beta}\left(
{\mu_{C}}^{a}
\right)
\right]
\nonumber\\
&\equiv&
\frac{g^2}{4\pi^2}\,
\left[
\frac{\pi^2}{3\,\beta^2}
-\frac{1}{2}\left(
{\mu_{C}}_{A}-{\mu_{C}}_{B}
\right)^2
+i\,\frac{\pi}{\beta}\left(
{\mu_{C}}_{A}-{\mu_{C}}_{B}
\right)
\right],
\end{eqnarray}
respectively.
Finally, it is worth to mention that it is possible to transform 
the real flavor chemical potential $\mu_{Q}$ 
to an imaginary chemical potential 
$\mu_{Q}\rightarrow i\,\frac{1}{\beta}\,{\theta_{fl}}_{Q}$. 
In this case, Eq.~(\ref{landau-freq-quark}) is reduced to
\begin{eqnarray}
\left({\omega}^2_{0\,Q (Q)}\right)_n
&=&
\frac{g^2}{4\pi^2}\,
\left[
\frac{\pi^2}{6\,\beta^2}
-\frac{1}{2}\frac{1}{\beta^2}
\left(
{\theta_{fl}}_{Q}+\theta_{n}
\right)^2
\right].
\end{eqnarray}
Hereinafter, the real chemical potentials for the flavor and color degrees 
of freedom will be considered in the discussion of the physical aspects 
of  the weakly interacting quark-gluon plasma 
above the critical point of deconfinement phase transition.
The quark self-energy with the external quark line indexes $i$ and $j$ 
is written as follows
\begin{eqnarray}
{ {\left.{\Sigma_{Q}}\right.}_{ij} }(p_0,\vec{p})
&=&
\left({\bf t}^{a}\right)_{in}\, \left({\bf t}^{b}\right)_{mj}\,
{ {\left.{\Sigma_{Q}}\right.}^{ab}_{nm} }\left(p_0,\vec{p}\right),
\nonumber\\
&=&
\left({\omega}^{2}_{0\,Q}\right)_{ij}
\,\left[
\int \frac{d\Omega}{4\pi}\,
\frac{\gamma_0-\vec{\gamma}\cdot\hat{k}}
{p_0-\hat{k}\cdot\vec{p}}
\right].
\end{eqnarray}
The quark plasma frequency that is labeled by 
the external quark line
(i.e. the external two legs of the internal quark-gluon loop) 
color indexes $i$ and $j$  is given by
\begin{eqnarray}
\left({\omega}^{2}_{0\,Q}\right)_{ij}&=&
\left({\bf t}^{a}\right)_{in} \left({\bf t}^{b}\right)_{mj}
\delta_{nm}\delta_{ab}
\left({\omega}^{2}_{0\,Q}\right)^{a}_n,
\nonumber\\
&=&
\left({\bf t}^{a}\right)_{in} \left({\bf t}^{b}\right)_{mj}
\delta_{nm}\delta_{ab}
\left[
\left({\omega}^{2}_{0\,Q}\right)_n
+\left({\omega}^{2}_{0\,Q}\right)^a
\right],
\end{eqnarray}
where the elements $\left({\bf t}^{a}\right)_{in}$ and
$\left({\bf t}^{b}\right)_{mj}$ are the vertexes which 
connect the external quark line with 
the internal quark and gluon loop segments.
The conservation of quark (line) color charge contracts 
the external color indexes $i$ and $j$ 
for the quark plasma frequency to
$\delta_{ij}\left({\omega}^{2}_{0\,Q}\right)_{i}$.
The adjoint color index $a$ is set 
to $a=\underbrace{(AB)}$ where $a=1\dots N^2_c-1$ 
while the fundamental-like indexes $A$ and $B$ 
run over $1\cdots N_c$.
For example, the adjoint color index set 
$a=(1,2,3,4,5,6,7,8)$ 
corresponds to
$a=\left(\underbrace{(11)}, \underbrace{(12)},
\underbrace{(13)},\underbrace{(21)},\underbrace{(22)},
\underbrace{(23)},\underbrace{(31)},\underbrace{(32)}\right)$.
The quark screening mass (i.e. Landau frequency) 
is defined as follows
\begin{eqnarray}
\left({\omega}^{2}_{0\,Q}\right)_{ij}
&=&
\left({\bf t}^{a}\right)_{in} \left({\bf t}^{b}\right)_{mj}
\delta_{nm}\delta_{ab}
\left[
\left({\omega}^{2}_{0\,Q (Q)}\right)_{n}
+
\left({\omega}^{2}_{0\,Q (G)}\right)^{a}
\right],
\nonumber\\
&\sim&
\frac{1}{2}
\left({\tau}^{\underbrace{(AB)}}\right)_{in} 
\left({\tau}^{\underbrace{(BA)}}\right)_{nj}
\left[
\left(
{\omega}^{2}_{0\,Q (Q)}
\right)_{n}
+
\left(
{\omega}^{2}_{0\,Q (G)}
\right)^{\underbrace{(AB)}}
\right].
\end{eqnarray}
The contraction over the fundamental color indexes 
$i\,j$ in the $SU(N_c)$ representation is reduced to 
\begin{eqnarray}
\left({\omega}^{2}_{0\,Q}\right)_{ij}
&=&
\left({\omega}^{2}_{0\,Q (Q)}\right)_{ij}
+
\left({\omega}^{2}_{0\,Q (G)}\right)_{ij},
\nonumber\\
&=&
\delta_{ij}\,\left(
\left({\omega}^{2}_{0\,Q (Q)}\right)_{j}
+
\left({\omega}^{2}_{0\,Q (G)}\right)_{j}
\right),
\nonumber\\
&=&\delta_{ij}\,\left({\omega}^{2}_{0\,Q}\right)_{j}.
\end{eqnarray}
The quark and gluon segments of 
the quark-gluon loop part of the quark self-energy 
gives the following Landau frequency elements, respectively,
\begin{eqnarray}
\left({\omega}^{2}_{0\,Q (Q)}\right)_{i}
&=&
\frac{1}{2N_c}
\left(
N_c\sum^{N_c}_{n=1}
\left({\omega}^{2}_{0\,Q}\right)_n
-\left({\omega}^{2}_{0\,Q} \right)_i
\right),
\end{eqnarray}
and
\begin{eqnarray}
\left({\omega}^{2}_{0\,Q (G)}\right)_{i}
&=&
\frac{1}{2N_c}
\left(
N_c\sum^{N_c}_{n=1}
\left({\omega}^{2}_{0\,Q}\right)^{\underbrace{(in)}}
-\left({\omega}^{2}_{0\,Q}\right)^{\underbrace{(ii)}}
\right).
\end{eqnarray}
Hence, the quark plasma frequency is given by
\begin{eqnarray}
\left({\omega}^{2}_{0\,Q}
\right)_{i}
&=&
\left({\omega}^{2}_{0\,Q (Q)}\right)_{i}
+
\left({\omega}^{2}_{0\,Q (G)}\right)_{i}.
\end{eqnarray}
The quark self-energy becomes
\begin{eqnarray}
{ {\left.{\Sigma_{Q}}\right.}_{ij} }(p_0,\vec{p})
&=&
\delta_{ij}\,
\left({\omega}^{2}_{0\,Q}\right)_{j}\,
\int \frac{d\Omega}{4\pi}
\left[
\frac{\gamma_0-\vec{\gamma}\cdot\hat{k}}
{p_0-\hat{k}\cdot\vec{p}}
\right].
\end{eqnarray}
Under certain circumstances, it possible to take 
the average over external color indexes to calculate the mean value
of the Landau frequency over the color charges. 
The color mean value of the quark plasma frequency is given by
\begin{eqnarray}
\left({\omega}^{2}_{0\,Q}\right)
&=&\frac{1}{N_{c}}\,\sum^{N_{c}}_{i=1}\,
\left({\omega}^{2}_{0\,Q}\right)_{i},
\nonumber\\
&=&
\frac{1}{2N_c}
\sum^{N_c}_{n=1}\sum^{N_c}_{m=1}
\left(1-\frac{\delta_{nm}}{N_c}\right)
\left[\left(
{\omega}^{2}_{0\,Q}\right)_{n}
+\left({\omega}^{2}_{0\,Q}\right)^{\underbrace{(nm)}}
\right].
\end{eqnarray}
%
%

\subsection{The imaginary part}
The analytic continuation of the 
momentum time-component
$p_0=p_{0}+i\eta$ develops an imaginary part 
for the space-like energy domain
$(p_{0}^2-\vec{p}^2)\le 0$.
This imaginary part describes 
the Landau-damping of the soft fermion field.
The analytic continuation of $p_{0}$ develops 
real and imaginary parts for the
quark self-energy as follows
\begin{eqnarray}
{\left.{\Sigma_{Q}}\right.}_{ij}\left(p_{0},\vec{p}\right)
&=&
\delta_{ij}\,
\left({\omega}^{2}_{0\,Q}\right)_{j}
\,
\left[\int \frac{d\Omega}{4\pi}\,
\frac{\left(\gamma_0-\vec{\gamma}\cdot\hat{k}\right)}
{\left(p_{0}-\hat{k}\cdot\vec{p}+i\eta\right)}
\right],
\nonumber\\
&=&
\Re{e}\,
\left(
{\left.{\Sigma_{Q}}\right.}_{ij}\left(p_{0},\vec{p}\right)
\right)
+
i\,\Im{m}\,
\left(
{\left.{\Sigma_{Q}}\right.}_{ij}\left(p_{0},\vec{p}\right)
\right).
\end{eqnarray}
The real and imaginary parts are not developed simultaneously 
but rather the real part is developed 
in the time-like energy domain
$\left(p_{0}^2-\vec{p}^{2}\right) > 0$
and then the real part is suppressed and 
is replaced by the imaginary part 
when the time-like energy turns to space-like energy.
Therefore, the real part is developed 
in the time-like energy domain 
$\vec{p}^{2}<p^{2}_{0}$. 
It is reduced to
\begin{eqnarray}
\Re{e}\,
\left(
{\left.{\Sigma_{Q}}\right.}_{ij}\left(p_{0},\vec{p}\right)
\right)
&=&\delta_{ij}\,
\left({\omega}^{2}_{0\,Q}\right)_{j}\,
\mbox{P}
\left[\int \frac{d\Omega}{4\pi}\,
\frac{\gamma_0-\vec{\gamma}\cdot\hat{k}}
{\left(p_{0}-\hat{k}\cdot\vec{p}\right)}
\right],
\nonumber\\
&=&
\delta_{ij}\,\left[ 
\Re{e}\,
\left(
{\left.{\Sigma_{Q}}\right.}_{j}\left(p_{0},\vec{p}\right)
\right)\right],
\nonumber\\
&~&  
~~~\left(
\mbox{with the time-like energy}~ p_{0}^{2}>\vec{p}^2
\right).
\end{eqnarray}
After evaluating the principal value integral, the real part becomes
\begin{eqnarray}
\Re{e}\,
\left(
{\left.{\Sigma_{Q}}\right.}_{j}\left(p_{0},\vec{p}\right) \right)
&=&
\left({\omega}^{2}_{0\,Q}\right)_{j}\,
\left[
\frac{1}{p_{0}}\,Q\left(\frac{p_{0}}{|\vec{p}|}\right)\,
\gamma_0
-\frac{1}{|\vec{p}|}
\left[-1+Q\left(\frac{p_{0}}{|\vec{p}|}\right)\right]\,
\vec{\gamma}\cdot\hat{p}
\right],
\nonumber\\
&~&  
~~~\left(
\mbox{with the time-like energy}~ p^2_{0}>\vec{p}^2
\right).
\end{eqnarray}
The imaginary part is developed in 
the space-like energy domain,
$\vec{p}^2\ge p^2_{0}$, 
and it is reduced to
\begin{eqnarray}
\Im{m}\,
\left( 
{\left.{\Sigma_{Q}}\right.}_{ij}\left(p_{0},\vec{p}\right) 
\right)
&=&
\delta_{ij}\,
{\left({\omega}^{2}_{0\,Q}\right)}_{j}\,
\left[
-\pi\int\frac{d\Omega}{4\pi}\,
\left(\gamma_0-\vec{\gamma}\cdot\hat{k}\right)
\,\delta\left(p_{0}-\hat{k}\cdot\vec{p}\right)
\right],
\nonumber\\
&=&\delta_{ij}\,
\left[
\Im{m}\,
\left( 
{\left.{\Sigma_{Q}}\right.}_{j}\left(p_{0},\vec{p}\right) 
\right)
\right],
\nonumber\\
&~&  
~~~\left(
\mbox{with the space-like energy}~
\vec{p}^2\ge p^2_{0}
\right).
\end{eqnarray}
After evaluating the integral over the Dirac-delta function,
the imaginary part becomes
\begin{eqnarray}
\Im{m}\,
\left(
{\left.{\Sigma_{Q}}\right.}_{ij}\left(p_{0},\vec{p}\right)
\right)
&=&
\delta_{ij}\,
{\left({\omega}^{2}_{0\,Q}\right)}_{j}\,
\left[
-\frac{\pi}{2|\vec{p}|}\,
\left(
\gamma_0-\frac{p_{0}}{|\vec{p}|}\vec{\gamma}\cdot\hat{p}
\right)\,
\theta\left(\frac{|\vec{p}|}{|p_{0}|}-1\right)
\right],
\nonumber\\
&~&  
~~~\left(\mbox{with the space-like energy} 
~ \vec{p}^2\ge p^2_{0}\right).
\end{eqnarray}

\subsection{Effective propagator for the soft quark}
In order to find the quark propagator inverse, 
it is more convenience to decompose the quark self-energy 
to the Foldy-Wouthuysen positive and negative energy 
components as follows
\begin{eqnarray}
{\left.{\Sigma_{Q}}\right.}_{ij}(p_{0},\vec{p})
&=&
\delta_{ij}\,
\left[
{\left.{\Sigma_{Q}^{(+)}}\right.}_{j}(p_{0},\vec{p})
\,
\gamma_0\,{\Lambda^{(+)}_{Q}}(\vec{k})
-
{\left.{\Sigma_{Q}^{(-)}}\right.}_{j}(p_{0},\vec{p})
\,
\gamma_0\,{\Lambda^{(-)}_{Q}}(\vec{k})
\right],
\end{eqnarray}
where the positive and negative energy components, 
respectively, read
\begin{eqnarray}
{\left.{\Sigma_{Q}^{(+)}}\right.}_{j}(p_{0},\vec{p})
&=&
{\left({\omega}^{2}_{0\,Q}\right)}_{j}\,
\left[
\frac{1}{|\vec{p}|}
+
\left(\frac{1}{p_{0}}-\frac{1}{|\vec{p}|}
\right)\,
Q\left(\frac{p_{0}}{|\vec{p}|}\right)
\right],
\end{eqnarray}
and
\begin{eqnarray}
{\left.{\Sigma_{Q}^{(-)}}\right.}_{j}(p_{0},\vec{p})
&=&
{\left({\omega}^{2}_{0\,Q}\right)}_{j}\,
\left[
\frac{1}{|\vec{p}|}
-
\left(\frac{1}{p_{0}}+\frac{1}{|\vec{p}|}
\right)\,
Q\left(\frac{p_{0}}{|\vec{p}|}\right)
\right].
\end{eqnarray}
The inverse of the effective soft quark propagator 
is calculated
by adding the quark self-energy to the inverse 
of the bar quark propagator as follows
\begin{eqnarray}
~~~ ~~~
{ {^{*}{\cal S}}^{-1}_{Q} }_{ij}(p_0,\vec{p})
&=&
{ {\cal S}^{-1}_{Q} }_{ij}(p_0,\vec{p})
+
i\,{ {\left.{\Sigma_{Q}}\right.}_{ij} }(p_{0},\vec{p}),
\nonumber\\
\mbox{or}
~~~
i\,{ {^{*}{\cal S}}^{-1}_{Q} }_{ij}(p_0,\vec{p})
&=&
i\,{ {\cal S}^{-1}_{Q} }_{ij}(p_0,\vec{p})
-
{ {\left.{\Sigma_{Q}}\right.}_{ij} }(p_{0},\vec{p}).
\end{eqnarray}
The bar quark propagator is decomposed
to the  Foldy-Wouthuysen positive and negative energy 
components as follows
\begin{eqnarray}
{ {\cal S}_{Q} }_{ij}(p_{0},\vec{p})
&=&
i\,\delta_{ij}
\left[
\frac{{\Lambda^{(+)}_{Q}}(\vec{k})\gamma_0}
{p_{0}-\epsilon_{Q}(\vec{k})}
+
\frac{{\Lambda^{(-)}_{Q}}(\vec{k})\gamma_0}
{p_{0}+\epsilon_{Q}(\vec{k})}
\right].
\end{eqnarray}
The inverse of the bar quark propagator 
is straightforward and it is reduced to
\begin{eqnarray}
{ {\cal S}^{-1}_{Q} }_{ij}(p_{0},\vec{p})&=&
-i\,\delta_{ij}
\left[
\left(p_{0}-\epsilon_{Q}(\vec{k})\right)
\gamma_0\,{\Lambda^{(+)}_{Q}}(\vec{k})
+
\left(p_{0}+\epsilon_{Q}(\vec{k})\right)
\gamma_0\,{\Lambda^{(-)}_{Q}}(\vec{k})
\right].
\end{eqnarray}
The effective soft quark propagator is found as follows, 
\begin{eqnarray}
{\left.{^{*}{\cal S}_{Q}}\right.}_{ij}\left(p_{0},\vec{p}\right)
&=&
i\,\delta_{ij}\left[
\frac{{\Lambda^{(+)}_{Q}}(\vec{p})\,\gamma_0}
{p_{0}-
\left(
{\epsilon_{Q}}(\vec{p})+
{ {\left.{\Sigma_{Q}^{(+)}}\right.}_{j} }(p_{0},\vec{p})
\right)}
+
\frac{{\Lambda^{(-)}_{Q}}(\vec{p})\,\gamma_0}
{p_{0}+
\left(
{\epsilon_{Q}}(\vec{p})+
{ {\left.{\Sigma_{Q}^{(-)}}\right.}_{j} }(p_{0},\vec{p})
\right)}
\right],
\nonumber\\
&=&
i\,\delta_{ij}\,\sum^{\pm}_{r}
\left(
{\left.{^*\Delta^{(r)}_{Q}}\right.}_{j} (p_{0},\vec{p})\right)
{h^{(r)}_{Q}} (\vec{p}),
\end{eqnarray}
where the following positive and negative energy projectors are introduced
\begin{eqnarray}
{h^{(r)}_{Q}}(\vec{p})={\Lambda^{(r)}_{Q}}(\vec{p})\,\gamma_0,
~~~~(\mbox{where}~ r=\pm).
\end{eqnarray}
The positive and negative energy decompositions, respectively, read,
\begin{eqnarray}
{\left.{^*\Delta^{(+)}_{Q}}\right.}_{j}\left(p_{0},\vec{p}\right)
&=&
\frac{1}{p_{0}-
\left({\epsilon_{Q}}(\vec{p})
+
{ {\left.{\Sigma_{Q}^{(+)}}\right.}_{j} }(p_{0},\vec{p})
\right)},
\end{eqnarray}
and
\begin{eqnarray}
{\left.{^*\Delta^{(-)}_{Q}}\right.}_{j}\left(p_{0},\vec{p}\right)
&=&
\frac{1}{p_{0}+
\left({\epsilon_{Q}}(\vec{p})
+
{ {\left.{\Sigma_{Q}^{(-)}}\right.}_{j} }\left(p_{0},\vec{p}\right)
\right)}.
\end{eqnarray}
In order to be able to work in the mixed-time representation
of the imaginary-time formalism of the thermal field theory, we need
to find all the poles of the effective soft quark propagator.
The positive and negative energy poles are found in the following
locations
\begin{eqnarray}
p_{0}={\overline{\omega}_{+}}_{j}:~~ 
\left(
{\overline{\omega}_{+}}_{j}-\left(\epsilon_{Q}(\vec{p})
+
{ {\left.{\Sigma_{Q}^{(+)}}\right.}_{j} }
({\overline{\omega}_{+}}_{j},\vec{p})\right)
\right)&=&0,
\nonumber\\
p_{0}={\overline{\omega}_{-}}_{j}:~~
\left( 
{\overline{\omega}_{-}}_{j}
+\left(\epsilon_{Q}(\vec{p})
+
{ {\left.{\Sigma_{Q}^{(-)}}\right.}_{j} }({\overline{\omega}_{-}}_{j},\vec{p})
\right)
\right)&=&0,
\end{eqnarray}
for the positive and negative Foldy-Wouthuysen energy components.
Therefore, positive and negative energy poles are written, 
respectively, as follows
\begin{eqnarray}
{\varepsilon^{(+)}_{Q}}_{j}(\vec{p})&=&
\left(\epsilon_{Q}(\vec{p})
+
{ {\left.{\Sigma_{Q}^{(+)}}\right.}_{j} }
({\overline{\omega}_{+}}_{j},\vec{p})\right),
~ 
{\overline{\omega}_{+}}_{j}={\varepsilon^{(+)}_{Q}}_{j}(\vec{p}),
~\mbox{and}~
p_{0}={\overline{\omega}_{+}}_{j},
\end{eqnarray}
and
\begin{eqnarray}
{\varepsilon^{(-)}_{Q}}_{j}(\vec{p})&=&
\left(\epsilon_{Q}(\vec{p})
+
{ {\left.{\Sigma_{Q}^{(-)}}\right.}_{j} }
({\overline{\omega}_{-}}_{j},\vec{p})\right),
~
{\overline{\omega}_{-}}_{j}=-{\varepsilon^{(-)}_{Q}}_{j}(\vec{p}),
~\mbox{and}~
p_{0}={\overline{\omega}_{-}}_{j}.
\end{eqnarray}
After introducing the flavor and color chemical potentials,
we can write the effective positive and negative energy poles
as follows,
\begin{eqnarray}
p_{0}\rightarrow\,
{\varepsilon^{*\,(+)}_{Q}}_{j}(\vec{p})&=&
{\varepsilon^{(+)}_{Q}}_{j}(\vec{p})-\mu_{Q}
-i\,\frac{\theta_j}{\beta},
\nonumber\\
p_{0}\rightarrow\,
-{\varepsilon^{*\,(-)}_{Q}}_{j}(\vec{p})&=&
-{\varepsilon^{(-)}_{Q}}_{j}(\vec{p})
-\mu_{Q}-i\,\frac{\theta_j}{\beta}.
\end{eqnarray}
The effective soft quark propagator in 
the mixed-time representation is reduced to
\begin{eqnarray}
i\,{^*\cal S}_{ij}(\tau,\vec{p})
&=&
\delta_{ij}\,\,\left[i\,{^*\cal S}_{j}(\tau,\vec{p})\right],
\nonumber\\
&=&
\delta_{ij}\,\left[
{\Lambda}^{(+)}_{Q}(\hat{p})\,\gamma_{0}\,
{\,^{*}\Delta}^{(+)}_{Q\,j}(\tau,\vec{p})
+
{\Lambda}^{(-)}_{Q}(\hat{p})\,\gamma_{0}\,
{\,^{*}\Delta}^{(-)}_{Q\,j}(\tau,\vec{p})\right],
\end{eqnarray}
where
\begin{eqnarray}
{\,^{*}\Delta}^{(+)}_{Q\,j}(\tau,\vec{p})
&=&
\left[
1-n_{F}\left(
{\varepsilon^{(+)}_{Q}}_{j}(\vec{p})
-\mu_{Q}-i\,\frac{\theta_j}{\beta}
\right)
\right]
\exp\left[-\tau\, 
\left(
{\varepsilon^{(+)}_{Q}}_{j}(\vec{p})
-\mu_{Q}-i\,\frac{\theta_j}{\beta}
\right)
\right],
\end{eqnarray}
and
\begin{eqnarray}
{\,^{*}\Delta}^{(-)}_{Q\,j}(\tau,\vec{p})&=&
n_{F}\left(
{\varepsilon^{(-)}_{Q}}_{j}(\vec{p})
+\mu_{Q}+i\,\frac{\theta_j}{\beta}
\right)
\exp\left[\tau\,
\left(
{\varepsilon^{(-)}_{Q}}_{j}(\vec{p})
+\mu_{Q}+i\,\frac{\theta_j}{\beta}
\right)
\right].
\end{eqnarray}
The mixed-time representation in the imaginary-time formalism
in the thermal field theory 
is represented in the following way
\begin{eqnarray}
{\cal S}\left(i\omega_n+\mu_{Q}+i\frac{\theta_{j}}{\beta}\right)
=\int^{\beta}_{0}d\tau e^{i\omega_n\tau}
{\cal S}^{>}_{j}(\tau), ~~~(\mbox{where}~~ \tau\ge 0),
\end{eqnarray}
and
\begin{eqnarray}
{\cal S}_{j}(\tau)=\frac{1}{\beta}\sum_{n}\exp[-i\omega_n\tau]
{\cal S}\left(
i\omega_n+\mu_{Q}+i\frac{\theta_{j}}{\beta}
\right).
\end{eqnarray}

\section{\label{section5} Gluon self-energy: Feynman diagrams}

\begin{figure} 
\includegraphics{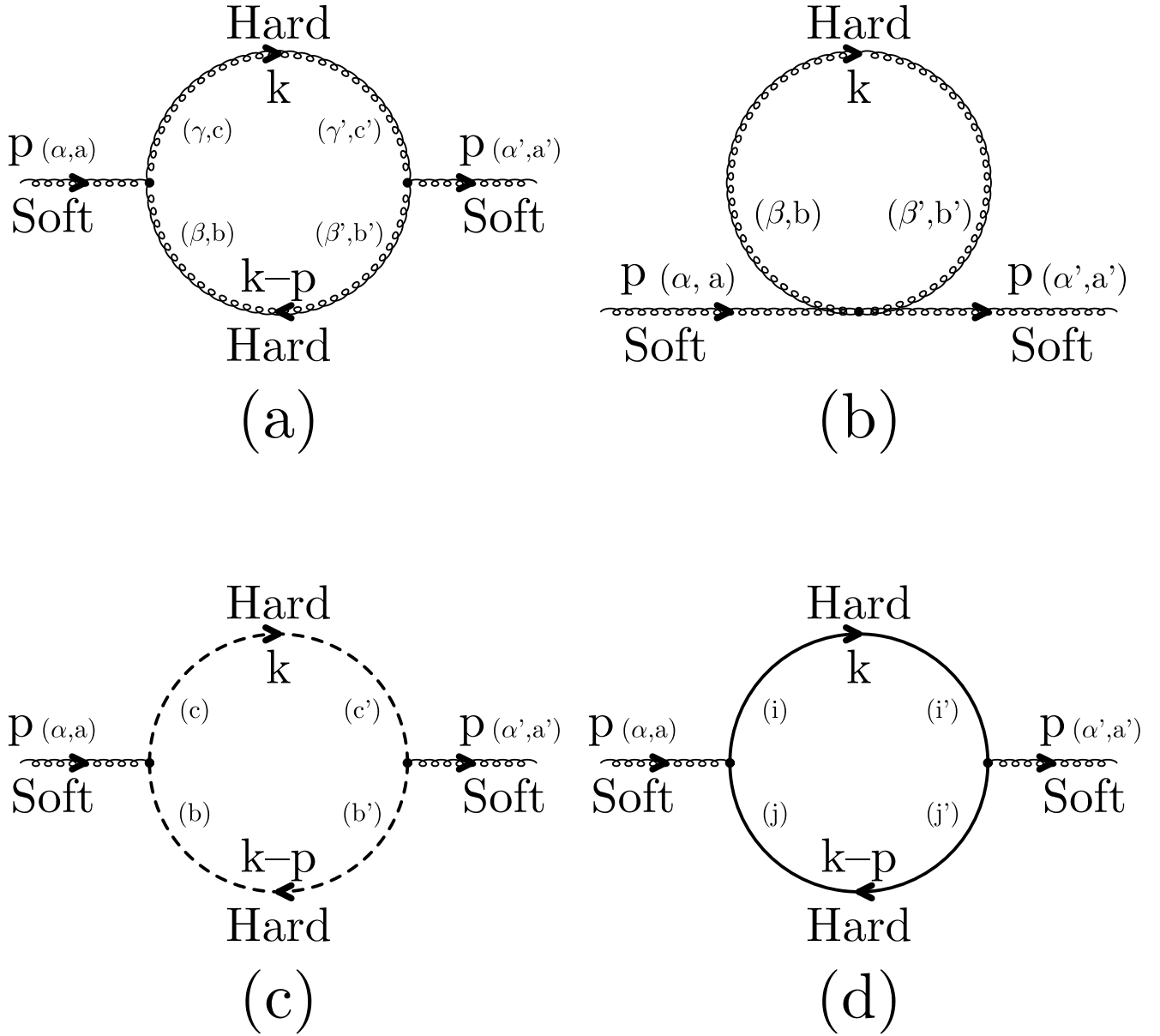}
\caption{\label{fig:gluoneff}
The soft gluon self-energy correction.} 
\end{figure}

The soft gluon polarization tensor, namely, 
${{\Pi}^{\alpha'\alpha}}^{a'a}(p_0,\vec{p})$
is calculated up to the one loop approximation 
and 
the Feynman diagrams for the gluon-loop, tadpole, ghost-loop 
and quark-loop are depicted 
in Figs.~(\ref{fig:gluoneff} a), (\ref{fig:gluoneff} b),
(\ref{fig:gluoneff} c) and (\ref{fig:gluoneff} d), respectively.
Hereinafter, the soft gluon polarization tensor 
is decomposed to
two parts which are the gluon's part, namely,
${{\Pi}^{\alpha'\alpha}_{(gp)}}^{a'a}(p_0,\vec{p})$
and the quark's part, namely, 
${{\Pi}^{\alpha'\alpha}_{(ql)}}^{a'a}(p_0,\vec{p})$.
The gluon's part of the soft gluon polarization tensor, 
${{\Pi}^{\alpha'\alpha}_{(gp)}}^{a'a}(p_0,\vec{p})$,
consists the gluon-loop, tadpole and ghost-loop
that are displayed 
in Figs.~(\ref{fig:gluoneff} a), (\ref{fig:gluoneff} b) 
and
(\ref{fig:gluoneff} c), respectively. 
Furthermore, the contributions of the gluon-loop, tadpole 
and ghost-loop to the soft gluon polarization tensor 
are represented by
${{\Pi}^{\alpha'\alpha}_{(g-l)}}^{a'a}(p_0,\vec{p})$,
${{\Pi}^{\alpha'\alpha}_{(tp)}}^{a'a}(p_0,\vec{p})$,
and
${{\Pi}^{\alpha'\alpha}_{(gh)}}^{a'a}(p_0,\vec{p})$, respectively.
On the other hand, the quark's part of  
the soft gluon polarization tensor, namely,
${{\Pi}^{\alpha'\alpha}_{(ql)}}^{a'a}(p_0,\vec{p})$,
consists only the quark-loop that is depicted 
in Fig.~(\ref{fig:gluoneff} d).
The color indexes are shown explicitly in the figures. 
These diagrams are evaluated in the HTL approximation. 
In the context of the HTL approximation
the external gluon line 
is assumed soft and carries soft momentum 
and energy of order $p\sim g\,T$ 
(i.e. $|\vec{p}|\sim g\,T$ and $p_{0}\sim g\,T$)
while the internal momentum is considered hard 
one and is of oder $k\sim T$. 
The hard internal and soft external momenta 
have the following relationship,
\begin{eqnarray}
p\sim g T \ll k\sim T ~\mbox{and}~ g\ll 1.
\end{eqnarray}
The polarization tensor 
in the HTL approximation is shown 
to be gauge independent.
It is straightforward to construct 
the gluon polarization tensor 
in the Feynman gauge where the final result 
is concluded to be gauge fixing independent.
It is known that in the four space-time dimensions, 
the gluon polarization tensor expression 
develops an ultraviolet divergence. 
Such divergence is eliminated by the gluon wave-function 
renormalization.
Moreover, the thermal contribution of 
the soft gluon polarization tensor 
\begin{eqnarray}
\left.
{ {\Pi}^{\alpha'\alpha} }^{a'a}(p_0,\vec{p})
\right|_{T}
&=&
{ {\Pi}^{\alpha'\alpha} }^{a'a}(p_0,\vec{p})
-
\left.
{ {\Pi}^{\alpha'\alpha} }^{a'a}(p_0,\vec{p})
\right|_{T=0},
\end{eqnarray}
has no ultraviolet divergence, 
since the statistical factors which are given by 
$n_{F}(\epsilon_{Q}(\vec{k}))$ for quark 
and $N_{G}(\epsilon_{G}(\vec{k}))$ for gluon
are exponential decreasing.
Hence, it is sufficient to evaluated the thermal soft gluon 
polarization in the standard 4 space-time dimensions 
and retaining only the thermal terms while dropping 
any term does not have an explicit temperature term. 
There is no need for any renormalization scheme 
in the present calculations.

\subsection{Gluon-loop kernel ${\Pi_{(g-l)}}\left(p,k\right)$}

The gluon loop part of the soft gluon polarization tensor
is depicted in Fig.~(\ref{fig:gluoneff} a). 
It is furnished as follows
\begin{eqnarray}
{\left.{ \Pi_{(g-l)\, \alpha'\alpha}}
\right.}^{a'a}(p,k)
&=&
{\left.{\Pi_{(g-l)}}_{\alpha'\alpha}\right.}^{a'a}(p_{0},\vec{p},k_{0},\vec{k}),
\nonumber\\
&=&
\frac{1}{2}\left(-i\,g\,f^{a'c'b'}\,
\Gamma^{(\mbox{out})}_{\alpha'\gamma'\beta'}(p,-k,-p+k)
\right)
\,{\left.{\cal G}^{\gamma'\gamma}\right.}^{c'c}(k_0,\vec{k})
\nonumber\\
&~&\times
\left(-i\, g\, f^{cab}\,
\Gamma^{(\mbox{in})}_{\gamma\alpha\beta}(-k,p,-p+k)\right)
\,{\left.{\cal G}^{\beta\beta'}\right.}^{b'b}(p_0-k_0,\vec{p}-\vec{k}),
\nonumber\\
&=&\frac{1}{2}
\left(+i\,g\, f^{a'c'b'}\,
\Gamma_{\alpha'\gamma'\beta'}\left(-p,k,p-k\right)\right)
\,{\left.{\cal G}^{\gamma'\gamma}\right.}^{c'c}(k_0,\vec{k})
\nonumber\\
&~&\times
\left(-i\,g\, f^{cab}\,
\Gamma_{\gamma\alpha\beta}(-k,p,-p+k)\right)
{\left.{\cal G}^{\beta\beta'}\right.}^{b'b}(p_0-k_0,\vec{p}-\vec{k}).
\label{gl-q1}
\end{eqnarray}
The inward-vertex 
$\Gamma^{(\mbox{in})}_{\gamma\alpha\beta}(-k,p,-p+k)=
\Gamma_{\gamma\alpha\beta}\left(-k,p,-p+k\right)$ 
has the inward momentum, 
while  the outward-vertex 
$\Gamma^{(\mbox{out})}_{\alpha'\gamma'\beta'}(p,-k,-p+k)=
-\Gamma_{\alpha'\gamma'\beta'}\left(-p,k,p-k\right)$ 
has an outward momentum. 
The gluon vertexes that are appearing in the gluon-loop 
are given by
\begin{eqnarray}
\Gamma_{\beta\gamma\alpha}\left(-k,p,-p+k\right)
&=&
\left[g_{\beta\gamma}(-p+2k)_{\alpha}
+g_{\gamma\alpha}(-k-p)_{\beta}
+g_{\alpha\beta}(2p-k)_{\alpha}
\right],
\nonumber\\
\Gamma_{\gamma'\beta'\alpha'}\left(-p,k,p-k\right)
&=&
-\left[
g_{\gamma'\beta'}(-2k+p)_{\alpha'}
+g_{\beta'\alpha'}(-2p+k)_{\gamma'}
+g_{\alpha'\gamma'}(p+k)_{\beta'}
\right],
\end{eqnarray}
for the in- and out-vertexes, respectively.
The gluon-loop kernel which is given 
by Eq.~(\ref{gl-q1}) is re-written as follows,
\begin{eqnarray}
{\left.{ \Pi_{(g-l)\, \alpha'\alpha}}
\right.}^{a'a}(p,k)
&=&
+\frac{g^2}{2}\,f^{a'c'b'}\, f^{abc}\, 
\Gamma_{\alpha'\gamma'\beta'}\left(-p,k,p-k\right) 
\,\Gamma_{\gamma\alpha\beta}\left(-k,p,-p+k\right)\,
g^{\gamma'\gamma'}\,g^{\beta\beta'}
\nonumber\\
&~&
\times
{\cal G}^{c'c}(k_0,\vec{k})\,
{\cal G}^{b'b}(p_0-k_0,\vec{p}-\vec{k}),
\nonumber\\
&=&-\frac{g^2}{2}({\bf T}^{a'})_{c'b'} ({\bf T}^{a})_{bc}
\Gamma_{\alpha'\gamma'\beta'}\left(-p,k,p-k\right)\, 
\Gamma_{\gamma\alpha\beta}\left(-k,p,-p+k\right)\, 
g^{\gamma'\gamma'}\,g^{\beta\beta'}
\nonumber\\
&~& 
\times
{\cal G}^{c'c}(k_0,\vec{k})\,
{\cal G}^{b'b}(p_0-k_0,\vec{p}-\vec{k}).
\label{gluon-loop-kernel1}
\end{eqnarray}
Moreover, under the assumption of the HTL approximation
where the hard internal momentum is supposed 
to be much bigger than the soft external gluon momentum 
by a ratio $p/k\,\sim\,g$
(i.e. $k\sim T\,\gg\, p\sim g\,T$), 
it is useful to introduce the following approximation,
\begin{eqnarray}
\Gamma_{\alpha'\gamma'\beta'}\left(-k,p,-p+k\right)
\,\Gamma_{\gamma\alpha\beta}\left(-p,k,p-k\right) 
\,g^{\gamma'\gamma'}\,g^{\beta\beta'}
\approx
10\,k_{\alpha}\,k_{\alpha'}\,+\,2\,g_{\alpha\alpha'}\,k^2.
\label{H-Loop-vertex}
\end{eqnarray}
The approximation, that is given by Eq.~(\ref{H-Loop-vertex}),
reduces the gluon-loop kernel which is given 
by Eq.~(\ref{gluon-loop-kernel1}) to
\begin{eqnarray}
{\left.{
\Pi_{(g-l)}^{\alpha'\,\alpha}
}\right.}^{a'a} (p,k)
&=&
-\frac{g^2}{2}\,({\bf T}^{a'})_{c'b'}\,({\bf T}^{a})_{bc}\,
\left[
10\,k^{\alpha}\,k^{\alpha'}\,+\,2\,g^{\alpha\,\alpha'}\,k^2
\right]
\nonumber\\
&~&
\times
{{\cal G}^{c'c}}(k_0,\vec{k})\,
{{\cal G}^{b'b}}(p_0-k_0,\vec{p}-\vec{k}).
\end{eqnarray}

\subsection{Tadpole diagram kernel  ${\Pi_{(tp)}}\left(p,k\right)$}

The gluon tadpole contributes to the soft gluon polarization tensor. 
Its Feynman diagram is depicted in Fig.~(\ref{fig:gluoneff} b). 
The gluon tadpole kernel is constructed by the four-gluons vertex 
and one gluon loop as follows,
\begin{eqnarray}
{\left.{
\Pi_{(tp)}^{\alpha\alpha'}
}\right.}^{a'a}(p,k)
&=&
{\left.{
\Pi_{(tp)}^{\alpha\alpha'}
}\right.}^{a'a}
(p_{0},\vec{p},k_{0},\vec{k}),
\nonumber\\
&=&
\frac{1}{2}\,g^2\,
{\left.{\Gamma^{\alpha'\alpha}_{\beta'\beta}}\right.}^{a'a\,c'c}
 \,g^{\beta'\beta}\,
{\cal G}^{c'c}(k_0,\vec{k}),
\nonumber\\
&=&
\frac{1}{2}\, g^2\,
{\left.{\Gamma^{\alpha'\alpha}_{\beta'\beta}}\right.}^{a'c'\,ca}
\,g^{\beta'\beta}\,
{\cal G}^{c'c}(k_0,\vec{k}).
\end{eqnarray}
The four-gluons vertex that is contracted 
by $g^{\beta'\beta}$ is reduced to
\begin{eqnarray}
g^{\beta'\beta}
{\left.{\Gamma^{\alpha'\alpha}_{\beta'\beta}}\right.}^{a'c'\,ca}
&=&
\left[
f^{a'c'e}\,f^{eca}\,
\left(g_{\alpha\alpha'}\,\delta_{\beta\beta}-g_{\alpha\alpha'}\right)
-
f^{a'ae}\,f^{ec'c}\,
\left(
g_{\alpha\alpha'}\,\delta_{\beta\beta}-g_{\alpha\alpha'}
\right)
\right],
\nonumber\\
&=&
(\delta_{\beta\beta}-1)
g_{\alpha\alpha'}\left[f^{a'c'e}\,f^{eca}\,+\,f^{a'ae}\,f^{ecc'}\right],
\nonumber\\
&=& 
2\,(\mbox{dim}_{4}-1)\, g_{\alpha\alpha'}\,
\left[f^{a'c'e}\,f^{aec}\right],
\nonumber\\
&=&
-6\, g_{\alpha\alpha'}\, 
({\bf T}^{a'})_{c'e'}\, ({\bf T}^{a})_{ec}\,\delta_{ee'}.
\end{eqnarray}
The dimension of the configuration space 
is set to 4 $(i.e. \mbox{dim}=4)$.
Therefore, the kernel of the gluon tadpole 
is reduced to
\begin{eqnarray}
{\left.{\Pi_{(tp)}^{\alpha'\alpha}}\right.}^{a'a}(p,k)
&\equiv&
-3\, g^2\, ({\bf T}^{a'})_{c'b'}\, ({\bf T}^{a})_{bc}\,
{\cal G}^{c'c}(k_0,\vec{k})
\,g_{\alpha\alpha'}\,\delta^{b'b},
\end{eqnarray}
where $p$ and $k$ are the soft external 
and the hard internal gluon momenta, 
respectively.
\subsection{Ghost-loop diagram kernel ${ \Pi_{(gh)} }(p,k)$}

The Feynman diagram for the ghost-loop 
is depicted in Fig.~(\ref{fig:gluoneff} c). 
The ghost-loop's kernel as a function 
of the soft external 
and hard internal momenta, $p$ and $k$, 
respectively, 
is constructed as follows
\begin{eqnarray}
{\left.{ \Pi_{(gh)}^{\alpha'\alpha} }\right.}^{a'a}(p,k)
&=&
{\left.{ \Pi_{(gh)}^{\alpha'\alpha} }\right.}^{a'a}
(p_{0},\vec{p},k_{0},\vec{k}),
\nonumber\\
&=&
\left(-i\, g\, f^{b'a'c'}\, k_{\alpha'}\right)
\,
{{\cal G}^{c'c}}(k_0,\vec{k})
\,
\left(-i\, g\, f^{cab}\, (k-p)_{\alpha}\right)
\nonumber\\
&~&~~~~
\times
{{\cal G}^{b'b}}(k_0-p_0,\vec{k}-\vec{p}).
\end{eqnarray}
It is approximated to
\begin{eqnarray}
{\left.{ \Pi_{(gh)}^{\alpha'\alpha} }\right.}^{a'a}(p,k)
&\approx&
\left(-i\, g\, f^{b'a'c'}\, k_{\alpha'}\right)\,
{{\cal G}^{c'c}}(k_0,\vec{k})
\,
\left(-i\, g\, f^{cab}\, k_{\alpha}\right)\,
{\cal G}^{b'b}(k_0-p_0,\vec{k}-\vec{p}),
\nonumber\\
&=& -g^2\, f^{a'c'b'}\,f^{abc}\,k_{\alpha'}\,k_{\alpha}\,
{{\cal G}^{c'c}}(k_0,\vec{k})\,
{{\cal G}^{b'b}}(k_0-p_0,\vec{k}-\vec{p}).
\label{ghost-l1}
\end{eqnarray}
In terms of the adjoint ${\bf T}^{a}$ and ${\bf T}^{a'}$ 
color matrices,
the kernel of the ghost-loop 
that is given by Eq.~(\ref{ghost-l1}) becomes, 
\begin{eqnarray}
{\left.{ \Pi_{(gh)}^{\alpha'\alpha} }\right.}^{a'a}
(p,k)
&=& g^2\, ({\bf T}^{a'})_{c' b'}\, ({\bf T}^{a})_{b c}\, 
\left[
k^{\alpha'}\,k^{\alpha}\,
{\cal G}^{c'c}(k_0,\vec{k})\,
{\cal G}^{b'b}(k_0-p_0,\vec{k}-\vec{p})
\right].
\end{eqnarray}

\subsection{Gluon's part  of the gluon's self-energy:
${\Pi_{(gp)}}(p_{0},\vec{p})=
{\Pi_{(g-l)}}(p_{0},\vec{p})+{\Pi_{(tp)}}
(p_{0},\vec{p})+{\Pi_{(gh)}}(p_{0},\vec{p})$}

The gluon's part of the gluon's self-energy consists of 
the gluon-loop, ghost-loop and tadpole interactions.
These interactions are depicted in
Figs.~(\ref{fig:gluoneff} a), (\ref{fig:gluoneff} b) 
and (\ref{fig:gluoneff} c), 
respectively.
The energy of the gluon's part is calculated by
integrating the kernel over the hard internal momentum 
$k$ as follows,
\begin{eqnarray}
{\left.{ \Pi_{(gp)}^{\alpha'\alpha} }\right.}^{a'a}(p)
&=&
{\left.{ \Pi_{(gp)}^{\alpha'\alpha} }\right.}^{a'a}
(p_{0},\vec{p}),
\nonumber\\
&=&\int \frac{d k_0}{2\pi}
\int \frac{d^3\vec{k}}{(2\pi)^3}\,
{\left.{ \Pi_{(gp)}^{\alpha'\alpha} }\right.}^{a'a}(p,k).
\end{eqnarray}
The kernel is expressed in terms 
of the both soft external 4-momentum $p$
and hard internal 4-momentum $k$  
and it reads,
\begin{eqnarray}
{\left.{ \Pi_{(gp)}^{\alpha'\alpha} }\right.}^{a'a}(p,k)
&=&
{\left.{ \Pi_{(g-l)}^{\alpha'\alpha} }\right.}^{a'a}(p,k)
+
{\left.{ \Pi_{(tp)}^{\alpha'\alpha} }\right.}^{a'a}(p,k)
+
{\left.{ \Pi_{(gh)}^{\alpha'\alpha} }\right.}^{a'a}(p,k).
\end{eqnarray}
The soft external momentum line, namely, 
$p$ is labeled by the external adjoint 
indexes $a'\,a$ 
while the hard internal momentum lines are labeled 
by the internal adjoint color indexes $b'\,b$ 
for one line segment and $c'\,c$ for the other line segment.
The external adjoint color indexes $a'\,a$ 
are coupled to the internal color indexes through the elements 
of the adjoint  color matrices (i.e. vertexes)
$({\bf T}^{a'})_{c'b'}$ and $({\bf T}^{a})_{bc}$.
Hence, the total kernel of the gluon's part 
of the gluon's self-energy is reduced to
\begin{eqnarray}
{\left.{ \Pi_{(gp)}^{\alpha'\alpha} }\right.}^{a'a}(p,k)
&=& 
({\bf T}^{a'})_{c'b'}\, ({\bf T}^{a})_{bc}\,
g^{2}
\,\left[
-\left(
4 k_{\alpha'}\,k_{\alpha} + g_{\alpha' \alpha}\, k^2
\right)\,
{\cal G}^{c' c}(k_0,\vec{k})\,
{\cal G}^{b' b}(k_0-p_0,\vec{k}-\vec{p})
\right.
\nonumber\\
&~& \left.
~~~~
- 3 g_{\alpha'\alpha}\,
{\cal G}^{c'c}(k_0,\vec{k})\,\delta^{b b'}
\right].
\label{gluon-part-total-gp}
\end{eqnarray}
It is useful to note the following relation
for the integration over the time-component 
of the hard internal momentum $k$, 
\begin{eqnarray}
({\bf T}^{a'})_{c'b'}\, ({\bf T}^{a})_{bc}\,
\int \frac{d k_0}{2\pi}
{\cal G}^{c'c}(k_0,\vec{k})\,\delta^{b b'}
&\equiv&
-
({\bf T}^{a'})_{c'b'}\, ({\bf T}^{a})_{bc}\,
\nonumber\\
&~& \times
\int \frac{d k_0}{2\pi} k^2\, 
{\cal G}^{c'c}(k_0,\vec{k})\,
{\cal G}^{b'b}(k_0-p_0,\vec{k}-\vec{p}),
\nonumber\\
&\equiv&
\frac{1}{2}\,
({\bf T}^{a'})_{c'b'}\, ({\bf T}^{a})_{bc}\,
\nonumber\\
&~& \times
\int \frac{d k_0}{2\pi}
\left[
{\cal G}^{c'c}(k_0,\vec{k})\,\delta^{b b'}
+
{\cal G}^{b'b}(k_0,\vec{k})\,\delta^{c c'}
\right].
\end{eqnarray}
The last line in the preceding integral is 
to indicate the symmetry 
of the term
$({\bf T}^{a'})_{c'b'}\, ({\bf T}^{a})_{bc}\,\delta^{b' b}\delta^{c' c}$
over the internal adjoint 
indexes under the sum of  the repeated indexes 
those are considered 
in Eq.~(\ref{gluon-part-total-gp}).
Therefore, the gluon's part of the gluon's self-energy 
is reduced to
\begin{eqnarray}
{\left.{ \Pi_{(gp)}^{\alpha'\alpha} }\right.}^{a'a}(p)
&=& 
{\left.{ \Pi_{(gp)}^{\alpha'\alpha} }\right.}^{a'a}
(p_{0},\vec{p}),
\nonumber\\
&=&
-4 ({\bf T}^{a'})_{c'b'}\, ({\bf T}^{a})_{bc}\,
g^2
\int \frac{d^3 \vec{k}}{(2\pi)^3}
\left[
\int \frac{d k_0}{2\pi}
k_{\alpha'}\,k_{\alpha}\,
{\cal G}^{c'c}(k_0,\vec{k})
\,
{\cal G}^{b'b}(k_0-p_0,\vec{k}-\vec{p})
\right.
\nonumber\\
&~&
~~~\left. ~~~
+\frac{1}{2} 
g_{\alpha' \alpha}\,
\int \frac{d k_0}{2\pi}
{\cal G}^{c' c}(k_0,\vec{k})\,\delta^{b b'}
\right],
\nonumber\\
&=&
-4 ({\bf T}^{a'})_{c'b'}\, ({\bf T}^{a})_{bc}\,
g^2
\int \frac{d^3 \vec{k}}{(2\pi)^3}
\int \frac{d k_0}{2\pi}
\left[
k_{\alpha'}\,k_{\alpha}
-\frac{1}{2}\,g_{\alpha' \alpha}\, k^{2}
\right]
\nonumber\\
&~&~~~~~~\times
{\cal G}^{c'c}(k_0,\vec{k})
\,
{\cal G}^{b'b}(k_0-p_0,\vec{k}-\vec{p}).
\end{eqnarray}
The notations for the gluon part of the gluon's 
self-energy is simplified to
\begin{eqnarray}
{\left.{ \Pi_{(gp)}^{\alpha'\alpha} }\right.}^{a'a}
\left(p_0,\vec{p}\right)
&=&
-4({\bf T}^{a'})_{c'b'}\, ({\bf T}^{a})_{bc}\,
\left(
{\left[k_{\alpha'}k_{\alpha} 
{\Pi}_{{\cal G}{\cal G}}\right]}^{b'b\,cc'} (p_0,\vec{p})
+\frac{1}{2}
g_{\alpha'\alpha}
\left[
{\Pi}_{{\cal G}}
\right]^{b'b\,cc'}(p_0,\vec{p})
\right).
\nonumber\\
\label{self-gluon-gpt1}
\end{eqnarray}
The two terms inside the parenthesis on 
the right hand side of Eq.~(\ref{self-gluon-gpt1})
is labeled by the internal adjoint color indexes 
$b'\,b$ and $c'c$. 
Moreover, the first term 
on the right hand side 
of Eq.~(\ref{self-gluon-gpt1})
is defined by
\begin{eqnarray}
{\left[k_{\alpha'}k_{\alpha} 
{\Pi}_{{\cal G}{\cal G}}\right]}^{b'b\,cc'}(p_{0},\vec{p})
&=&
g^{2}\,\int \frac{d^3 \vec{k}}{(2\pi)^3}
\int \frac{d k_0}{2\pi}
k_{\alpha'}\,k_{\alpha}\,
{\cal G}^{c'c}(k_0,\vec{k})\,
{\cal G}^{b'b}(k_0-p_0,\vec{k}-\vec{p}).
\label{self-gluon-gpt1-a}
\end{eqnarray}
While the second term 
on the right hand side 
of Eq.~(\ref{self-gluon-gpt1})
is written as follows,
\begin{eqnarray}
\left[{\Pi}_{{\cal G}}\right]^{b'b\,cc'}(p_0,\vec{p})
&=&
\int\frac{d^{3}\vec{k}}{(2\pi)^3}
\,\int\frac{d k_0}{2\pi}
{{\Pi}_{\cal G}}^{b'b\,cc'}(p,k),
\nonumber\\
&=&
-{\left[k^2
\Pi_{{\cal G}{\cal G}}
\right]}^{b'b\,cc'}(p_0,\vec{p}),
\end{eqnarray}
where
\begin{eqnarray}
{\left[ {\Pi}_{{\cal G}}\right]}^{b'b\,cc'}(p_{0},\vec{p})
&=&
g^{2}\,\int \frac{d^3 \vec{k}}{(2\pi)^3}\,
\int \frac{d k_0}{2\pi}
{\cal G}^{c'c}(k_0,\vec{k})\,\delta^{b b'},
\end{eqnarray}
and
\begin{eqnarray}
{\left[k^2\Pi_{{\cal G}{\cal G}}\right]}^{b'b\,cc'}
(p_0,\vec{p})
&=&
g^2\int\frac{d^{3}\vec{k}}{(2\pi)^3}\,
\int\frac{d k_0}{2\pi}\,
k^2\,{\cal G}^{b'b}(k_0,\vec{k})
\,{\cal G}^{c'c}(k_0-p_0,\vec{k}-\vec{p}).
\end{eqnarray}
The second term inside the parenthesis 
on the right hand side 
of Eq.~(\ref{self-gluon-gpt1}) shall be evaluated 
at first 
because it is easier than the first one.
The term, namely,
${\left[k^2\Pi_{{\cal G}{\cal G}}
\right]}^{b'b\,cc'}(p_0,\vec{p})$
is written the mixed-time representation 
of the imaginary-time formalism as follows
\begin{eqnarray}
{\left[k^2\Pi_{{\cal G}{\cal G}}\right]}^{b'b\,cc'}
(p_{0},\vec{p},\vec{k})
&=&
- g^2\,\int \frac{d k_0}{2\pi}\,
\int^{\beta}_{0} d\tau\, \int^{\beta}_{0} d\tau'\,
e^{\left[ k_{0} - i\frac{1}{\beta}\phi^{b}\right]\tau}\,
e^{\left[ (k_0-p_0) - i\frac{1}{\beta}\phi^{c}\right]\tau'}\,
\delta(\tau)
\nonumber\\
&~&
\times
{\cal G}^{c'c}(\tau',\vec{k}-\vec{p})\,\delta^{bb'}.
\end{eqnarray}
It is evaluated as follows,
\begin{eqnarray}
{\left[k^2\Pi_{{\cal G}{\cal G}}\right]}^{b'b\,cc'}
(p_{0},\vec{p},\vec{k})
&\equiv& 
- g^2\,{\cal G}^{cc'}(0,\vec{k}-\vec{p})
\,\delta^{bb'},
\nonumber\\
&\approx&
-\frac{g^2}{2\epsilon_{G}(\vec{k})} 
\left[
1+N_G\left(
\epsilon_{G}(\vec{k})-
i\frac{1}{\beta}\phi^{c}\right)
\right.
\nonumber\\
&~& ~~~ ~~~ ~~~
\left.
+ N_G\left(\epsilon_{G}(\vec{k})+
i\frac{1}{\beta}\phi^{c}\right)
\right]\,\delta^{bb'}\,\delta^{cc'}.
\label{kk-Pi-pk}
\end{eqnarray}
After integrating Eq.~(\ref{kk-Pi-pk}) 
over the hard internal momentum $\vec{k}$, 
we get
\begin{eqnarray}
{\left[k^2\Pi_{{\cal G}{\cal G}}\right]}^{b'b\,cc'}
(p_0,\vec{p})
&=&
\int\frac{d^3\vec{k}}{(2\pi)^3}\,
{\left[k^2\Pi_{{\cal G}{\cal G}}\right]}^{bc}
(p_{0},\vec{p},\vec{k})
\,\delta^{b'b}\,\delta^{c c'},
\nonumber\\
&=&
{\left[k^2\Pi_{{\cal G}{\cal G}}\right]}^{bc}(p_0,\vec{p})
\,\delta^{b'b}\,\delta^{c c'},
\end{eqnarray}
where
\begin{eqnarray}
{\left[k^2\Pi_{{\cal G}{\cal G}}\right]}^{bc}(p_0,\vec{p})
&\approx&
- g^2\,\int\frac{d^3\vec{k}}{(2\pi)^3}\,
\frac{1}{2\epsilon_{G}(\vec{k})}
\left[
N_G\left(
\epsilon_{G}(\vec{k})-
i\frac{1}{\beta}\phi^{c}\right)
+
N_G\left(\epsilon_{G}(\vec{k})+
i\frac{1}{\beta}\phi^{c}\right)
\right]
\nonumber\\
&~& + ~\left(\mbox{non-thermal part}\right).
\end{eqnarray}
The thermal part of the above equation is given 
by dropping the non-thermal part.
Therefore, the second term on the right hand side 
of Eq.~(\ref{self-gluon-gpt1}) is reduced to
\begin{eqnarray}
{\left[\Pi_{{\cal G}}\right]}^{bc}
(p_0,\vec{p})&=&
- {\left[k^2\Pi_{{\cal G}{\cal G}}\right]}^{bc}(p_0,\vec{p}),
\nonumber\\
&\approx&
\frac{1}{4}\, \left(
{\bf m}^2_{D\,(G)}\right)^{bc}.
\end{eqnarray}
The Debye mass 
$\left({\bf m}^2_{D\,(G)}\right)^{bc}$
which is contracted by
$({\bf T}^{a'})_{c b}\, 
({\bf T}^{a})_{b c}\,
\left(
{\bf m}^2_{D\,(G)}\right)^{bc}=
({\bf T}^{a'})_{c b}\, ({\bf T}^{a})_{b c}\,
\left(
{\bf m}^2_{D\,(G)}\right)^{b}$  
is reduced to
\begin{eqnarray}
\left(
{\bf m}^2_{D\,(G)}\right)^{bc} 
&=&
g^2\int \frac{d|\vec{k}|}{2\pi^2}
|\vec{k}|
\left[
N_G\left(
\epsilon_{G}(\vec{k})-
i\frac{1}{\beta}\phi^{b}\right)
+
N_G\left(\epsilon_{G}(\vec{k})+
i\frac{1}{\beta}\phi^{b}\right)
\right.
\nonumber\\
&~&\left.
+
N_G\left(
\epsilon_{G}(\vec{k})-
i\frac{1}{\beta}\phi^{c}\right)
+
N_G\left(\epsilon_{G}(\vec{k})+
i\frac{1}{\beta}\phi^{c}\right)
\right],
\nonumber\\
&=&
2\,g^{2}\int \frac{d|\vec{k}|}{2\pi^2}
|\vec{k}|
\left[
N_G\left(
\epsilon_{G}(\vec{k})-
i\frac{1}{\beta}\phi^{b}\right)
+
N_G\left(\epsilon_{G}(\vec{k})+
i\frac{1}{\beta}\phi^{b}\right)
\right].
\label{Debye-symmetrize-1}
\end{eqnarray}
It is natural to symmetrize the adjoint color indexes 
$b$ and $c$  
under the sum of all possible configurations.
Moreover, the first term on the right hand side 
of Eq.~(\ref{self-gluon-gpt1}) 
can be evaluated in a straightforward way.
Eq.~(\ref{self-gluon-gpt1-a}) is written as follows
\begin{eqnarray}
{\left[k_{\alpha}k_{\alpha'}
\Pi_{{\cal G}{\cal G}}
\right]}^{b'b\,cc'}(p_0,\vec{p})
&=&
\int \frac{d^3\vec{k}}{(2\pi)^3}\,
{\left[k_{\alpha}k_{\alpha'}
\Pi_{{\cal G}{\cal G}}
\right]}^{b'b\,cc'}(p_{0},\vec{p},\vec{k}),
\nonumber\\
&=&
\int \frac{d^3\vec{k}}{(2\pi)^3}\,
\int \frac{d k_0}{2\pi}\,
{\left[k_{\alpha}k_{\alpha'}
\Pi_{{\cal G}{\cal G}}
\right]}^{b'b\,cc'}(p_{0},\vec{p},k_{0},\vec{k}),
\label{kk-Pi-pk1}
\end{eqnarray}
where
\begin{eqnarray}
{\left[k_{\alpha}k_{\alpha'}
\Pi_{{\cal G}{\cal G}}
\right]}^{b'b\,cc'}
(p_{0},\vec{p},k_{0},\vec{k})
&=&
g^2\,k_{\alpha'}\,k_{\alpha}\,
{\cal G}^{b'b}(k_0,\vec{k})
\,{\cal G}^{c'c}(k_0-p_0,\vec{k}-\vec{p}).
\end{eqnarray}
In the format of the mixed-time formalism,
Eq.~(\ref{kk-Pi-pk1}) is written as follows
\begin{eqnarray}
{\left[k_{\alpha}k_{\alpha'}
\Pi_{{\cal G}{\cal G}}
\right]}^{b'b\,cc'}(p_{0},\vec{p},\vec{k})
&=&
g^2\,
\int \frac{d k_0}{2\pi}
\,\int^{\beta}_{0} d\tau \int^{\beta}_{0} d\tau'
e^{\left[ k_0 - i\frac{1}{\beta}\phi^{b}\right]\tau}
e^{\left[ (k_0-p_0) - i\frac{1}{\beta}\phi^{c}\right]\tau'}
\nonumber\\
&~&
\times
\left[
k_{\alpha}k_{\alpha'}
{\cal G}\right]^{b'b}(\tau,\vec{k})
\,
{\cal G}^{c'c}(\tau',\vec{k}-\vec{p}),
\nonumber\\
&=&
g^2\,
\int^{\beta}_{0} d\tau
e^{
-\left( p_0 -i\frac{1}{\beta}\phi^{b}
+i\frac{1}{\beta}\phi^{c}
\right)\left(\beta-\tau\right)}
\nonumber\\
&~&
\times
{\left[k_{\alpha} k_{\alpha'} {\cal G}\right]}^{b'b}
(\tau,\vec{k})\,
{\cal G}^{cc'}(\beta-\tau,\vec{k}-\vec{p}),
\nonumber\\
&=&
g^2\,
\int^{\beta}_{0} d\tau
e^{
\left( p_0 -i\frac{1}{\beta}\phi^{a}
\right)\tau}
{\left[k_{\alpha} k_{\alpha'} {\cal G}
\right]}^{b'b}(\tau,\vec{k})
\,{\cal G}^{cc'}(\beta-\tau,\vec{k}-\vec{p}).
\label{kk-gg-master}
\end{eqnarray}
In the above equation (i.e. Eq.~(\ref{kk-gg-master})), 
the following relation is considered 
\begin{eqnarray}
\int \frac{d k_0}{2\pi}
\int^{\beta}_{0} d\tau \int^{\beta}_{0} d\tau'
e^{\left( k_0 - i\frac{1}{\beta}\phi^{b}\right)
\left(\tau+\tau'\right)}
&=&
\frac{1}{\beta}
\sum_{n}^{\mbox{even}}
\int^{\beta}_{0} d\tau \int^{\beta}_{0} d\tau'
e^{i\omega_n
\left(\tau+\tau'\right)},
\nonumber\\
&=&
\int^{\beta}_{0} d\tau \int^{\beta}_{0} d\tau'
\delta(\tau+\tau'-\beta),
\end{eqnarray} 
besides the following identity  
\begin{eqnarray}
\exp\left[
-\left( p_0 -i\frac{1}{\beta}\phi^{b}
+i\frac{1}{\beta}\phi^{c}
\right)\beta\right]
&=&
\exp\left[
-\left( p_0 -i\frac{1}{\beta}\phi^{a}
\right)\beta\right],
\nonumber\\
&=&1,
\end{eqnarray}
is adopted in the calculation.
The term 
${\left[k_{\alpha} k_{\alpha'} 
{\cal G}\right]}^{b'b}(\tau,\vec{k})$,
which is appeared in Eq.~(\ref{kk-gg-master}),
is defined in the context of the mixed-time representation. 
It needs a rather more sophisticated work to be calculated.
It is decomposed to the Lorentz components, namely, 
zero, zero-vector and finally vector-vector 
components as follows,
\begin{eqnarray}
{\left.\left[k_0 {\cal G}\right]\right.}^{b b'}(\tau,\vec{k})
&=&
~\frac{1}{2}
\left[
1+N_G\left(\epsilon_{G}(\vec{k})-i\frac{1}{\beta}\phi^{b}
\right)\right]
e^{-\tau\left(\epsilon_{G}(\vec{k})-i\frac{1}{\beta}\phi^{b}
\right)}\,\delta^{bb'}
\nonumber\\
&~&
-\frac{1}{2}
\left[
N_G\left(\epsilon_{G}(\vec{k})+i\frac{1}{\beta}\phi^{b}
\right)
\right]
e^{\tau\left(\epsilon_{G}(\vec{k})+i\frac{1}{\beta}\phi^{b}
\right)}\,\delta^{bb'},
\end{eqnarray}
\begin{eqnarray}
\left[k_0\vec{k}_j {\cal G}\right]^{b b'}(\tau,\vec{k})
&=&
~\frac{\vec{k}_j}{2}
\left[
1+N_G\left(\epsilon_{G}(\vec{k})-i\frac{1}{\beta}\phi^{b}
\right)\right]
e^{-\tau\left(\epsilon_{G}(\vec{k})-i\frac{1}{\beta}\phi^{b}
\right)}\,\delta^{b'b}
\nonumber\\
&~&
-\frac{\vec{k}_j}{2}
\left[
N_G\left(\epsilon_{G}(\vec{k})+i\frac{1}{\beta}\phi^{b}
\right)
\right]
e^{\tau\left(\epsilon_{G}(\vec{k})+i\frac{1}{\beta}\phi^{b}
\right)}\,\delta^{bb'},
\end{eqnarray}
and
\begin{eqnarray}
\left[\vec{k}_i\vec{k}_j {\cal G}\right]^{b b'}(\tau,\vec{k})
&=&
~\frac{\vec{k}_i\vec{k}_j}{2\epsilon_{G}(\vec{k})}
\left[
1+N_G\left(\epsilon_{G}(\vec{k})-i\frac{1}{\beta}\phi^{b}
\right)\right]
e^{-\tau\left(\epsilon_{G}(\vec{k})-i\frac{1}{\beta}\phi^{b}
\right)}\,
\delta^{bb'}
\nonumber\\
&~&
+\frac{\vec{k}_i\vec{k}_j}{2\epsilon_{G}(\vec{k})}
\left[
N_G\left(\epsilon_{G}(\vec{k})+i\frac{1}{\beta}\phi^{b}
\right)
\right]
e^{\tau\left(\epsilon_{G}(\vec{k})+i\frac{1}{\beta}\phi^{b}
\right)}\,\delta^{bb'},
\end{eqnarray}
respectively.
The term 
${\cal G}^{c c'}(\beta-\tau,\vec{k}-\vec{p})$, 
which is appeared in the last line of 
Eq.~(\ref{kk-gg-master}), 
is reduced to
\begin{eqnarray}
{\cal G}^{c c'}&=&
{\cal G}^{c c'}(\beta-\tau,\vec{k}-\vec{p}),
\nonumber\\
&=&
~\frac{1}{2\epsilon_{G}(\vec{k}-\vec{p})}
\left[
1+N_G\left(\epsilon_{G}(\vec{k}-\vec{p})-i\frac{1}{\beta}\phi^{c}
\right)\right]
e^{-\left(\beta-\tau\right)\left(
\epsilon_{G}(\vec{k}-\vec{p})-
i\frac{1}{\beta}\phi^{c}
\right)}\,
\delta^{cc'}
\nonumber\\
&~&
+\frac{1}{2\epsilon_{G}(\vec{k}-\vec{p})}
\left[
N_G\left(\epsilon_{G}(\vec{k}-\vec{p})+i\frac{1}{\beta}\phi^{c}
\right)
\right]
e^{\left(\beta-\tau\right)\left(
\epsilon_{G}(\vec{k}-\vec{p})+
i\frac{1}{\beta}\phi^{c}
\right)}\,\delta^{cc'}.
\label{kk-gg-sec-g1}
\end{eqnarray}
It can be parametrized 
in the following way,
\begin{eqnarray}
{\cal G}^{cc'}&=&
{\cal G}^{c c'}(\beta-\tau,\vec{k}-\vec{p}),
\nonumber\\
&=&
~\frac{1}{2\epsilon_{G}(\vec{k}-\vec{p})}
\left[
N_G\left(\epsilon_{G}(\vec{k}-\vec{p})-i\frac{1}{\beta}\phi^{c}
\right)\right]
e^{\tau\left(
\epsilon_{G}(\vec{k}-\vec{p})-
i\frac{1}{\beta}\phi^{c}
\right)}\,\delta^{c c'}
\nonumber\\
&~&
+\frac{1}{2\epsilon_{G}(\vec{k}-\vec{p})}
\left[
1+N_G\left(\epsilon_{G}(\vec{k}-\vec{p})
+i\,\frac{1}{\beta}\phi^{c}
\right)
\right]
e^{-\tau\left(
\epsilon_{G}(\vec{k}-\vec{p})+
i\,\frac{1}{\beta}\phi^{c}\right)}\,
\delta^{cc'}.
\end{eqnarray}
In the similar manner, the term 
$\left[k_{0}{\cal G}\right]^{c c'}
\left(\beta-\tau,\vec{k}-\vec{p}\right)$
is parametrized as follows
\begin{eqnarray}
\left[k_{0}{\cal G}\right]^{c c'}
\left(\beta-\tau,\vec{k}-\vec{p}\right)&=&
~\frac{1}{2}
\left[
N_G\left(\epsilon_{G}(\vec{k}-\vec{p})-i\frac{1}{\beta}\phi^{c}
\right)\right]
e^{\tau\left(
\epsilon_{G}(\vec{k}-\vec{p})-
i\frac{1}{\beta}\phi^{c}
\right)}\,\delta^{c c'}
\nonumber\\
&~&
-\frac{1}{2}
\left[
1+N_G\left(\epsilon_{G}(\vec{k}-\vec{p})
+i\,\frac{1}{\beta}\phi^{c}
\right)
\right]
e^{-\tau\left(
\epsilon_{G}(\vec{k}-\vec{p})+
i\,\frac{1}{\beta}\phi^{c}\right)}\,
\delta^{cc'}.
\end{eqnarray}
%
%
%
%
%
%
The term, namely, 
$\left[k_{\alpha'}k_{\alpha}\Pi_{{\cal G}{\cal G}}
\right]^{b'b\,cc'}(p_0,\vec{p})$
is decomposed to the Foldy-Wouthuysen 
positive and negative energy components:
\begin{eqnarray}
\left[k_{\alpha'}k_{\alpha}
\Pi_{{\cal G}{\cal G}}\right]^{b'b\,cc'}(p_0,\vec{p})
&=&
\sum_{r=\pm}\sum_{s=\pm} 
\left[k_{\alpha'}k_{\alpha}\Pi^{(rs)}_{{\cal G}{\cal G}}
\right]^{b'b\,cc'}(p_0,\vec{p}).
\end{eqnarray}
The positive-positive and negative-negative energy components 
of the Foldy-Wouthuysen decomposition are reduced to
\begin{eqnarray}
\left[k_{0}k_{0}\Pi^{(++)}_{{\cal G}{\cal G}}
\right]^{b c}(p_0,\vec{p},\vec{k})
&\approx&
\frac{g^2}{4}
\frac{
\left[
N_G\left(\epsilon_{G}(\vec{k})-i\frac{1}{\beta}\phi^{b}
\right)
-N_G\left(\epsilon_{G}(\vec{k}-\vec{p})-i\frac{1}{\beta}\phi^{c}
\right)\right]}
{p_0 -\left(\epsilon_{G}(\vec{k})
-\epsilon_{G}(\vec{k}-\vec{p})\right)},
\label{k0k0-positive-positive}
\end{eqnarray}
and
\begin{eqnarray}
\left[k_{0}k_{0}\Pi^{(--)}_{{\cal G}{\cal G}}
\right]^{b c}(p_0,\vec{p},\vec{k})
&\approx&
-\frac{g^2}{4}
\frac{
\left[
N_G\left(\epsilon_{G}(\vec{k})+i\frac{1}{\beta}\phi^{b}
\right)
-N_G\left(\epsilon_{G}(\vec{k}-\vec{p})+i\frac{1}{\beta}\phi^{c}
\right)\right]}
{p_0 + \left(\epsilon_{G}(\vec{k})
-\epsilon_{G}(\vec{k}-\vec{p})\right)},
\label{k0k0-negative-negative1}
\end{eqnarray}
respectively, where the following approximation 
\begin{eqnarray}
\epsilon_{G}(\vec{k})\,/\,\epsilon_{G}(\vec{k}-\vec{p})\,
&\approx&\, 1,
\label{k0k0-PI-approx1}
\end{eqnarray}
is adopted.
Since, the terms $\left[k_{0}k_{0}\Pi^{(rs)}_{{\cal G}{\cal G}}
\right]^{b c}(p_0,\vec{p},\vec{k})$ 
are located under the integral 
$\int \frac{d^{3}\vec{k}}{(2\pi)^{3}}$, 
the symmetry over the polar integration
$\hat{k}\cdot\vec{p}$ can be considered.
The negative-negative energy component is reduced to
\begin{eqnarray}
\left[k_{0}k_{0}\Pi^{(--)}_{{\cal G}{\cal G}}
\right]^{b c}(p_0,\vec{p})
&=&
\int \frac{d^{3}\vec{k}}{(2\pi)^{3}}
\left[k_{0}k_{0}\Pi^{(--)}_{{\cal G}{\cal G}}
\right]^{b c}(p_0,\vec{p},\vec{k})
\nonumber\\
&\approx&
-\frac{g^2}{4}
\int \frac{d^{3}\vec{k}}{(2\pi)^{3}}
\frac{
\left[
N_G\left(\epsilon_{G}(\vec{k})+i\frac{1}{\beta}\phi^{b}
\right)
-N_G\left(\epsilon_{G}(\vec{k}+\vec{p})+i\frac{1}{\beta}\phi^{c}
\right)\right]}
{p_0 + \left(\epsilon_{G}(\vec{k})
-\epsilon_{G}(\vec{k}+\vec{p})\right)}.
\nonumber\\
\label{k0k0-negative-negative2}
\end{eqnarray}
In the subsequent calculations, 
it is useful to consider the following
approximations
\begin{eqnarray}
\epsilon_{G}(\vec{k})
-\epsilon_{G}(\vec{k}-\vec{p})&\approx&
\hat{k}\cdot\vec{p}\,\sim\, g\,T\,\cos(\theta),
\nonumber\\
\epsilon_{G}(\vec{k})+\epsilon_{G}(\vec{k}-\vec{p})
&\approx& 2\,|\vec{k}|\,\sim\, 2\,T,
\label{k0k0-PI-Taylor1}
\end{eqnarray}
for the HTL approximation
with a soft external momentum $p$ 
and a hard internal momentum $k$.
The other energy components 
of the Foldy-Wouthuysen 
decomposition, namely,
$(+-)$, $(-+)$, $(--)$ 
can be calculated in a similar way.
Using the approximations that are given 
in Eqns.~(\ref{k0k0-PI-approx1}) and (\ref{k0k0-PI-Taylor1}), 
the sum of 
the positive-positive and negative-negative 
energy kernel components, 
with the internal loop adjoint color indexes $b$ and $c$ 
which are given by Eqns.~(\ref{k0k0-positive-positive}) and
~(\ref{k0k0-negative-negative2}), is reduced to
\begin{eqnarray}
\begin{array}{l}
\left(\left[k_{0}k_{0}\Pi^{(++)}_{{\cal G}{\cal G}}
\right]^{b c}(p_0,\vec{p},\vec{k})
+
\left[k_{0}k_{0}\Pi^{(--)}_{{\cal G}{\cal G}}
\right]^{b c}(p_0,\vec{p},\vec{k})
\right)
\\
~~~ = ~
\frac{g^2}{4}
\int \frac{d^{3} \vec{k}}{(2\pi)^{3}}
\left[\frac{1}{p_{0}-\hat{k}\cdot\vec{p}}\right]
\left[
N_G\left(\epsilon_{G}(\vec{k})-i\frac{1}{\beta}\phi^{b}
\right)
-
N_G\left(\epsilon_{G}(\vec{k})+i\frac{1}{\beta}\phi^{b}
\right) \right]
\\
~~~ ~ + ~
\frac{g^2}{4}
\int \frac{d^{3} \vec{k}}{(2\pi)^{3}}
\left[\frac{1}{p_{0}-\hat{k}\cdot\vec{p}}\right]
\left[
N_G\left(\epsilon_{G}(\vec{k}-\vec{p})-i\frac{1}{\beta}\phi^{c}
\right)
-
N_G\left(\epsilon_{G}(\vec{k}+\vec{p})+i\frac{1}{\beta}\phi^{c}
\right) \right].
\end{array}
\nonumber\\
\end{eqnarray}
It is approximated to
\begin{eqnarray}
\begin{array}{l}
\left(\left[k_{0}k_{0}\Pi^{(++)}_{{\cal G}{\cal G}}
\right]^{b c}(p_0,\vec{p})
+
\left[k_{0}k_{0}\Pi^{(--)}_{{\cal G}{\cal G}}
\right]^{b c}(p_0,\vec{p})
\right)
\\
~~ = 
\frac{g^2}{4}
\int \frac{d \Omega_{k} }{4\pi}
\left[\frac{1}{p_{0}-\hat{k}\cdot\vec{p}}\right]
\int\frac{d |\vec{k}|}{2\pi^{2}}
|\vec{k}|^{2}
\left[
N_G\left(\epsilon_{G}(\vec{k})-i\frac{1}{\beta}\phi^{b}
\right)
-
N_G\left(\epsilon_{G}(\vec{k})+i\frac{1}{\beta}\phi^{b}
\right) \right]
\\
~~~ ~ - 
\frac{g^2}{4}
\int \frac{d \Omega_{k} }{4\pi}
\left[\frac{1}{p_{0}-\hat{k}\cdot\vec{p}}\right]
\int\frac{d |\vec{k}|}{2\pi^{2}}
|\vec{k}|^{2}
\left[
N_G\left(\epsilon_{G}(\vec{k})-i\frac{1}{\beta}\phi^{c}
\right)
-
N_G\left(\epsilon_{G}(\vec{k})+i\frac{1}{\beta}\phi^{c}
\right) \right]
\\
~~~ ~ - 
\frac{g^2}{4}
\left[1-
{p_{0}}\int \frac{d\Omega_{k}}{4\pi}
\frac{1}{p_{0}-\hat{k}\cdot\vec{p}}\right]
\int\frac{d |\vec{k}|}{2\pi^{2}}
|\vec{k}|^{2}
\frac{d}{d|\vec{k}|}
\left[
N_G\left(\epsilon_{G}(\vec{k})-i\frac{1}{\beta}\phi^{c}
\right)
+
N_G\left(\epsilon_{G}(\vec{k})+i\frac{1}{\beta}\phi^{c}
\right) \right].
\end{array}
\nonumber\\
\label{k0k0-PI-positive-positive-neg-neg}
\end{eqnarray}
In a similar way, 
the sum of positive-negative and negative-positive 
Foldy-Wouthuysen energy components 
is reduced to
\begin{eqnarray}
\begin{array}{l}
\left(
\left[k_{0}k_{0}\Pi^{(+-)}_{{\cal G}{\cal G}}
\right]^{b c}(p_0,\vec{p})
+
\left[k_{0}k_{0}\Pi^{(-+)}_{{\cal G}{\cal G}}
\right]^{b c}(p_0,\vec{p})
\right)
\\
~~~ = 
-\frac{g^2}{4}
\int\frac{d |\vec{k}|}{2\pi^{2}}
\frac{|\vec{k}|}{2}
\left[
N_G\left(\epsilon_{G}(\vec{k})-i\frac{1}{\beta}\phi^{b}
\right)
+
N_G\left(\epsilon_{G}(\vec{k})+i\frac{1}{\beta}\phi^{b}
\right) \right]
\\
~~~ ~~~ 
-\frac{g^2}{4}
\int\frac{d |\vec{k}|}{2\pi^{2}}
\frac{|\vec{k}|}{2}
\left[
N_G\left(\epsilon_{G}(\vec{k})-i\frac{1}{\beta}\phi^{c}
\right)
+
N_G\left(\epsilon_{G}(\vec{k})+i\frac{1}{\beta}\phi^{c}
\right) \right].
\end{array}
\nonumber\\
\label{k0k0-PI-positive-neg-neg-positive}
\end{eqnarray}
The gluon's part of the Debye mass
due to the Feynman diagrams of   
the gluon loop, tadpole and ghost loop, 
reads
\begin{eqnarray}
\left({\bf m}^2_{D\,(G)}\right)^{b}
&=&
-g^2\,\int \frac{d|\vec{k}|}{2\pi^2}
\,|\vec{k}|^2\,
\frac{d}{d|\vec{k}|}
\left[
N_G\left(\epsilon_{G}(\vec{k})-i\frac{1}{\beta}\phi^{b}\right)
+
N_G\left(\epsilon_{G}(\vec{k})+i\frac{1}{\beta}\phi^{b}\right)
\right],
\nonumber\\
&=&
2\,g^2\,\int \frac{d|\vec{k}|}{2\pi^2}
\,|\vec{k}|\,
\left[
N_G\left(\epsilon_{G}(\vec{k})-i\frac{1}{\beta}\phi^{b}\right)
+
N_G\left(\epsilon_{G}(\vec{k})+i\frac{1}{\beta}\phi^{b}\right)
\right].
\end{eqnarray}
The additional terms those appear 
in Eq.~(\ref{k0k0-PI-positive-positive-neg-neg}) 
can be parametrized by
\begin{eqnarray}
\left(\Delta\,{\bf m}^2_{D\,(G)}\right)^{b}
&=&
\,g^2\,\int \frac{d|\vec{k}|}{2\pi^2}
\,|\vec{k}|^{2}
\,
\left[
N_G\left(\epsilon_{G}(\vec{k})-i\frac{1}{\beta}\phi^{b}\right)
-
N_G\left(\epsilon_{G}(\vec{k})+i\frac{1}{\beta}\phi^{b}\right)
\right].
\end{eqnarray}
The sum over all the Foldy-Wouthuysen
energy components of the composite 
$\left[k_{0}k_{0}\Pi_{{\cal G}{\cal G}}
\right]^{b c}(p_0,\vec{p})$ 
becomes
\begin{eqnarray}
\left[k_{0}k_{0}\Pi_{{\cal G}{\cal G}}
\right]^{b c}(p_0,\vec{p})
&=&
-\frac{1}{8}\left[
\frac{1}{2}\left({\bf m}^{2}_{D\,(G)}\right)^{b}
+
\frac{1}{2}\left({\bf m}^{2}_{D\,(G)}\right)^{c}
\right]
\nonumber\\
&~& +
\frac{1}{8}\left[
2\,-\,2\,p_0\,\int \frac{d\Omega}{4\pi}
\frac{1}
{\left(p_0-\hat{k}\cdot\vec{p}\right)}
\right]\left({\bf m}^{2}_{D\,(G)}\right)^{c}
\nonumber\\
&~& +
\frac{1}{4}
\int \frac{d\Omega}{4\pi}
\frac{1}
{\left(p_0-\hat{k}\cdot\vec{p}\right)}
\left[
\left(\Delta{\bf m}^{2}_{D\,(G)}\right)^{b}
-
\left(\Delta{\bf m}^{2}_{D\,(G)}\right)^{c}
\right].
\end{eqnarray}
The same calculations are performed 
for the other Lorentz components
such as
$\left[k_{0}k_{i}
\Pi_{{\cal G}{\cal G}}
\right]^{b c}(p_0,\vec{p})$
and 
$\left[k_{i}k_{j}\Pi_{{\cal G}{\cal G}}
\right]^{b c}(p_0,\vec{p})$ components.
For instance the term 
$\left[k_{i}k_{j}\Pi_{{\cal G}{\cal G}}
\right]^{b c}(p_0,\vec{p})$
is reduced to
\begin{eqnarray}
\left[k_{i} k_{j}\Pi_{{\cal G}{\cal G}}
\right]^{b c}(p_0,\vec{p})
&=&
\frac{1}{8}\left[
\frac{2}{3}\left({\bf m}^{2}_{D\,(G)}\right)^{c}
+
\frac{1}{3}
\left(
\frac{1}{2}\left({\bf m}^{2}_{D\,(G)}\right)^{b}
+
\frac{1}{2}\left({\bf m}^{2}_{D\,(G)}\right)^{c}
\right)
\right]
\,\delta_{ij}
\nonumber\\
&~& 
-\frac{2\,p_{0}}{8}
\,
\int \frac{d\Omega}{4\pi}
\frac{\hat{k}_{i}\hat{k}_{j}}
{\left(p_0-\hat{k}\cdot\vec{p}\right)}
\,\left({\bf m}^{2}_{D\,(G)}\right)^{c}
\nonumber\\
&~& 
+\frac{1}{4}
\int \frac{d\Omega}{4\pi}
\frac{\hat{k}_{i}\hat{k}_{j}}
{\left(p_0-\hat{k}\cdot\vec{p}\right)}
\left[
\left(\Delta{\bf m}^{2}_{D\,(G)}\right)^{b}
-
\left(\Delta{\bf m}^{2}_{D\,(G)}\right)^{c}
\right],
\label{space-space-lorentz-polarization-1}
\end{eqnarray}
where the following tensor identity 
has been considered in the above calculation
\begin{eqnarray}
\int\frac{d\Omega}{4\pi}\left(3\hat{k}_i\hat{k}_j
-\delta_{ij}\right)=0.
\end{eqnarray}
The Lorentz momentum components of 
$\hat{k}_{\alpha'}\,\hat{k}_{\alpha}$
are written as follows
\begin{eqnarray}
\hat{k}_{\alpha'}\,\hat{k}_{\alpha}
&=&\left(
1,\hat{k}_{i},\hat{k}_j,\hat{k}_{i}\hat{k}_{j}
\right).
\end{eqnarray}
The general result can be obtained 
by the same calculations. 
The result reads
\begin{eqnarray}
\left[k_{\alpha'} k_{\alpha}\Pi_{{\cal G}{\cal G}}
\right]^{b c}(p_0,\vec{p})
&=&
2\,\delta^{0}_{\alpha'}\,\delta^{0}_{\alpha}
\left(\frac{1}{8}\right)\left[
\frac{4}{3}\left({\bf m}^{2}_{D\,(G)}\right)^{c}
-\frac{1}{3}\left\{
\frac{1}{2}\left({\bf m}^{2}_{D\,(G)}\right)^{b}
+
\frac{1}{2}\left({\bf m}^{2}_{D\,(G)}\right)^{c}
\right\}
\right]
\nonumber\\
&~&
-\,g_{\alpha'\alpha}\,
\left(\frac{1}{8}\right)
\left[
\frac{2}{3}\left({\bf m}^{2}_{D\,(G)}\right)^{c}
+
\frac{1}{3}
\left\{
\frac{1}{2}\left({\bf m}^{2}_{D\,(G)}\right)^{b}
+
\frac{1}{2}\left({\bf m}^{2}_{D\,(G)}\right)^{c}
\right\}
\right]
\nonumber\\
&~& 
-\frac{2\,p_{0}}{8}
\,
\int \frac{d\Omega}{4\pi}
\frac{\hat{k}_{\alpha'}\hat{k}_{\alpha}}
{\left(p_0-\hat{k}\cdot\vec{p}\right)}
\,
\left({\bf m}^{2}_{D\,(G)}\right)^{c}
\nonumber\\
&~& +
\frac{1}{4}
\int \frac{d\Omega}{4\pi}
\frac{\hat{k}_{\alpha'}\hat{k}_{\alpha}}
{\left(p_0-\hat{k}\cdot\vec{p}\right)}
\left[
\left(\Delta{\bf m}^{2}_{D\,(G)}\right)^{b}
-
\left(\Delta{\bf m}^{2}_{D\,(G)}\right)^{c}
\right].
\label{lorentz-lorentz-polarization-1}
\end{eqnarray}
The term $\left(\Delta{\bf m}^{2}_{D\,(G)}\right)^{b}$ 
generates
a quadratic order temperature term (i.e. $\propto T^{3}$).
This term also appears in 
Refs.~\cite{ Hidaka:2008dr,Hidaka:2009hs,Hidaka:2009ma}.
Although these quadratic temperature
terms emerge in the internal loop segments 
with internal adjoint color indexes, they cancel each other 
in the final result  for the loop 
with external adjoint color indexes 
(see Eq.~(\ref{adjoint-quadratic-symmetry}))
when the sum over the internal adjoint color 
indexes is considered.

By using the symmetry 
over the adjoint color indexes
$b$ and $c$ that is given by 
Eqns.~(\ref{adjoint-symmetry-b-and-c}) and 
(\ref{adjoint-quadratic-symmetry}),
the Lorentz-Lorentz polarization tensor 
(e.g. Eq.~(\ref{lorentz-lorentz-polarization-1})) 
is contracted as follows
\begin{eqnarray}
\begin{array}{l}
\left({\bf T}^{a'}\right)_{c b}
\left({\bf T}^{a}\right)_{b c}
\,
\left[k_{\alpha'} k_{\alpha}\Pi_{{\cal G}{\cal G}}
\right]^{b c}(p_0,\vec{p})
\\
~~~ ~~~
=
\frac{1}{8}\,
\left({\bf T}^{a}\right)_{c b}
\left({\bf T}^{a}\right)_{b c}
\,
\left({\bf m}^{2}_{D\,(G)}\right)^{c}
\,
\left[
\left(
2\delta^{0}_{\alpha'}\,\delta^{0}_{\alpha}
-g_{\alpha'\alpha}  
\right)
-
2\,p_{0}\,
\int \frac{d\Omega}{4\pi}
\frac{\hat{k}_{\alpha'}\hat{k}_{\alpha}}
{\left(p_0-\hat{k}\cdot\vec{p}\right)}
\right].
\end{array}
\nonumber\\
\label{lorentz-lorentz-polarization-final}
\end{eqnarray}
Hence after eliminating the redundant terms, 
it is simplified to
\begin{eqnarray}
\left[k_{\alpha'} k_{\alpha}\Pi_{{\cal G}{\cal G}}
\right]^{b c}(p_0,\vec{p})
&=&
\frac{1}{8}
\,
\left({\bf m}^{2}_{D\,(G)}\right)^{c}
\,
\left[
\left(
2\delta^{0}_{\alpha'}\,\delta^{0}_{\alpha}
-g_{\alpha'\alpha}  
\right)
-
2\,p_{0}\,
\int \frac{d\Omega}{4\pi}
\frac{\hat{k}_{\alpha'}\hat{k}_{\alpha}}
{\left(p_0-\hat{k}\cdot\vec{p}\right)}
\right].
\nonumber\\
\label{lorentz-lorentz-polarization-final2}
\end{eqnarray}
The quadratic temperature terms 
which are generated in the internal gluon 
loop are eliminated when the summation 
over the internal adjoint color indexes is considered.
These quadratic terms simply disappear 
from the gluon polarization tensor 
with the external line legs those are labeled 
by the adjoint  indexes $a'\,a$. 
The 3- gluons vertex is anti-symmetry over 
the interchange of two adjoint color indexes.
Since the internal gluon loop has two 3-gluon vertexes, 
it is symmetry over 
the interchange of two adjoint color indexes.
The internal loop gluon polarization 
tensor becomes
\begin{eqnarray}
{\Pi_{(gp)\,\alpha'\alpha}}^{b c}(p_{0},\vec{p})
&=&
-4\,
\left(
\left[k_{\alpha'}\,k_{\alpha}
\Pi_{{\cal G}{\cal G}}
\right]^{b c}(p_0,\vec{p})
-\frac{1}{2}\,g_{\alpha'\alpha}\,
\left[k^2\Pi_{{\cal G}{\cal G}}
\right]^{b c}(p_0,\vec{p})
\right),
\nonumber\\
&=&
-4\,
\left(
\left[k_{\alpha'}\,k_{\alpha}
\Pi_{{\cal G}{\cal G}}
\right]^{b c}(p_0,\vec{p})
+\frac{1}{2}\,g_{\alpha'\alpha}\,
\left[\Pi_{{\cal G}}\right]^{b c}
(p_0,\vec{p})
\right),
\nonumber\\
&=&
\left({\bf m}^{2}_{D\,(G)}\right)^{bc}
\left[
-\delta^{0}_{\alpha'}\,\delta^{0}_{\alpha}
+ p_0\,\int \frac{d\Omega}{4\pi}
\frac{\hat{k}_{\alpha'}\,\hat{k}_{\alpha}}
{\left(p_0-\hat{k}\cdot\vec{p}\right)}
\right].
\end{eqnarray}

When the color charges couple with the quark and gluon
kinetic degree of freedom, the Debye mass 
has a color charge.
The gluon's part of the Debye mass 
is labeled by an adjoint color index. 
The external adjoint color indexes are given 
by $a'\,a$
while the internal segments for the gluon loop 
are given by the adjoint color indexes 
$b'\,b$ and $c'\,c$. 
The internal adjoint color indexes are reduced 
to $b\,c$ by the contraction 
of $\delta^{c'\,c}$ and $\delta^{b'\,b}$ 
and the summation 
over all the possible configurations.  
The integration by parts reduces the gluon's part 
of the Debye mass to the following result,
\begin{eqnarray}
\left({\bf m}^2_{D\,(G)}\right)^{b}
&=&
2 g^2\,\int \frac{d|\vec{k}|}{2\pi^2}|\vec{k}|
\left[
N_G\left(\epsilon_{G}(\vec{k})-i\frac{1}{\beta}\phi^{b}\right)
+
N_G\left(\epsilon_{G}(\vec{k})+i\frac{1}{\beta}\phi^{b}\right)
\right],
\nonumber\\
&=&
\frac{g^2}{\pi^2}\,
\left[
\frac{\pi^2}{3\,\beta^2}
-\frac{1}{2}\left(i\,\frac{1}{\beta}\phi^{b}\right)^2
+i\,\frac{\pi}{\beta}\left(i\,\frac{1}{\beta}\phi^{b}\right)
\right],
\nonumber\\
&\rightarrow&
\frac{g^2}{\pi^2}\,
\left[
\frac{\pi^2}{3\,\beta^2}
-\frac{1}{2}\left({\mu_{C}}^{b\equiv\underbrace{(AB)}}\right)^2
+i\,\frac{\pi}{\beta}\left({\mu_{C}}^{b\equiv\underbrace{(AB)}}\right)
\right],
\nonumber\\
&\rightarrow&
\frac{g^2}{\pi^2}\,
\left[
\frac{\pi^2}{3\,\beta^2}
-\frac{1}{2}\left(
{\mu_{C}}_{A}-{\mu_{C}}_{B}\right)^2
+i\,\frac{\pi}{\beta}\left(
{\mu_{C}}_{A}-{\mu_{C}}_{B}\right)
\right].
\label{Debye-symmetrize-2}
\end{eqnarray}
It is interesting to note that when
the fundamental color chemicals become equal
${\mu_{C}}_{A}\,\approx\,{\mu_{C}}_{B}$,
the adjoint color chemical becomes negligible
${\mu_{C}}^{\underbrace{(AB)}}=
\left({\mu_{C}}_{A}-{\mu_{C}}_{B}\right)
\approx 0$ and Eq.~(\ref{Debye-symmetrize-2})
is reduced simply to
$\left({\bf m}^2_{D\,(G)}\right)^{b c}=g^2\, T^{2}/3$.
When the fundamental color potentials are not equal
(i.e. ${\mu_{C}}_{A}\neq{\mu_{C}}_{B}$), 
Eq.~(\ref{Debye-symmetrize-2}) develops 
an imaginary part and this leads to nontrivial results.
The symmetrization over the internal adjoint color 
indexes, namely, $b$ and $c$ is given by
\begin{eqnarray}
({\bf T}^{a'})_{c b}\, ({\bf T}^{a})_{b c}\,
\left({\bf m}^2_{D\,(G)}\right)^{bc}
&=&
({\bf T}^{a'})_{c b}\, ({\bf T}^{a})_{b c}\,
\left[\left({\bf m}^2_{D\,(G)}\right)^{b}
+
\left({\bf m}^2_{D\,(G)}\right)^{c}\right]/2,
\nonumber\\
&=&
({\bf T}^{a'})_{c b}\, ({\bf T}^{a})_{b c}\,
\left({\bf m}^2_{D\,(G)}\right)^{b},
\end{eqnarray}
which is identical 
to Eq.~(\ref{Debye-symmetrize-1}). 
The term 
$\left(\Delta{\bf m}^2_{D\,(G)}\right)^{b}$ 
is determined as follows
\begin{eqnarray}
\left(\Delta{\bf m}^2_{D\,(G)}\right)^{b}
&=&
\,g^2\,\int \frac{d|\vec{k}|}{2\pi^2}
\,|\vec{k}|^{2}
\,
\left[
N_G\left(\epsilon_{G}(\vec{k})-i\frac{1}{\beta}\phi^{b}\right)
-
N_G\left(\epsilon_{G}(\vec{k})+i\frac{1}{\beta}\phi^{b}\right)
\right],
\nonumber\\
&=&
\frac{g^{2}\, T^{3}}{\pi^{2}}\,
\left[
\mbox{Li}_{3}\left(e^{i\phi^{b}}\right)
-
\mbox{Li}_{3}\left(e^{-i\phi^{b}}\right)
\right].
\end{eqnarray}
The term with Poly-Logarithm functions can be simplified 
as follows 
\begin{eqnarray}
\left[
\mbox{Li}_{3}\left(e^{i\phi^{b}}\right)
-
\mbox{Li}_{3}\left(e^{-i\phi^{b}}\right)
\right]
&=&
-\frac{1}{6}(i\phi^{b})\left[(i\phi^{b})-i\pi\right]
\left[(i\phi^{b})-2i\pi\right],
\nonumber\\
&\equiv&
-\frac{1}{6}\left(\frac{{\mu_{C}}^{b}}{T}\right)
\left[\left(\frac{{\mu_{C}}^{b}}{T}\right)-i\pi\right]
\left[\left(\frac{{\mu_{C}}^{b}}{T}\right)-2i\pi\right].
\end{eqnarray}
Furthermore, 
the term $\left(\Delta{\bf m}^2_{D\,(G)}\right)^{b}$
is found to be eliminated from the final results
of the gluon polarization tensor with external line legs,
namely, 
${\Pi_{(gp)\,\alpha'\alpha}}^{a'a}(p_{0},\vec{p})
=\left({\bf T}^{a'}\right)_{c b}
\left({\bf T}^{a}\right)_{b c}
{\Pi_{(gp)\,\alpha'\alpha}}^{b c}(p_{0},\vec{p})$.

The gluon's part of the gluon polarization tensor
with the soft external momentum $p$ 
and the external adjoint color indexes 
$a'\,a$
becomes  
\begin{eqnarray}
{\left.{\Pi_{(gp)\,\alpha'\alpha}}\right.}^{a'a} \left(p_0,\vec{p}\right)
&=&
{\left.{\Pi_{(g-l)\,\alpha'\alpha}}\right.}^{a' a}\left(p_0,\vec{p}\right)
+
{\left.{\Pi_{(tp)\,\alpha'\alpha}}\right.}^{a' a}\left(p_0,\vec{p}\right)
+
{\left.{\Pi_{(gh)\,\alpha'\alpha}}\right.}^{a' a}\left(p_0,\vec{p}\right),
\nonumber\\
&=&
\left({\bf T}^{a'}\right)_{c'b'}
\left({\bf T}^{a}\right)_{bc}
\, \left[
{\Pi_{(gp)\,\alpha'\alpha}}^{b c}(p_{0},\vec{p}) 
\right] \,
\delta^{b'b}\,\delta^{c'c},
\nonumber\\
&=&
\left({\bf m}^2_{D\,(G)}\right)^{a'a}
\left[
-\delta^{0}_{\alpha'}\,\delta^{0}_{\alpha}
+ p_0\,\int \frac{d\Omega}{4\pi}
\frac{\hat{k}_{\alpha'}\,\hat{k}_{\alpha}}
{\left(p_0-\hat{k}\cdot\vec{p}\right)}
\right].
\label{gluon-polarization-gp-total1}
\end{eqnarray}
The gluon's part of the screening Debye mass 
is given by 
\begin{eqnarray}
\left({\bf m}^2_{D\,(G)}\right)^{a'a}
&=& \left({\bf m}^2_{D\,(G)}\right)^{a}\,\delta^{a'a},
\nonumber\\
&=&
\left({\bf T}^{a'}\right)_{c'b'}
\left({\bf T}^{a}\right)_{bc}
\left({\bf m}^2_{D\,(G)}\right)^{b c}
\,\delta^{b' b}\,\delta^{c' c},
\end{eqnarray}
where the two external gluon legs 
are labeled by the adjoint color indexes $a'\,a$
while the hard  internal loop segments are labeled by 
$b'\,b$ and $c'\,c$.

\subsection{Quark-loop diagram kernel ${\Pi_{ql}}(p,k)$}
The quark-loop part of the gluon-self energy is depicted
in Fig.~(\ref{fig:gluoneff} d). 
The structure of the quark-loop 
in the gluon's self-energy is quite different from 
the gluon's part that is given by the gluon-loop, 
tadpole and the ghost-loop.
The interaction kernel of the quark-loop is a function
of the soft external momentum $p$ 
and the hard internal momentum $k$.
The two soft external gluon legs are identified by
the adjoint color indexes $a'\,a$ 
and the Lorentz polarization indexes, 
namely, $\alpha'\,\alpha$. 
It is given by 
\begin{eqnarray}
{ {\Pi^{\alpha'\alpha}_{(ql)}} }^{a'a}(p,k)
&=&
{ {\Pi^{\alpha'\alpha}_{(ql)}} }^{a'a}
(p_{0},\vec{p},k_{0},\vec{k}),
\nonumber\\
&=&
\sum^{N_{F}}_{Q}
{ {\Pi^{\alpha'\alpha}_{(ql)\,Q}} }^{a'a}(p,k),
\label{quark-loop-self1a}
\end{eqnarray}
where the sum is taken over the flavors 
and the kernel of quark-loop polarization tensor 
is given by 
\begin{eqnarray}
{ {\Pi^{\alpha'\alpha}_{(ql)\,Q}} }^{a'a}(p,k)&=&
{ {\Pi^{\alpha'\alpha}_{(ql)\,Q}} }^{a'a}
(p_{0},\vec{p},k_{0},\vec{k}),
\nonumber\\
&=&
\mbox{tr}\left[
\left(
-g\,\gamma^{\alpha'}\,{\bf t}^{a'\dagger}_{i'j'} 
\right)\,
i\,
{{\cal S}_{Q}}_{j'j}(k_0,\vec{k})
\right.
\nonumber\\
&~& 
~~~ ~~~\times\,\left.
\left(
-g\,\gamma^{\alpha}\,{\bf t}^{a}_{ji}
\right)\,
i\,
{{\cal S}_{Q}}_{i i'}
(p_0-k_0,\vec{p}-\vec{k})
\right].
\label{quark-loop-self1b}
\end{eqnarray}
Eq.~(\ref{quark-loop-self1b}) is reduced to
\begin{eqnarray}
{ {\Pi^{\alpha'\alpha}_{(ql)\,Q}} }^{a'a}(p,k)
&=&
{\bf t}^{a'\dagger}_{i'j'}\,{\bf t}^{a}_{ji}
\,\left[
{\left.{ \Pi^{\alpha'\alpha}_{(ql)\,Q} }\right.}_{j'j\,ii'}(p,k)
\right],
\nonumber\\
&\equiv&
{\bf t}^{a'}_{i'j'}\,{\bf t}^{a}_{ji}
\,\left[
{ \left.{\Pi^{\alpha'\alpha}_{(ql)\,Q} }
\right.}_{j'j\,ii'}(p,k)\right].
\label{gluon-quark-loop1}
\end{eqnarray}
In terms of the internal loop fundamental color indexes, 
it is reduced to
\begin{eqnarray}
{\left.{\Pi^{\alpha'\alpha}_{(ql)\,Q}}\right.}_{j'j\,ii'}(p,k)
&=&
{\left.{\Pi^{\alpha'\alpha}_{(ql)\,Q}}\right.}_{j'j\,ii'}
(p_{0},\vec{p},k_{0},\vec{k}),
\nonumber\\
&=&
g^2\,\mbox{tr}\left[
\gamma^{\alpha'}\,
i\,
{{\cal S}_{Q}}_{j'j}(k_0,\vec{k})
\,\gamma^{\alpha}\,
i\,
{{\cal S}_{Q}}_{ii'}(p_0-k_0,\vec{p}-\vec{k})
\right].
\label{kernel-ql-decompose}
\end{eqnarray}
The kernel of quark-loop for the gluon polarization tensor
${\left.{\Pi^{\alpha'\alpha}_{(ql)\,Q}}\right.}_{j'j\,ii'}(p,k)$
is identified by the internal fundamental color indexes 
$i'\,i$ and $j'\,j$ for the two internal quark segments
of the quark loop. 
The external adjoint color indexes 
$a'\,a$ for the two soft external gluon legs
are coupled to the internal quark fundamental  
color indexes for the two hard internal segments
(i.e. the upper and lower arcs)
of the internal quark-loop
through the fundamental matrices elements 
${\bf t}^{a'}_{i'j'}$ and ${\bf t}^{a}_{ji}$.
The trace (i.e. tr) is over the Dirac matrices which
appear in the quark propagators. 
The summation is considered over all the given flavors. 
It will be a good approximation to take the sum only 
over the up, down and strange flavors 
since the heavy flavors are suppressed strongly
because of their large masses. 

\subsection{Quark's part of gluon self-energy: 
$\Pi_{ql}\left(p_{0},\vec{p}\right)$}

The quark loop in the gluon polarization tensor 
is written as follows,
\begin{eqnarray}
{\left.{\Pi^{\alpha'\alpha}_{(ql)}}\right.}^{a' a}(p_{0},\vec{p})
&=&
\int \frac{d^3\vec{k}}{(2\pi)^3} \int \frac{d k_0}{2\pi}
\,\sum^{N_F}_{Q}{\left.{\Pi^{\alpha'\alpha}_{(ql)\,Q}}\right.}^{a' a}
(p_{0},\vec{p},k_{0},\vec{k}),
\nonumber\\
&=&
\sum^{N_F}_{Q}
{\bf t}^{a'\dagger}_{i'j'}{\bf t}^{a}_{ji}
\left[
{\left.{\Pi^{\alpha'\alpha}_{(ql)\,Q}}\right.}_{j'j\, ii'}
(p_0,\vec{p})\right],
\nonumber\\
&\equiv&
\sum^{N_F}_{Q}
{\bf t}^{a'}_{i'j'}{\bf t}^{a}_{ji}
\left[
{\left.{\Pi^{\alpha'\alpha}_{(ql)\,Q}}\right.}_{j'j\, ii'}
(p_0,\vec{p})\right],
\label{gluon-ql-energy1}
\end{eqnarray}
where 
${\left.{\Pi^{\alpha'\alpha}_{(ql)\,Q}}\right.}^{a' a}
(p_{0},\vec{p},k_{0},\vec{k})$
is the colored kernel of the quark loop self-energy 
that is defined by Eqns.~(\ref{quark-loop-self1a}) 
and (\ref{quark-loop-self1b}).
It is shown in Eq.~(\ref{gluon-ql-energy1}) that 
the quark-loop is given in terms of the soft external 
gluon momentum $p$, the Lorentz polarization indexes 
$\alpha'\,\alpha$ and the adjoint color indexes $a'\,a$ as well.
Furthermore, the quark-loop part of the gluon polarization tensor, 
which carries external adjoint color indexes $a'\,a$, 
is decomposed to the components 
those are identified by 
the fundamental color indexes 
$i'\,i$ and $j'\,j$. 
Those  fundamental color indexes correspond 
the upper and lower segments of the internal quark loop.
As mention below Eq.~(\ref{kernel-ql-decompose}),
the adjoint color indexes 
$a'\,a$ for the two soft external gluon legs
are coupled to the internal quark fundamental color indexes
for the two hard internal quark segments 
(i.e. the upper and lower arcs)
of the internal quark-loop
through the elements of fundamental matrices
${\bf t}^{a'}_{i'j'}$ and ${\bf t}^{a}_{ji}$. 
In the imaginary-time formalism,
the quark-loop that is identified by the internal quark-loop's
fundamental color indexes $i'\,i$ and $j'\,j$ 
can be written in terms 
of the mixed-time representation as follows,
\begin{eqnarray}
\left[
{\left.{\Pi^{\alpha'\alpha}_{(ql)\,Q}}\right.}_{j'j\, ii'}
(p_{0},\vec{p},k_{0},\vec{k})\right]
&=&
\int^{\beta}_{0} d\tau\int^{\beta}_{0} d\tau'
e^{\left[k_0-\mu_Q-i\frac{1}{\beta}\theta_j\right]\tau}
e^{\left[(k_0-p_0)-\mu_Q-i\frac{1}{\beta}\theta_i\right]\tau'}
\nonumber\\
&~&
~~~\times
\left[
{\left.{\Pi^{\alpha'\alpha}_{(ql)\,Q}}\right.}_{j'j\, ii'}
(\tau,\vec{k};\tau',\vec{k}-\vec{p})\right].
\label{quark-loop-kernel-p-k1a}
\end{eqnarray}
The quark-loop kernel becomes a function of 
the mixed-time variables  $\tau$ and $\tau'$. 
It is reduced to
\begin{eqnarray}
\left[
{\left.{\Pi^{\alpha'\alpha}_{(ql)\,Q}}\right.}_{j'j\, ii'}
(\tau,\vec{k};\tau',\vec{k}-\vec{p})
\right]
&=&
g^2\,
\mbox{tr}\left[\gamma^{\alpha'}\,
i\,
{{\cal S}_{Q}}_{j'j}(\tau,\vec{k})\,
\gamma^{\alpha}\,
i\,
{{\cal S}_{Q}}_{ii'}
(\tau',\vec{k}-\vec{p})\right].
\label{quark-loop-kernel-p-k1b}
\end{eqnarray}
The integration of Eq.~(\ref{quark-loop-kernel-p-k1a})
over the momentum time-component $k_0$ 
gives
\begin{eqnarray}
\left[
{\left.{\Pi^{\alpha'\alpha}_{(ql)\,Q}}\right.}_{j'j\, ii'}
(p_0,\vec{p},\vec{k})\right]
&=&
\int\frac{d k_{0}}{2\pi}
\left[
{\left.{\Pi^{\alpha'\alpha}_{(ql)\,Q}}\right.}_{j'j\, ii'}
(p_{0},\vec{p},k_{0},\vec{k})\right],
\nonumber\\
&=&
-\int^{\beta}_{0} d\tau 
e^{-\left[p_0+i\,\frac{1}{\beta}(\theta_i-\theta_j)\right](\beta-\tau)}
\nonumber\\
&~&~\times
\left[
{\left.{\Pi^{\alpha'\alpha}_{(ql)\,Q}}\right.}_{j'j\, ii'}
(\tau,\vec{k};\beta-\tau,\vec{k}-\vec{p})\right],
\label{quark-loop-kernel-p-k1c}
\end{eqnarray}
where following identity 
\begin{eqnarray}
\int\frac{d k_0}{2\pi}
e^{\left[k_0 - \mu_Q-i\frac{1}{\beta}\theta_j\right]\tau}
e^{\left[(k_0-p_0) 
- \mu_Q-i\frac{1}{\beta}\theta_i\right]\tau'}
&=&
\int\frac{d k_0}{2\pi}
e^{\left[k_0 - \mu_Q-i\frac{1}{\beta}\theta_j\right](\tau+\tau')}
e^{-\left[p_0 + i\frac{1}{\beta}(\theta_i-\theta_j)\right]\tau'},
\nonumber\\
&=&
\frac{1}{\beta}\sum_n e^{i\omega_n(\tau+\tau')}
e^{-\left[p_0 + i\frac{1}{\beta}(\theta_i-\theta_j)\right]\tau'},
\nonumber\\
&=&
-\delta\left(\tau+\tau'-\beta\right)
e^{-\left[p_0 + i\frac{1}{\beta}(\theta_i-\theta_j)\right]\tau'},
\label{quark-loop-kernel-identity1}
\end{eqnarray}
is considered.
Furthermore, the following useful identity
\begin{eqnarray}
\exp\left(
\left[p_0 + i\frac{1}{\beta}(\theta_i-\theta_j)\right]\beta
\right)=1,
\end{eqnarray}
is considered because of the Matsubara definition 
for the gluon time-like energy, namely,
\begin{eqnarray}
p_0+i\frac{1}{\beta}(\theta_i-\theta_j)&=&
i\,\frac{2m\pi}{\beta},
\nonumber\\
&\equiv& p_0+i\frac{1}{\beta}\phi^{a}, 
~~~\left(\mbox{given that}~ a=\underbrace{(ij)}\right).
\end{eqnarray}
The standard Foldy-Wouthuysen energy decomposition 
simplifies the calculation 
of Eq.~(\ref{quark-loop-kernel-p-k1c}). 
Moreover, Eq.~(\ref{quark-loop-kernel-p-k1b}) is decomposed 
into the positive and negative energy components 
as follows
\begin{eqnarray}
\left[
{\left.{\Pi^{\alpha'\alpha}_{(ql)\,Q}}\right.}_{j'j\, ii'}
(\tau,\vec{k};\beta-\tau,\vec{k}-\vec{p})
\right]
&=&\sum_{r=\pm}\sum_{s=\pm}
\left[
{\left.{\Pi^{(rs)\,\alpha'\alpha}_{(ql)\,Q}}\right.}_{ji}
(\tau,\vec{k},\vec{p})\right]\,\delta_{j'j}\,\delta_{i'i}.
\end{eqnarray}
Using the preceding Foldy-Wouthuysen decomposition, 
Eq.~(\ref{quark-loop-kernel-p-k1c}) is decomposed 
to the positive and negative energy components 
as follows
\begin{eqnarray}
\left[
{\left.{\Pi^{\alpha'\alpha}_{(ql)\,Q}}\right.}_{j'j\, ii'}
(p_0,\vec{p},\vec{k})
\right]
&=&
-\int^{\beta}_{0} d\tau
e^{-\left[p_0+i\frac{1}{\beta}
\left(\theta_i-\theta_j\right)\right]
\left(\beta-\tau\right)}
\nonumber\\
&~&~\times
\sum^{\pm}_{r}\sum^{\pm}_{s}
\left[
{\left.{\Pi^{(rs)\,\alpha'\alpha}_{(ql)\,Q}}\right.}_{ji}
(\tau,\vec{k},\vec{p})\right]\,\delta_{j'j}\,\delta_{i'i},
\nonumber\\
&=&
\sum^{\pm}_{r}\sum^{\pm}_{s}
\left[
{\left.{\Pi^{(rs)\,\alpha'\alpha}_{(ql)\,Q}}\right.}_{ji}
(p_0,\vec{p},\vec{k})\right]\,\delta_{jj'}\,\delta_{ii'}.
\label{ql-mixed-representation-tau}
\end{eqnarray}
The positive-positive component, namely, 
${\Pi^{(++)\,\alpha'\alpha}_{(ql)\,Q}}_{ji}(\tau,\vec{k},\vec{p})$
is reduced to
\begin{eqnarray}
\left[
{\Pi^{(++)\,\alpha'\alpha}_{(ql)\,Q}}(\tau,\vec{k},\vec{p})
\right]
&=&
g^2\,
{ \Lambda_{Q}^{(++)}}^{\alpha'\alpha}\left(\vec{k},\vec{p}\right)
\left[
1-n_{F}\left(
\epsilon_Q(\vec{k})-\mu_Q-i\frac{1}{\beta}\theta_j
\right)
\right] 
\nonumber\\
&~&\times
\left[
1-n_{F}
\left(\epsilon_Q(\vec{k}
-\vec{p})-\mu_Q-i\frac{1}{\beta}\theta_i\right)
\right]
\nonumber\\
&~&
\times
\exp\left(-\tau 
\left[\epsilon_Q(\vec{k})
-\mu_Q-i\frac{1}{\beta}\theta_j\right]
\right)
\nonumber\\
&~&
\times
\exp\left(
-(\beta-\tau) 
\left[\epsilon_Q(\vec{k}-\vec{p})-\mu_Q-i\frac{1}{\beta}\theta_i\right]
\right),
\end{eqnarray}
where the double Foldy-Wouthuysen energy projector is given by
\begin{eqnarray}
{ \Lambda_{Q}^{(rs)} }^{\alpha'\alpha}(\vec{k},\vec{p})&=&
\mbox{tr}\left[
\gamma^{\alpha'}\,{\Lambda_{Q}^{(r)}}(\vec{k})\,\gamma_0
\gamma^{\alpha}\,{\Lambda_{Q}^{(s)}}(\vec{k}-\vec{p})\,\gamma_0
\right], ~(\mbox{where}~ r=\pm,\, s=\pm).
\end{eqnarray}
The double Foldy-Wouthuysen energy projectors are identified by
the double Lorentz polarization indexes $\alpha'\,\alpha$ and
the double positive and negative energy channels.
The Foldy-Wouthuysen energy component, namely, $(++)$ 
of the quark-loop part of the gluon polarization tensor
in the mixed-time representation reads,
\begin{eqnarray}
\left[
{\left.{\Pi^{(++)\,\alpha'\alpha}_{(ql)\,Q}}\right.}_{ji}
(\tau,\vec{k},\vec{p})
\right]
&=&
g^2\,
{ \Lambda_{Q}^{(++)\,\alpha'\alpha} }\left(\vec{k},\vec{p}\right)
\left[
1-n_{F}\left(
\epsilon_Q(\vec{k})-\mu_Q-i\frac{1}{\beta}\theta_{j}
\right)
\right] 
\nonumber\\
&~&\times
\left[
1-n_{F}\left(
\epsilon_Q(\vec{k}-\vec{p})-\mu_Q-i\frac{1}{\beta}\theta_{i}
\right)
\right]\nonumber\\
&~&
\times
\exp\left(-\tau 
\left[\epsilon_Q(\vec{k})-\mu_Q-i\frac{1}{\beta}\theta_{j}\right]
\right)
\nonumber\\
&~&
\times
\exp\left(
-(\beta-\tau) 
\left[\epsilon_Q(\vec{k}-\vec{p})-\mu_Q-i\frac{1}{\beta}\theta_{i}\right]
\right).
\end{eqnarray}
The results for Foldy-Wouthuysen energy components, 
namely,
$(+-)$, $(-+)$ and $(--)$
are obtained in a similar manner.
The approximation of the double
Foldy-Wouthuysen energy projectors
in the limit $(|\vec{k}|\gg |\vec{p}|)$ simplifies
the calculations drastically.
They are identified by the Lorentz polarization 
indexes $\alpha'\,\alpha$.
The time-time double
Foldy-Wouthuysen energy projectors
are approximated  to
\begin{eqnarray}
{ \Lambda_{Q}^{(rs)\,0 0} }(\vec{k},\vec{p})&\equiv&
{ \Lambda_{Q\,0 0}^{(rs)} }(\vec{k},\vec{p})
\approx
\left\{\left(1 + r s\right)\right\},
\end{eqnarray}
where $\left\{r=\pm, s=\pm\right\}$ 
are the positive and negative energy components.
The time-space components are given by
\begin{eqnarray}
{ \Lambda_{Q}^{(rs)\,0 m} }(\vec{k},\vec{p})&\equiv&
{ \Lambda_{Q\,0 m}^{(rs)} }(\vec{k},\vec{p})\approx
-\left(s+r\right)\,\hat{k}_{m}.
\end{eqnarray}
Furthermore, the space-space components
are approximated to
\begin{eqnarray}
{ \Lambda_{Q}^{(rs)\,n m} }(\vec{k},\vec{p})&\equiv&
{ \Lambda_{Q\,n m}^{(rs)} }(\vec{k},\vec{p})\approx
\left[\delta_{ij}\left(1-r\,s\right)
+ 2 r\,s\,\hat{k}_{n}\,\hat{k}_{m}\right].
\end{eqnarray}
Therefore, the Lorentz polarization components, 
namely, $\alpha'\,\alpha$
of the double Foldy-Wouthuysen energy 
projectors are reduced to
\begin{eqnarray}
\left\{
{ \Lambda_{Q}^{(rs)\,00} }(\vec{k},\vec{p})
\right\}&\approx&\left\{2, 0, 0, 2\right\},
\nonumber\\
\left\{
{ \Lambda_{Q}^{(rs)\,0 m} }(\vec{k},\vec{p})\right\}
&\approx&
\left\{-2\hat{k}_{m},0,0,2\hat{k}_{m}\right\},
\nonumber\\
\left\{ {\Lambda_{Q}^{(rs)\,n m} }(\vec{k},\vec{p})\right\}
&\approx&
\left\{
2\hat{k}_{n}\,\hat{k}_{m},
2\left(\delta_{nm}-\hat{k}_{n}\,\hat{k}_{m}\right),
2\left(\delta_{nm}-\hat{k}_{n}\,\hat{k}_{m}\right),
2\hat{k}_{n}\,\hat{k}_{m}
\right\},
\end{eqnarray}
for the positive and negative 
$\left\{(++), (+-), (-+), (--)\right\}$ 
energy components, respectively.
%
%
%
Evaluating the integral 
in Eq.~(\ref{ql-mixed-representation-tau})
over the time variable $\tau$ in order to find
the inverse transformation of the mixed-time representation, 
the positive-positive energy component 
of the quark-loop part of 
the gluon polarization tensor is reduced to 
\begin{eqnarray}
{\left.{\Pi^{(++)\,\alpha'\alpha}_{(ql)\,Q}}\right.}_{ji}
(p_0,\vec{p})
&=&
\int \frac{d^3\vec{k}}{(2\pi)^{3}}
\left[
{\left.{\Pi^{(++)\,\alpha'\alpha}_{(ql)\,Q}}\right.}_{ji}
(p_0,\vec{p},\vec{k})
\right],
\end{eqnarray}
where
\begin{eqnarray}
{\left.{\Pi^{(++)\,\alpha'\alpha}_{(ql)\,Q}}\right.}_{ji}
(p_0,\vec{p},\vec{k})
&=&
-g^2\,{ \Lambda_{Q}^{(rs)\alpha'\alpha} }(\vec{k},\vec{p})
\nonumber\\
&\times&
\left[
\frac{n_{F}\left(
\epsilon_{Q}(\vec{k})-\mu_Q-i\frac{\theta_j}{\beta}
\right)
-n_{F}\left(
\epsilon_{Q}(\vec{k}-\vec{p})-\mu_Q-i\frac{\theta_i}{\beta}
\right)}
{\,p_0-\left(\epsilon_{Q}(\vec{k})-\epsilon_{Q}(\vec{k}-\vec{p})
\right)}
\right].
\label{gluon-gl-positive-positive-cal1}
\end{eqnarray}
The results for the components $(+-)$, $(-+)$ 
and $(--)$
are obtained in a similar manner.
Using the symmetry argument under the sum
of the repeated index, we have the following result,
\begin{eqnarray}
{\bf t}^{a'}_{ij}\,{\bf t}^{a}_{ji}\,
\left[
n_{F}\left(
\epsilon_{Q}(\vec{k})-\mu_Q-i\frac{\theta_j}{\beta}
\right)
-n_{F}\left(
\epsilon_{Q}(\vec{k})-\mu_Q-i\frac{\theta_i}{\beta}
\right)
\right]&=&0.
\end{eqnarray}
Since in the HTL approximation, 
we have $\vec{p}<<\vec{k}$ (i.e. $p/k\,\sim\, g$),
then by evaluating the integration over $\vec{k}$,
Eq.~(\ref{gluon-gl-positive-positive-cal1})
is approximated to
\begin{eqnarray}
{\left.{\Pi^{(++)\,n m}_{(ql)\,Q}}\right.}_{ji}
(p_0,\vec{p})
&=&
\int \frac{d|\vec{k}|}{2\pi^2} |\vec{k}|^2\,
\int \frac{d\Omega_k}{4\pi} 
\left[
{\left.{\Pi^{(++)\,n m}_{(ql)\,Q}}\right.}_{ji}
(p_0,\vec{p},\vec{k})
\right],
\nonumber\\
&=&
- 2 g^2
\int \frac{d\Omega_k}{4\pi}\,\hat{k}_{n}\hat{k}_{m}\,
\frac{\hat{k}\cdot\vec{p}}
{p_0-\hat{k}\cdot\vec{p}}
\nonumber\\
&~&~~~\times\,
\left(\int 
\frac{d|\vec{k}|}{2\pi^2} |\vec{k}|^2
\,\frac{d}{d|\vec{k}|}
\left[
n_{F}\left(
\epsilon_{Q}(\vec{k})-\mu_Q-i\frac{\theta_j}{\beta}
\right)\right]\right),
\nonumber\\
&\equiv&
\frac{1}{2}\left[
(\cdots\theta_{i}\cdots)+(\cdots\theta_{j}\cdots)
\right],
\end{eqnarray}
for space-space, namely, $n\,m$ Lorentz component.
The last line indicates the symmetrization 
over the color indexes $i$ and $j$.
The symmetry over the polar integration leads to
\begin{eqnarray}
\int \frac{d\Omega_k}{4\pi}
\left(\delta_{nm}-\hat{k}_{n}\hat{k}_{m}\right)
\frac{\hat{k}\cdot\vec{p}}
{p_0+\hat{k}\cdot\vec{p}}
&=&
-\int \frac{d\Omega_k}{2\pi}
\left(\delta_{nm}-\hat{k}_{n}\hat{k}_{m}\right)
\frac{\hat{k}\cdot\vec{p}}
{p_0-\hat{k}\cdot\vec{p}}.
\end{eqnarray}
Moreover, a similar result is obtained 
for $(--)$ component
but replacing
$n_{F}\left(
\epsilon_{Q}(\vec{k})-\mu_Q-i\frac{\theta_j}{\beta}
\right)$
with 
$n_{F}\left(
\epsilon_{Q}(\vec{k})+\mu_Q+i\frac{\theta_j}{\beta}
\right)$. 
The same approximation is considered in calculating 
the $(+-)$ and $(-+)$ components. 
The result for the sum over $(+-)$ and $(-+)$ components
is reduced to 
\begin{eqnarray}
\sum^{r=\pm}_{s=-r}
{\left.{\Pi^{(rs)\,n m}_{(ql)\,Q}}\right.}_{ji}
(p_0,\vec{p})&=&
\left[{\left.{\Pi^{(+-)\,n m}_{(ql)\,Q}}\right.}_{ji}
(p_0,\vec{p})
+
{\left.{\Pi^{(-+)\,n m}_{(ql)\,Q}}\right.}_{ji}
(p_0,\vec{p})\right],
\nonumber\\
&=&
\int \frac{d|\vec{k}|}{2\pi^2}  |\vec{k}|^2
\int \frac{d\Omega_k}{4\pi} 
\left[
{\left.{\Pi^{(+-)\,n m}_{(ql)\,Q}}\right.}_{ji}
(p_0,\vec{p},\vec{k})
+
{\left.{\Pi^{(-+)\,n m}_{(ql)\,Q}}\right.}_{ji}
(p_0,\vec{p},\vec{k})\right],
\nonumber\\
&\approx&
2 g^2\int \frac{d\Omega_k}{4\pi} 
\left(\delta_{nm}-\hat{k}_n\hat{k}_m\right)
\left(
\int \frac{d|\vec{k}|}{2\pi^2}
|\vec{k}|
\right.
\nonumber\\
&\times&
\left.
\left[1+
n_{F}\left(
\epsilon_{Q}(\vec{k})-\mu_Q-i\frac{\theta_j}{\beta}
\right)
+
n_{F}\left(
\epsilon_{Q}(\vec{k})+\mu_Q+i\frac{\theta_j}{\beta}
\right)
\right]
\right).
\label{positive-negative-total-screen1}
\end{eqnarray}
By dropping the first term inside the square bracket
which is a non-thermal one 
and retaining only the thermal terms,  
Eq.~(\ref{positive-negative-total-screen1}) 
becomes,
\begin{eqnarray}
\sum^{r=\pm}_{s=-r}
{\left.{\Pi^{(rs)\,n m}_{(ql)\,Q}}\right.}_{ji}
(p_0,\vec{p})
&\approx&
-g^2\int \frac{d\Omega_k}{4\pi} 
\left(\delta_{nm}-\hat{k}_n\,\hat{k}_m\right)
\left(
\int \frac{d|\vec{k}|}{2\pi^2}
|\vec{k}|^2
\,
\frac{d}{d|\vec{k}|} 
\right.
\nonumber\\
&\times&
\left.
\left[
n_{F}\left(
\epsilon_{Q}(\vec{k})-\mu_Q-i\frac{\theta_j}{\beta}
\right)
+
n_{F}\left(
\epsilon_{Q}(\vec{k})+\mu_Q+i\frac{\theta_j}{\beta}
\right)
\right]
\right).
\end{eqnarray}
Furthermore, by imposing the tensor identity
\begin{eqnarray}
\int d\Omega_k \left(\delta_{nm}-3\hat{k}_n\hat{k}_m\right)
=0,
\end{eqnarray}
and the relation
\begin{eqnarray}
\frac{\hat{k}\cdot\vec{p}}{p_{0}-\hat{k}\cdot\vec{p}}
&=&
1-\frac{p_{0}}{p_{0}-\hat{k}\cdot\vec{p}},
\end{eqnarray}
the quark-loop part, namely, ${\Pi_{ql}}(p_0,\vec{p})$ 
of the gluon polarization tensor is reduced to
\begin{eqnarray}
\left[
{\left.{\Pi^{n m}_{(ql)}}\right.}_{ji}(p_0,\vec{p})
\right]
&=&
\sum^{n_{F}}_{Q=1} 
\left[
{\left.{\Pi^{n m}_{(ql)\,Q}}\right.}_{ji}(p_0,\vec{p})
\right],
\nonumber\\
&\approx&
-2 g^2\,\sum^{n_{F}}_{Q=1}\,
\int \frac{d|\vec{k}|}{2\pi^2}
|\vec{k}|^2
\frac{d}{d|\vec{k}|}
\left[
n_{F}\left(
\epsilon_{Q}(\vec{k})-\mu_Q-i\frac{\theta_j}{\beta}
\right)
\right.
\nonumber\\
&~&~~~
\left.
+
n_{F}\left(
\epsilon_{Q}(\vec{k})+\mu_Q+i\frac{\theta_j}{\beta}
\right)
\right]
\left(
p_{0}\,
\int \frac{d\Omega_k}{4\pi}
\frac{\hat{k}_{n}\hat{k}_{m}}{p_0-\hat{k}\cdot\vec{p}}
\right).
\label{ql-polarization-ij1}
\end{eqnarray}
Furthermore,
Eq.~(\ref{ql-polarization-ij1}) is re-expressed as follows
\begin{eqnarray}
\left[
{\left.{\Pi^{n m}_{(ql)}}\right.}_{ji}(p_0,\vec{p})
\right]
&\approx&
\left(
{\bf m}^{2}_{D\,(Q)}
\right)_{ij} 
\left(p_0
\int \frac{d\Omega_k}{4\pi}
\frac{\hat{k}_{n}\hat{k}_{m}}{p_0-\hat{k}\cdot\vec{p}}
\right).
\end{eqnarray}
The quark-loop part of the Debye mass is symmetrized 
as follows
\begin{eqnarray}
\left(
{\bf m}^{2}_{D\,(Q)}
\right)_{ij}&\equiv&
\frac{1}{2}
\left[ \left({\bf m}^{2}_{D\,(Q)}\right)_{i}
+
\left({\bf m}^{2}_{D\,(Q)}\right)_{j} \right].
\end{eqnarray}
It should be noted here
that the fundamental color indexes, namely, 
$i$ and $j$ run
over all color charges in the final calculation.
The summation over the fundamental color index $i$ 
is identical to that over $j$.
Hence, the quark-loop part of the Debye mass, 
with the fundamental color index $i$,
is reduced to
\begin{eqnarray}
\left(
{\bf m}^{2}_{D\,(Q)}\right)_{i}
&=& 
\sum^{n_{F}}_{Q=1}
\left({{\bf m}^{2}_{D\,(Q)}}_{Q}\right)_{i},
\label{Debye-quark-part-total}
\end{eqnarray}
where
\begin{eqnarray}
\left(
{{\bf m}^{2}_{D\,(Q)}}_{Q}\right)_{i}&=&
4g^2
\int\frac{d|\vec{k}|}{2\pi^2}
|\vec{k}|
\left[
n_{F}\left(
\epsilon_{Q}(\vec{k})-\mu_Q-i\frac{\theta_{i}}{\beta}
\right)
+
n_{F}\left(
\epsilon_{Q}(\vec{k})+\mu_Q+i\frac{\theta_{i}}{\beta}
\right)\right],
\nonumber\\
&=&
\frac{2g^2}{\pi^2}\,
\left[
\frac{\pi^2}{6\,\beta^2}
+\frac{1}{2}
\left(
\mu_{Q}+i\,\frac{1}{\beta}\theta_{i}
\right)^2
\right].
\label{Debye-quark-part}
\end{eqnarray}
When the real fundamental color potential is considered, namely,
$i\,\frac{1}{\beta}\theta_{i}\rightarrow {\mu_{C}}_{i}$,
Eq.~(\ref{Debye-quark-part-total}) is reduced to
\begin{eqnarray}
\left({{\bf m}^{2}_{D\,(Q)}}\right)_{i}
&\rightarrow&
\frac{2g^2}{\pi^2}\,\sum^{N_{F}}_{Q=1}
\left[
\frac{\pi^2}{6\,\beta^2}
+\frac{1}{2}
\left(\mu_{Q}+{\mu_{C}}_{i}\right)^2
\right].
\end{eqnarray}
The quark-loop part of the gluon polarization tensor
with the Lorentz polarization indexes, namely,
$\alpha\,\alpha'$ 
and the internal fundamental color indexes, namely,
$i$ and $j$ is reduced to
\begin{eqnarray}
\left[
{\left.{\Pi^{\alpha'\alpha}_{(ql)}}\right.}_{ji}(p_0,\vec{p})
\right]
&=& 
\sum^{N_F}_{Q=1}\left[
{\left.{\Pi^{\alpha'\alpha}_{(ql)\,Q}}\right.}_{ji}(p_0,\vec{p})
\right],
\nonumber\\
&=&
\left({\bf m}^2_{D\,(Q)}\right)_{ji}
\left[
-\delta_{0\alpha'}\delta_{0\alpha}
+ 
p_0\,
\int \frac{d\Omega_k}{4\pi}
\frac{\hat{k}_{\alpha'}\,\hat{k}_{\alpha}}
{p_0-\hat{k}\cdot\vec{p}}
\right].
\end{eqnarray}
Moreover, the quark-loop part 
with the Lorentz polarization indexes $\alpha\,\alpha'$ 
and the external adjoint color indexes, namely, 
$a$ and $a'$ for 
the two external gluon legs  is transformed as follows,
\begin{eqnarray}
\left[
{\left.{\Pi^{\alpha' \alpha}_{(ql)}}\right.}^{a' a}(p_0,\vec{p})
\right]
&=&
{\bf t}^{a'}_{i'j'}\,{\bf t}^{a}_{ji}\,\delta_{j'j}\delta_{i'i}\,
\left[
{\left.{\Pi^{\alpha'\alpha}_{(ql)}}\right.}_{ji}(p_0,\vec{p})
\right],
\nonumber\\
&=&
\left({\bf m}^{2}_{D\,(Q)}\right)^{a'\,a}
\left[
-\delta_{0\alpha'}\,\delta_{0\alpha}
+
p_0\,\int \frac{d\Omega_k}{4\pi}
\frac{\hat{k}_{\alpha'}\,\hat{k}_{\alpha}}
{p_0-\hat{k}\cdot\vec{p}}
\right].
\label{gluon-polarization-ql-total1}
\end{eqnarray}
Therefore, the quark-loop part of the gluon Debye mass 
with the external gluon legs, which are labeled by 
the adjoint color indexes $a\,a'$, 
is related to the Debye mass with internal fundamental 
color indexes $i\,j$ in the following way
\begin{eqnarray}
\left({\bf m}^{2}_{D\,(Q)}\right)^{a'a}&=&
{\bf t}^{a'}_{ij}{\bf t}^{a}_{ji}
\,
\left({\bf m}^{2}_{D\,(Q)}\right)_{ij},
\nonumber\\
&=&
{\bf t}^{a'}_{ij}{\bf t}^{a}_{ji}
\,
\left({\bf m}^{2}_{D\,(Q)}\right)_{i}, \,\,
(\mbox{or}~i\rightarrow j).
\end{eqnarray}

\subsection{The Gluon Polarization tensor 
and the Debye mass}
The total gluon polarization tensor, namely,
${{\Pi}^{\alpha'\alpha}}^{a'a}\left(p_0,\vec{p}\right)$ 
is found by adding the contribution of the
gluon's part  which consists 
the gluon loop, ghost loop and tadpole
and the contribution of the quark loop all together.
The contribution of the gluon's part 
is given by 
Eq.~(\ref{gluon-polarization-gp-total1})
while the contribution from the quark loop 
is given by 
Eq.~(\ref{gluon-polarization-ql-total1}).
Hence, the total gluon self-energy 
from the all Feynman diagrams 
which are displayed
in Figs.~(\ref{fig:gluoneff} a), 
(\ref{fig:gluoneff} b), (\ref{fig:gluoneff} c) 
and 
(\ref{fig:gluoneff} d)
is given by
\begin{eqnarray}
{{\Pi}^{\alpha'\alpha}}^{a'a}\left(p_0,\vec{p}\right)
&=&
\left[
{\Pi_{(gp)}^{\alpha'\alpha}}^{a'a}
\left(p_0,\vec{p}\right)
+
{\Pi_{(ql)}^{\alpha'\alpha}}^{a'a}\left(p_0,\vec{p}\right)
\right].
\end{eqnarray} 
Hence, the (total) gluon polarization tensor 
is reduced to
\begin{eqnarray}
{{\Pi}_{\alpha'\alpha}}^{a'a}
\left(p_0,\vec{p}\right)
&=&
\left({\bf m}^2_{D}\right)^{a'a}
\left[
-\delta_{0\alpha'}\delta_{0\alpha}
+ p_0\int \frac{d\Omega_k}{4\pi}
\frac{\hat{k}_{\alpha'}\hat{k}_{\alpha}}
{p_0 - \hat{k}\cdot\vec{p}}
\right].
\end{eqnarray}
The gluon Debye mass with the external adjoint color
indexes $a$ and $a'$ for the two external gluon legs  
is given by
\begin{eqnarray}
\left({\bf m}^2_{D}\right)^{a'a}&=&
\left[
\left({\bf m}^2_{D\,(Q)}\right)^{a'a}
+
\left({\bf m}^2_{D\,(G)}\right)^{a'a}
\right],
\end{eqnarray}
where the first term on the right hand side comes 
from the quark-loop while the second one is contribution 
of the gluon's part of the gluon polarization tensor.
The gluon Debye mass with the external adjoint color
indexes $a$ and $a'$ is related 
to the internal loop Debye mass elements 
with the internal fundamental color indexes $i\,j$
and the internal adjoint color indexes $b\,c$ 
in the following way,
\begin{eqnarray}
\left({\bf m}^2_{D}\right)^{a'\,a}
&=&
\left[
{{\bf t}^{a'}}_{ij}\,{{\bf t}^{a}}_{ji}
\left({\bf m}^2_{D\,(Q)}\right)_{ij}
+
{{\bf T}^{a'}}_{cb}\,{{\bf T}^{a}}_{bc}
\left({\bf m}^2_{D\,(G)}\right)_{bc}
\right].
\end{eqnarray}
It should be noted that 
$\left({\bf m}^2_{D\,(Q)}\right)_{ij}\equiv
\left({\bf m}^2_{D\,(Q)}\right)_{i}$
and 
$\left({\bf m}^2_{D\,(G)}\right)_{bc}\equiv
\left({\bf m}^2_{D\,(G)}\right)_{b}$. 
Furthermore, the adjoint color index $b$ can be replaced 
by the double fundamental-like color indexes 
by the notation structure
$b=\underbrace{(n\,m)}$.
The summation over the repeated indexes 
and considering the symmetry of the fundamental and adjoint 
indexes $i\,j$ and $b\,c$, respectively, 
contract the double adjoint color indexes
and lead to following result
\begin{eqnarray}
\left({\bf m}^2_{D}\right)^{a'a}
&=&
\delta^{a'a}
\left({\bf m}^2_{D}\right)^{a}.
\end{eqnarray}
Hence,  the gluon Debye mass is identified 
by only one external adjoint color 
index $a$ in the following way,
\begin{eqnarray}
\left({\bf m}^2_{D}\right)^{a}&=&
\left[
N_c\left({\bf m}^2_{D\,(G)}\right)^{a}
+\frac{1}{2}
\left({\bf m}^2_{D\,(Q)}\right)^{a}
\right],
\end{eqnarray}
where 
\begin{eqnarray}
\left({\bf m}^2_{D\,(G)}\right)^{a}&=&
\frac{\sqrt{2}}{N_{c}}\sum^{N_c}_{ij}
{{\bf t}^{a}}_{in}
\left(
{\bf m}^2_{D\,(G)}
\right)^{\underbrace{(nj)}},
\end{eqnarray}
and 
\begin{eqnarray}
\left({\bf m}^2_{D\,(Q)}\right)^{a}&=&
\sqrt{2}\sum^{N_c}_{ij}
{{\bf t}^{a}}_{ij}
\left(
{\bf m}^2_{D\,(Q)}
\right)_{i}.
\end{eqnarray}
Moreover, the double adjoint color indexes 
for the gluon polarization tensor 
can be contracted 
and the soft gluon polarization tensor 
is reduced to
\begin{eqnarray}
{{\Pi}_{\alpha'\alpha}}^{a'a}
\left(p_0,\vec{p}\right)
&=&
\delta^{a'\,a}\, 
{{\Pi}_{\alpha'\alpha}}^{a},
\label{gluon-polarization-index-a1}
\end{eqnarray}
where
\begin{eqnarray}
{{\Pi}_{\alpha'\alpha}}^{a}\left(p_0,\vec{p}\right)
&=&
\left({\bf m}^2_{D}\right)^{a}
\left[
-\delta_{0\alpha'}\delta_{0\alpha}
+ p_0\int \frac{d\Omega_k}{4\pi}
\frac{\hat{k}_{\alpha'}\hat{k}_{\alpha}}
{p_0 - \hat{k}\cdot\vec{p}}
\right].
\label{gluon-polarization-index-a2}
\end{eqnarray}
The final result of the soft gluon polarization tensor
is gauge fixing independent.
The gluon polarization tensor 
${{\Pi}_{\alpha'\alpha}}^{a'a}\left(p_0,\vec{p}\right)$
is transverse and is satisfying the relation
$p^{\alpha'}\,
{{\Pi}_{\alpha'\alpha}}^{a'a}(p_{0},\vec{p})=0$.
The soft gluon polarization tensor can be written 
with respect to the Lorentz polarization indexes 
$\alpha'\,\alpha=\left\{0\,0,0\,i,i\,j\right\}$, 
respectively, for the time-time,
time-space and space-space Lorentz polarization components.
The time-time component of the gluon polarization
reads 
\begin{eqnarray}
{{\Pi}_{00}}^{a}\left(p_0,\vec{p}\right)
&=&
\left({\bf m}^2_{D}\right)^{a}
\left[
-1
+p_0\int \frac{d\Omega_k}{4\pi}
\frac{1}
{p_0-\hat{k}\cdot\vec{p}}
\right],
\nonumber\\
&=&
-\left({\bf m}^2_{D}\right)^{a}
\left[
1
-\frac{1}{2}\,p_{0}\,\int^{1}_{-1} d x
\frac{1}
{p_0-|\vec{p}|x}
\right],
\end{eqnarray}
while time-space component reads,
\begin{eqnarray}
{{\Pi}_{0i}}^{a}\left(p_0,\vec{p}\right)
&=&
\left({\bf m}^2_{D}\right)^{a}
\left[
-p_0\hat{p}_i\int \frac{d\Omega_k}{4\pi}
\frac{(\hat{k}\cdot\hat{p})}
{p_0-\hat{k}\cdot\vec{p}}
\right],
\nonumber\\
&=&
\left({\bf m}^2_{D}\right)^{a}
\left[
-p_0\hat{p}_i \frac{1}{2}
\int^{1}_{-1} dx
\frac{x}
{p_0-|\vec{p}|x}
\right].
\end{eqnarray}
In order to compute the space-space 
Lorentz polarization components 
with space indexes, namely, $i\,j$,
the following transformation
\begin{eqnarray}
\vec{k}=\sum_{i}
\left(\vec{k}\cdot\hat{p}_i\right)\hat{p}_i,
\end{eqnarray}
is introduced.
In the spherical coordinate, it reads
\begin{eqnarray}
\vec{k}&=&\left(\vec{k}\cdot\hat{p}_{r}\right)\hat{p}_{r}
+\left(\vec{k}\cdot\hat{p}_{\theta}\right)\hat{p}_{\theta}
+\left(\vec{k}\cdot\hat{p}_{\phi}\right)\hat{p}_{\phi},
\\
&\equiv&
\left(\vec{k}\cdot\hat{p}\right)\hat{p}
+\left(\vec{k}\cdot\hat{p}_{\perp}\right)\hat{p}_{\perp}.
\end{eqnarray}
Subsequently, the transverse projector becomes
\begin{eqnarray}
\hat{p}_{\perp i}\hat{p}_{\perp j}\rightarrow
\left(\delta_{ij}-\hat{p}_{i}\hat{p}_{j}\right).
\end{eqnarray}
Therefore, the space-space component 
with space polarization indexes, namely, $i\,j$ 
is reduced to
\begin{eqnarray}
{{\Pi}_{ij}}^{a}\left(p_0,\vec{p}\right)
&=&
\left(\hat{p}_{i}\hat{p}_{j}\right)
\left({\bf m}^2_{D}\right)^{a}
\left[
\frac{p_0}{2}
\int^{1}_{-1} dx
\frac{x^2}{p_0-|\vec{p}|x}\right]
\nonumber\\
&~&
\,+\,
\left(\delta_{ij}-\hat{p}_{i}\hat{p}_{j}\right)
\left({\bf m}^2_{D}\right)^{a}
\left[
\frac{1}{2}\frac{p_0}{2}
\int^{1}_{-1} dx
\frac{1-x^2}
{p_0-|\vec{p}|x}
\right].
\end{eqnarray}
The gluon polarization tensor
is decomposed to longitudinal (electric) 
and transverse (magnetic) components
as follows
\begin{eqnarray}
{{\Pi}_{ij}}^{a}\left(p_0,\vec{p}\right)
&=&
\left(\delta_{ij}-\hat{p}_{i}\hat{p}_{j}\right)
{\Pi_{T}}^{a}(p_0,\vec{p})
-\left(\hat{p}_{i}\hat{p}_{j}\right)
\frac{p^2_0}{\vec{p}^2}
{\Pi_{L}}^{a}(p_0,\vec{p}),
\end{eqnarray}
where
\begin{eqnarray}
{\Pi_{L}}^{a}(p_0,\vec{p})
&=&
-\left({\bf m}^2_{D}\right)^{a}
\frac{\vec{p}^2}{p^2_0}
\left[
\frac{p_0/|\vec{p}|}{2}
\int^{1}_{-1} dx
\frac{x^2}
{\left(p_0/|\vec{p}|\right)\,-\,x}\right],
\end{eqnarray}
and
\begin{eqnarray}
{\Pi_{T}}^{a}(p_0,\vec{p})
&=&
\left({\bf m}^2_{D}\right)^{a}
\left[
\frac{1}{2}\frac{p_0/|\vec{p}|}{2}
\int^{1}_{-1} dx
\frac{1-x^2}
{\left(p_0/|\vec{p}|\right)\,-\,x}
\right]. 
\end{eqnarray}

\subsection{Real and imaginary parts
of the gluon polarization tensor}

The real and imaginary parts of the gluon
polarization tensor are essential to determine
the Debye screening and Landau damping phenomena, 
respectively.
The Debye screening means that the range 
of the gauge interaction is reduced by the factor 
$e^{-\sqrt{{\left({\bf m}^2_{D}\right)}^{a}}\,r}$ 
where ${\left({\bf m}^2_{D}\right)}^{a}$ is square 
Debye mass. 
This corresponds that the gluon is acquiring 
an effective mass in order to soften the infrared 
behavior of the gluon static electric propagator 
component 
$\frac{-1}{p^2}\rightarrow 
\frac{-1}{p^2+{\left({\bf m}^2_{D}\right)}^{a}}$.
In this case, the the Debye mass acts 
as an infrared cutoff. 
It is known that the static magnetic field is not associated 
with an infrared cutoff and subsequently the transverse 
component is not screened.
%

The Landau damping stems from the imaginary part of 
the gluon polarization tensor. 
It is the mechanical energy that is transfered 
from the chromo-field to the plasma constituent particles.
Furthermore, the imaginary part is associated 
with the decay rate of the gluon where the resulting energy 
is absorbed by the plasma constituents and 
subsequently, the chromo-field is damped. 
The Landau damping in QCD is non-trivial
phenomena due to the nonlinear effects 
besides  the Landau damping 
is operative in both soft gluons and quarks.
The gluons with soft momentum and 
hard thermal loop self-energy are dynamically 
screened  by the Landau damping.  

The analytic continuation of the gluon polarization tensor
is introduced as follows 
\begin{eqnarray}
\frac{1}{x\,-\,x_{0}\,\pm\,i\,\eta}&=&
P\left(\frac{1}{x\,-\,x_{0}}\right)
\mp\,i\,\pi\,\delta\left(x\,-\,x_{0}\right).
\end{eqnarray}
where the first term on the right hand side  is the principal value
while the second term is the Dirac delta function.
Hence, the longitudinal and transverse components
of the soft gluon polarization tensor are written, 
respectively,  as follows
\begin{eqnarray}
{\Pi_{L}}^{a}
(p_0,\vec{p})
&=&
-\left({\bf m}^2_{D}\right)^{a}
\frac{\vec{p}^2}{p^2_0}
\left(
\frac{1}{2}\frac{p_0}{|\vec{p}|}
\int^{1}_{-1} dx
x^2
\left[
P\left(
\frac{1}{\frac{p_0}{|\vec{p}|}\,-\,x}
\right)
-i\,\pi\,\delta
\left(\frac{p_0}{|\vec{p}|}\,-\,x\right)
\right]\right),
\end{eqnarray}
and
\begin{eqnarray}
{\Pi_{T}}^{a}
(p_0,\vec{p})
&=&
\left({\bf m}^2_{D}\right)^{a}
\left(
\frac{1}{2}\frac{1}{2} \frac{p_0}{|\vec{p}|}
\int^{1}_{-1} dx 
\left(1-x^2\right)
\left[
\frac{1}{ \frac{p_0}{|\vec{p}|} \,-\,x}
-i\,\pi\,\delta
\left(\frac{p_0}{|\vec{p}|}\,-\,x\right)
\right]\right). 
\end{eqnarray}
Since the variable $x$ is restricted to the range 
$-1\,\le\,x\,\le\,1$,
the constraint for the principal value to avoid the singularity
is the time-like energy region $p^{2}_{0} > \vec{p}^2$
while the constraint to develop the Dirac delta function 
is the space-like energy region $\vec{p}^2\ge p^{2}_{0}$.
Therefore, the real and imaginary parts virtually 
can not be generated simultaneously 
but rather they can be dominated either by a real 
part in the time-like energy domain or by an imaginary 
part for the space-like energy domain. 
The real part is suppressed in the time-like energy.

At first, when the energy runs over the time-like 
energy domain 
(i.e. physical domain $p_0 > |\vec{p}|$),
the longitudinal and transverse gluon polarization 
tensor components turn to their real parts, 
respectively, as follows,
\begin{eqnarray}
{\Pi_{L}}^{a}(p_0,\vec{p})&=&
\Re{e}\,
{\Pi_{L}}^{a}(p_0,\vec{p}),
~\left(
\mbox{time-like energy}~ {p}^{2}_{0}\,>\,{\vec{p}}^{2}
\right),
\nonumber\\
\Re{e}\,
{\Pi_{L}}^{a}(p_0,\vec{p})
&=&
\left({\bf m}^2_{D}\right)^{a}
\left[1-Q_0\left(\frac{p_0}{|\vec{p}|}
\right)\right],
\end{eqnarray}
and
\begin{eqnarray}
{\Pi_{T}}^{a}(p_0,\vec{p})&=&
\Re{e}\, 
{\Pi_{T}}^{a}(p_0,\vec{p}),
~\left(
\mbox{time-like energy}~ {p}^{2}_{0}\,>\,{\vec{p}}^{2}
\right),
\nonumber\\
\Re{e}\, 
{\Pi_{T}}^{a}(p_0,\vec{p})
&=&
\frac{1}{2}\left({\bf m}^2_{D}\right)^{a}
\left[\frac{p^2_0}{\vec{p}^2}+\left(1-
\frac{p^2_0}{\vec{p}^2}\right)
Q_0\left(\frac{p_0}{|\vec{p}|}\right)
\right].
\end{eqnarray}
On the other hand, when the energy switches 
to run over the space-like energy
domain (i.e. $0\,\le\, p_0\,\le\, |\vec{p}|$), 
the gluon polarization tensor components turn 
to be dominated by the imaginary parts 
and subsequently they are reduced solely 
to their imaginary parts in the following way,
\begin{eqnarray}
{\Pi_{L}}^{a}\left(p_0,\vec{p}\right)
&=&
i\,\Im{m}\,
{\Pi_{L}}^{a}\left(p_0,\vec{p}\right),
~\left(
\mbox{space-like energy}~{p}^{2}_{0}\,\le\,{\vec{p}}^2
\right),
\nonumber\\
\Im{m}\, 
{\Pi_{L}}^{a}\left(p_0,\vec{p}\right)
&=&
\left({\bf m}^2_{D}\right)^{a}
\left(\frac{\pi p_0}{2|\vec{p}|}\right)
\theta\left(|\vec{p}|-p_{0}\right),
\end{eqnarray}
and
\begin{eqnarray}
{\Pi_{T}}^{a}\left(p_0,\vec{p}\right)&=&
i\,\Im{m}\, 
{\Pi_{T}}^{a}\left(p_0,\vec{p}\right),
~\left(
\mbox{space-like energy}~{p}^{2}_{0}\,\le\,{\vec{p}}^{2}
\right),
\nonumber\\
\Im{m}\,
 {\Pi_{T}}^{a}\left(p_0,\vec{p}\right)
&=&
-\frac{1}{2}\left({\bf m}^2_{D}\right)^{a}
\left(1-\frac{p^2_0}{\vec{p}^2}\right)
\left(\frac{\pi p_0}{2|\vec{p}|}\right)
\theta\left(|\vec{p}|-p_0\right),
\end{eqnarray}
for the longitudinal and transverse 
components, respectively.

\subsection{Effective Gluon propagator}
The effective gluon propagator 
with adjoint color indexes $a'\,a$
and Lorentz polarization indexes 
$\mu\,\nu$ 
for the gluon's two external legs
is represented as follows
\begin{eqnarray}
\,{{^*{\cal G}}_{\mu\nu}}^{a'a}(p_0,\vec{p})
&=&
\delta^{a'a}
\,{{^*{\cal G}}_{\mu\nu}}^{a}(p_0,\vec{p}).
\end{eqnarray}
The adjoint color indexes $a'\,a$
are contracted by the Kronecker delta $\delta^{a'\,a}$. 
The gluon propagator is gauge dependent. 
The gluon propagator in the Coulomb gauge fixing, 
where the electric and magnetic components 
can be separated, reads
\begin{eqnarray}
\,{{^*{\cal G}_{C}}_{\mu\nu}}^{a}(p_0,\vec{p})
&=&
\,{{^*{\cal G}_{C}}_{\mu\nu}}^{a}(p_0,\vec{p})
-\xi_{C}\,\frac{p_{\mu} p_{\nu}}{\vec{p}^2}\,\frac{1}{\vec{p}^2}.
\end{eqnarray}
The gluon propagator is decomposed to 
\begin{eqnarray}
\,{{^*{\cal G}_{C}}_{00}}^{a}(p_0,\vec{p})
&=&
\,{^*{\cal G}_{L}}^{a}(p_0,\vec{p}),
\nonumber\\
\,{{^*{\cal G}_{C}}_{ij}}^{a}(p_0,\vec{p})
&=&(\delta_{ij}-\hat{p}_i\hat{p}_j)\,
\,{^*{\cal G}_{T}}^{a}(p_0,\vec{p}),
\end{eqnarray}
in the strict Coulomb gauge 
where $\xi_{C}=0$.
The effective gluon propagator 
$\,{{^*{\cal G}_{C}}_{\mu\,\nu}}^{a}(p_0,\vec{p})$
is determined by finding 
the inverse of the following quantity,
\begin{eqnarray}
\,{{^*{\cal G}_{C}^{-1}}_{\mu\nu}}^{a}(p_0,\vec{p})
&=& {{{\cal G}_{C}^{-1}}_{\mu\nu}}^{a}(p_0,\vec{p})
+{\Pi_{\mu\nu}}^{a}(p_{0},\vec{p}),
\end{eqnarray}
where ${{{\cal G}_{C}^{-1}}_{\mu\nu}}^{a}(p_0,\vec{p})$
is the gluon propagator in the Coulomb gauge.
The longitudinal part becomes,
\begin{eqnarray}
\,{^*{\cal G}_{L}}^{a}(p_0,\vec{p})&=&
\frac{-1}
{\vec{p}^2+{\Pi_{L}}^{a}(p_0,\vec{p})},
\nonumber\\
&=&
-\frac{1}{\vec{p}^2}\,+\,
\left[
\frac{{\Pi_{L}}^{a}(p_0,\vec{p})}{\vec{p}^2}
\right]
\frac{1}
{\vec{p}^2+{\Pi_{L}}^{a}(p_0,\vec{p})},
\end{eqnarray}
while the transverse component is reduced to
\begin{eqnarray}
\,{^*{\cal G}_{T}}^{a}(p_0,\vec{p})&=&
\frac{-1}
{p^2_0-\vec{p}^2-{\Pi_{T}}^{a}(p_0,\vec{p})}.
\end{eqnarray}
In the time-like energy domain (i.e. $p^2_{0}\,>\,\vec{p}^2$),
it is possible to write the longitudinal 
and transverse components of the gluon propagator 
near  the mass shell residues, respectively, as follows
\begin{eqnarray}
{{\cal G}_{L}}^{a}(p_0,\vec{p})
&=&
\frac{-1}
{\left(
\vec{p}^2+{\Pi_{L}}^{a}(p_0,\vec{p})
\right)},
\nonumber\\
&\approx&
-\frac{{z_{L}}^{a}(p_0,|\vec{p}|)}
{p^2_0-\left({\overline{\omega}_{L}}^{a}\right)^2},
\nonumber\\
&~&~~~ 
(\mbox{mass-shell residues:}~
p^2_0\approx 
\left({\overline{\omega}_{L}}^{a}\right)^2,
\end{eqnarray}
where
the longitudinal residue 
${\overline{\omega}_{L}}^{a}$
is the solution of
\begin{eqnarray}
{\overline{\omega}_{L}}^{a}:~~
\vec{p}^2+
{\Pi_{L}}^{a}({\overline{\omega}_{L}}^{a},\vec{p})
&=&0,
\end{eqnarray}
and
\begin{eqnarray}
\,{^*{\cal G}_{T}}^{a}(p_0,\vec{p})
&\approx&
-\frac{{z_{T}}^{a}(\vec{p})}
{\left(
p^2_0-
\left({\overline{\omega}_{T}}^{a}\right)^2
\right)},
\nonumber\\
&~&~~~ 
(\mbox{mass-shell residues:}~
p^2_0\approx 
\left({\overline{\omega}_{T}}^{a}\right)^2,
\end{eqnarray}
where
the transverse residue 
${\overline{\omega}_{T}}^{a}$
is the solution of
\begin{eqnarray}
{\overline{\omega}_{T}}^{a}:~~
\left(
{\overline{\omega}_{T}}^{a}\right)^2-\vec{p}^2
-\Pi^a_{T}({\overline{\omega}_{T}}^{a},\vec{p})&=&0.
\end{eqnarray}
The longitudinal and transverse pre-factors
${z_{L}}^{a}(\vec{p})$
and
${z_{T}}^{a}(\vec{p})$, respectively, 
are positive functions.
In the time-like energy domain and in the 
limit $|\vec{p}|\rightarrow 0$ 
or $p_0/|\vec{p}|\rightarrow \infty$, 
the longitudinal and transverse polarization 
components are reduced,  respectively,  to
\begin{eqnarray}
\lim_{p_0/|\vec{p}|\rightarrow \mbox{large}}
\Pi_{L}(p_0,\vec{p})&=&
-\frac{1}{3}\frac{\vec{p}^2}{p^2_0}
{\left({\bf m}^2_{\cal G}\right)}^{a},
\nonumber\\
\lim_{p_0/|\vec{p}|\rightarrow \mbox{large}}
\Pi_{T}(p_0,\vec{p})&=&
\frac{1}{3}{\left({\bf m}^2_{\cal G}\right)}^{a}.
\end{eqnarray}
Hence, the effective longitudinal and transverse 
gluon propagator components  become
\begin{eqnarray}
\lim_{\vec{p}^2<<p^2_0}
\,{^{*}{\cal G}_{L}}^{a}(p_0,\vec{p})
&\approx&
-\frac{1}{\vec{p}^2-\frac{\vec{p}^2}{p^2_0}
{\left({\bf m}^2_{\cal G}\right)}^a/3},
\nonumber\\
&=&
-\frac{p^2_0}{\vec{p}^2}\frac{1}{p^2_0-
{\left({\bf m}^2_{\cal G}\right)}^a/3},
~~(\mbox{time-like energy domain}),
\end{eqnarray}
and
\begin{eqnarray}
\lim_{\vec{p}^2<<p^2_0}
\,{^{*}{\cal G}_{T}}^{a}(p_0,\vec{p})&\approx&
-\frac{1}{p^2_0-{\left({\bf m}^2_{\cal G}\right)}^a/3},
~~(\mbox{time-like energy domain}),
\end{eqnarray}
respectively.
For example, the transverse gluon propagator component 
is re-expressed in the time-like energy domain 
as follows
\begin{eqnarray}
{^*{\cal G}_{T}}^{a}(p_0,\vec{p})
&=&
\frac{ {z_{T}}^{a}(\vec{p}) }
{2\,{\overline{\omega}_{T}}^{a}}
\left[
\frac{1}{{\overline{\omega}_{T}}^{a}-p_{0}}
-
\frac{1}{-{\overline{\omega}_{T}}^{a}-p_{0}}
\right],
\nonumber\\
&=&
\frac{\pi
\,
{z_{T}}^{a}(\vec{p})
}{{\overline{\omega}_{T}}^{a}}
\,
\int \frac{d\xi}{2\pi}
\,
\left[
\frac{
\delta\left(\xi-{\overline{\omega}_{T}}^{a}\right)
}{\xi-p_{0}}
-
\frac{
\delta\left(\xi+{\overline{\omega}_{T}}^{a}\right)
}{\xi-p_{0}}
\right].
\end{eqnarray}
The transverse poles 
are restricted to the time-like dispersion relation
${\overline{\omega}_{T}}^{a}>|\vec{p}|$.
It is can be written in the form of the spectral density 
formalism in the following way
\begin{eqnarray}
{^*{\cal G}_{T}}^{a}(p_0,\vec{p})
&=&
\int \frac{d\xi}{2\pi}
\frac{{\left(\,{^*\rho_{T}}^{a}
(\xi,|\vec{p}|)\right)}}
{\xi-p_{0}}.
\end{eqnarray}
The spectral density for the time-like 
energy domain, namely, $p^{2}_{0}>\vec{p}^2$, 
is given by
\begin{eqnarray}
{^*\rho_{T}}^{a}
\left(\xi,|\vec{p}|\right)
&=&
\frac{\pi}{{\overline{\omega}_{T}}^{a}}
\,
{z_{T}}^{a}(\vec{p})
\,
\left[
\delta\left(\xi-{\overline{\omega}_{T}}^{a}\right)
-
\delta\left(\xi+{\overline{\omega}_{T}}^{a}\right)
\right].
\end{eqnarray}
Furthermore, the extrapolation from 
the time-like energy domain 
to the space-like energy domain is established by
the analytic continuation and setting
$p_{0}\rightarrow p_{0}+i\,\eta$ in following way
\begin{eqnarray}
{^*{\cal G}_{T}}^{a}
\left(p_0+i\eta,\vec{p}\right)
&=&
\int^{\infty}_{-\infty} \frac{d\xi}{2\pi}
{^*\rho_{T}}^{a}\left(\xi,|\vec{p}|\right)
\,
\frac{1}{\xi-p_{0}-i\,\eta},
\nonumber\\
&=&
\int^{\infty}_{-\infty} \frac{d\xi}{2\pi}
\,
{^*\rho_{T}}^{a}\left(\xi,|\vec{p}|\right)
\,\left[P\left(
\frac{1}{\xi-p_{0}}\right)
+i\,\pi\,\delta(\xi-p_{0})
\right],
\nonumber\\
&=&
\left[\int^{\infty}_{-\infty} \frac{d\xi}{2\pi}
\,
{^*\rho_{T}}^{a}\left(\xi,|\vec{p}|\right)
P\left(
\frac{1}{\xi-p_{0}}\right)\right]
+i\,\frac{1}{2}\,
{\left(\,{^*\rho_{T}}^{a}
(p_{0},|\vec{p}|)\right)}.
\end{eqnarray}
The first term on the right hand side 
which is enclosed by square 
brackets appears in the time-like energy domain 
and it is responsible
for the transverse Debye screening while  
the second term, which is an imaginary one, emerges 
in the space-like energy domain and it causes the Landau damping.  
The longitudinal and transverse gluon polarization tensors
are imaginary in the space-like energy domain  
$\vec{p}^2\ge {p}_{0}^{2}$. 
They are calculated as follows
\begin{eqnarray}
{^*\rho_{S}}^{a}\left(p_{0},|\vec{p}|\right)
&=&
2\,\Im{m}\,
{\left(\,{^*{\cal G}_{S}}^{a}(p_0+i\eta,\vec{p})\right)},
\end{eqnarray}
where the subscript $S=L,T$ refers to the longitudinal 
and the transverse components, respectively.
They are re-written as follows 
\begin{eqnarray}
{^*\rho_{S}}^{a}\left(p_{0},|\vec{p}|\right)
&=&
2\,\Im{m}\,
{\left(\,{^*{\cal G}_{S}}^{a}(p_0+i\eta,\vec{p})\right)}
\,\theta\left(\vec{p}^2-{p}_{0}^2\right),
\nonumber\\
&=&
{\beta_{S}}^{a}\left({p}_{0},\vec{p}\right)
\,\theta\left(\vec{p}^2-{p}_{0}^2\right),
~~~~(\mbox{with}~ S=L,T).
\end{eqnarray}
Hence, in the context of the spectral density formalism, 
the space-like 
longitudinal and transverse spectral densities 
become
\begin{eqnarray}
{^*\rho_{S}}^{a}\left(\xi,|\vec{p}|\right)
&=& {\beta_{S}}^{a}\left(\xi,\vec{p}\right)
\,\theta\left(\vec{p}^2-\xi^2\right),
\end{eqnarray}
where
\begin{eqnarray}
{\beta_{S}}^{a}\left(\xi,\vec{p}\right)
&=&
2\,\Im{m}\,
{\left(\,{^*{\cal G}_{S}}^{a}
(\xi+i\eta,\vec{p})\right)},
\end{eqnarray}
and $S=L,T$ correspond the longitudinal and transverse 
components, respectively.
Moreover, both longitudinal and transverse gluon propagators 
split to the Debye screening parts
and the Landau damping parts, respectively, 
as follows
\begin{eqnarray}
{^*{\cal G}_{L}}^{a}
\left(p_0+i\eta,\vec{p}\right)
&=&
-\frac{1}{\vec{p}^2}
+\left[\int^{\infty}_{-\infty} \frac{d\xi}{2\pi}
\,{\left(\,{^*\rho_{L}}^{a}(\xi,|\vec{p}|)\right)}
P\left(
\frac{1}{\xi-{p}_{0}}\right)  
\,\theta\left(\xi^2-\vec{p}^2\right)
\right]
\nonumber\\
&~&
+\int^{\infty}_{-\infty} \frac{d\xi}{2\pi}
\frac{\left[{\beta_{L}}^{a}\left(\xi,\vec{p}\right)
\,\theta\left(\vec{p}^2-\xi^2\right)\right]}
{ \xi-{p}_{0} },
\label{spectral1-prop-L}
\end{eqnarray}
and
\begin{eqnarray}
{^*{\cal G}_{T}}^{a}
\left(p_0+i\eta,\vec{p}\right)
&=&
\left[\int^{\infty}_{-\infty} \frac{d\xi}{2\pi}
\,{\left(\,{^*\rho_{T}}^{a}(\xi,|\vec{p}|)\right)}
P\left(
\frac{1}{\xi-{p}_{0}}\right)  
\,\theta\left(\xi^2-\vec{p}^2\right)
\right]
\nonumber\\
&~&
+\int^{\infty}_{-\infty} \frac{d\xi}{2\pi}
\frac{\left[{\beta_{T}}^{a}\left(\xi,\vec{p}\right)
\,\theta\left(\vec{p}^2-\xi^2\right)\right]}
{ \xi-{p}_{0} }.
\label{spectral1-prop-T}
\end{eqnarray}
The first integral on the right hand side 
of Eqns.~(\ref{spectral1-prop-L}) and (\ref{spectral1-prop-T})
is the Debye screening in the time-like energy domain
while the second integral is the Landau damping 
in the space-like energy domain.
It is possible to write the longitudinal and
transverse gluon propagator components 
in the terms of the spectral density formalism
in the following way
\begin{eqnarray}
{\,^{*}{\cal G}_{L}}^{a}\left(p_{0},\vec{p}\right)
&=&
-\frac{1}{|\vec{p}|^2}
+\int^{\infty}_{-\infty} 
\frac{d\xi}{2\pi}
\frac{
{\,^{*}{\rho_{L}}^a}\left(\xi,\vec{p}\right)
}{\xi-p_{0}},
\end{eqnarray}
and
\begin{eqnarray}
{\,^{*}{\cal G}_{T}}^{a}\left(p_{0},\vec{p}\right)
&=&
\int^{\infty}_{-\infty} 
\frac{d\xi}{2\pi}
\frac{
{\,^{*}{\rho_{T}}^a}\left(\xi,\vec{p}\right)
}{\xi-p_{0}},
\end{eqnarray}
respectively.
Furthermore, the longitudinal and transverse gluon 
propagator components are transformed to
\begin{eqnarray}
{\,^{*}{\cal G}_{L}}^{a}\left(\tau,\vec{p}\right)
&=&
-\frac{\sum_{l}\delta(\tau-l\beta)}{|\vec{p}|^2}
\nonumber\\
&~&
+\int^{\infty}_{-\infty} 
\frac{d\xi}{2\pi}
{\,^{*}{\rho_{L}}^a}
\left(\xi-i\,\frac{\phi^{a}}{\beta},\vec{p}\right)
\,e^{-\left(\xi-i\,\frac{\phi^{a}}{\beta}\right)\,\tau}\,
\left[
\theta(\tau)+N_{G}\left(\xi-i\,\frac{\phi^{a}}{\beta}\right)
\right],
\end{eqnarray}
and
\begin{eqnarray}
{\,^{*}{\cal G}_{T}}^{a}\left(\tau,\vec{p}\right)
&=&
~\int^{\infty}_{-\infty} 
\frac{d\xi}{2\pi}
{\,^{*}{\rho_{T}}^a}
\left(\xi-i\,\frac{\phi^{a}}{\beta},\vec{p}\right)
\,e^{-\left(\xi-i\,\frac{\phi^{a}}{\beta}\right)\,\tau}\,
\left[
\theta(\tau)+N_{G}\left(\xi-i\,\frac{\phi^{a}}{\beta}\right)
\right],
\end{eqnarray}
respectively, 
in the mixed-time representation 
of the imaginary-time formalism.

\section{\label{section6} The effective vertexes}

In order to study the effective hard thermal quark self-energy 
and the effective hard thermal gluon polarization tensor 
besides the other quantities in the ultra-relativistic 
heavy ion collisions, the effective $n$-quarks 
and $m$-gluons vertexes become essential in addition 
to the effective quark and gluon propagators.
The effective $n$-quarks and $m$-gluons vertexes 
can be calculated to any order 
of the coupling constant $g$.  
For only the sake of  simplicity, 
we shall limit the approximation up to the order $g^{n+m}$ 
for the vertexes with 
$n+m$ external gluons and quarks.
The calculation of the effective vertexes for $n$-quarks 
and $m$-gluons is demonstrated by calculating 
the effective quark-quark-gluon vertex 
and the effective 2-quarks and 2-gluons vertex. 
The Feynman diagrams those contribute 
to the effective quark-quark-gluon vertex
up to the order $g^{3}$ are displayed 
in Fig.~(\ref{effective3pointvertex})
while those contribute 
to the effective 2-quarks and 2-gluons vertex
up to the order $g^{4}$ are displayed 
in Fig.~(\ref{effective4pointvertex}).

\subsection{quark-quark-gluon vertex}
\begin{figure} 
\includegraphics{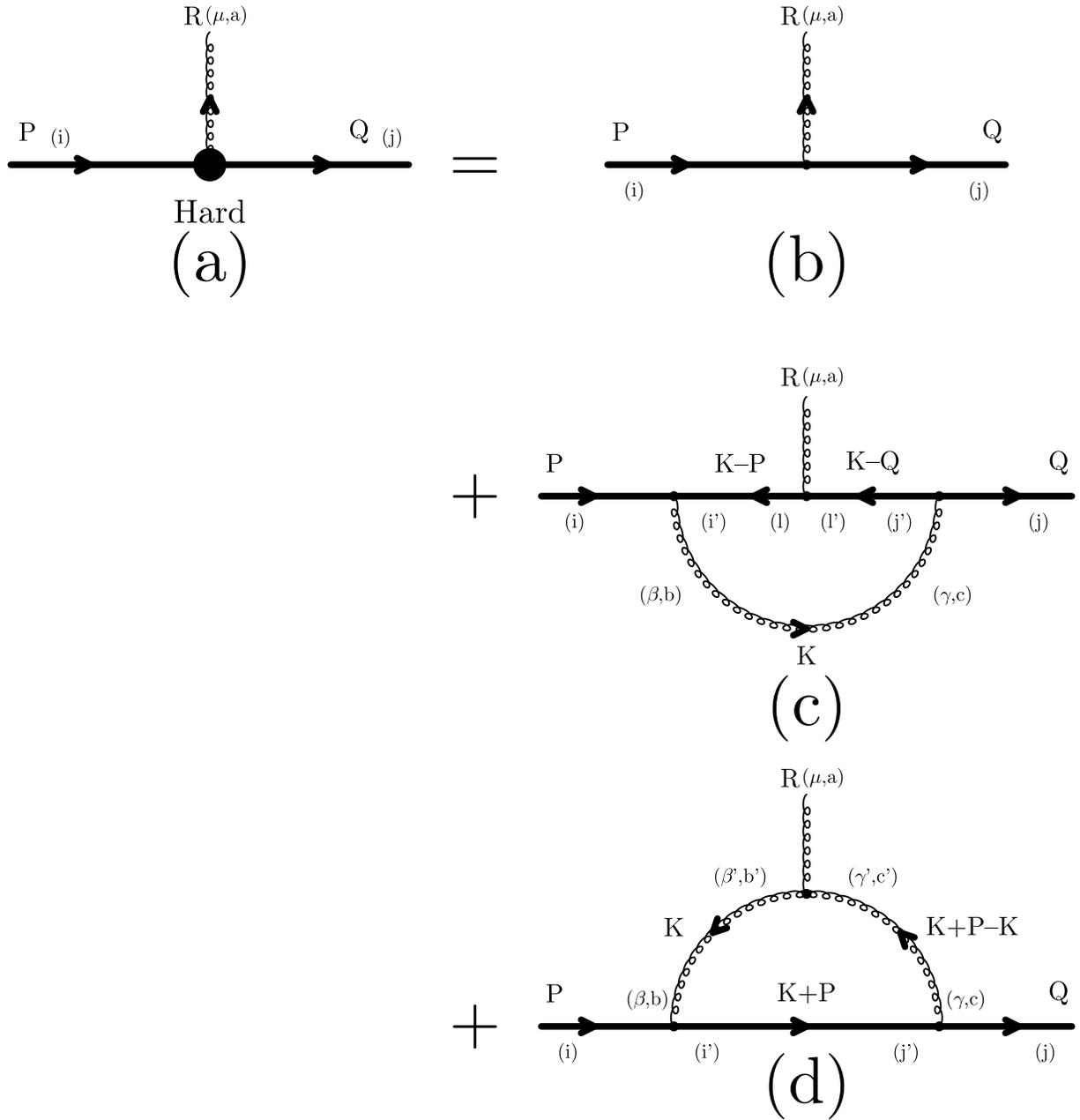}
\caption{\label{effective3pointvertex}
The effective hard quark-quark-gluon vertex.} 
\end{figure}

The effective quark-quark-gluon vertex can be calculated 
to any order of the interaction coupling $g$. 
Its effective interaction is depicted 
in Fig.~(\ref{effective3pointvertex} a).
In order to calculated the effective quark-quark-gluon vertex 
up to the order of $g^{3}$, the effective quark-quark-gluon vertex
is given by the sum of the interactions that are given 
in Figs.~(\ref{effective3pointvertex} b), 
(\ref{effective3pointvertex} c) and (\ref{effective3pointvertex} d).
The effective vertex is found by calculating the bar vertex 
and the corrections up to the order $g^{3}$.
The bar quark-quark-gluon vertex is the interaction 
up to order $g$ and its Feynman diagram 
is depicted in Fig.~(\ref{effective3pointvertex} b). 
It reads
\begin{eqnarray}
{\Gamma^{3(0)}}^{a\,\mu}&=&
- g \gamma^{\mu} {\bf t}^{a}_{ij}.
\end{eqnarray}
The next lowest order corrections to the bar vertex 
is of order $g^3$.  The first correction 
is displayed in Fig.~(\ref{effective3pointvertex} c). 
The kernel of the first correction is constructed 
as follows
\begin{eqnarray}
{\Gamma^{3(A)}}^{a\,\mu}_{ij}(P,Q;R)&=&
\int \frac{d^4 k}{(2\pi)^4}
\left[- g \gamma^{\beta} {\bf t}^{b}_{i'i}\right]\,
{\cal G}^{\beta\gamma}_{bc}(k)\,
\left[- g \gamma^{\gamma} {\bf t}^{c}_{jj'}\right]\,
i\,
{\cal S}_{j'l}(k-Q)\,
\nonumber\\
&~&
~~~ ~~~\times
\left[- g \gamma^{\mu} {\bf t}^{a}_{l'l}\right]\,
i\,
{\cal S}_{l'i'}(k-P),
\nonumber\\
&=&
-{g^{3}}\,{ {\cal T}^{3(A)} }^{a}_{ij\,[i'j'\,b]}\, 
{ {\cal V}^{3(A)} }^{\mu}_{[i'j'\,b]}(P,Q;R), 
\end{eqnarray}
where the momentum $R$ is assigned 
for the external gluon leg while the momenta 
$P$ and $Q$ are assigned for the two external quark legs.
The vertex decomposition with respect 
to the fundamental color indexes, namely, $i\,j$ 
for two quarks and the adjoint color indexes, namely, 
$a$ for one gluon is given by
\begin{eqnarray}
{ {\cal T}^{3(A)} }^{a}_{ij\,[i'j'\,b]}&=&
{\bf t}^{b}_{i'i} {\bf t}^{b}_{jj'} {\bf t}^{a}_{i'j'},
\end{eqnarray}
and
\begin{eqnarray}
{ {\cal V}^{3(B)} }^{\mu}_{[i'j'\,b]}(P,Q;R)
&=&
\int \frac{d^4 k}{(2\pi)^4}
\gamma^{\beta} 
{\cal G}^{\beta\gamma}_{bc}(k)\delta_{bc}
\gamma^{\gamma}\, 
i\,
{\cal S}_{j'}(k-Q)
\gamma^{\mu}\,
i\,
{\cal S}_{i'}(k-P).
\end{eqnarray}
The second term correction is illustrated 
by the Feynman diagram that is depicted 
in Fig.~(\ref{effective3pointvertex} d). 
This correction is of order $g^{3}$.
Its interaction vertex is constructed as follows
\begin{eqnarray}
{ \Gamma^{3(B)} }^{a\,\mu}_{ij}(P,Q;R)&=&
\int \frac{d^4 k}{(2\pi)^4}
\left[- g \gamma^{\beta} {\bf t}^{b}_{i'i}\right]\,
i\,
{\cal S}_{i'j'}(k+P)
\left[- g \gamma^{\gamma} {\bf t}^{c}_{j'j}\right]
{\cal G}^{\gamma'\gamma}_{cc'}(k+P-Q)
\nonumber\\
&~&
~~~\times\left[
-i g f_{ab'c'}
\Gamma^{\mu\beta'\gamma'}
\left(-R,-k,k+P-Q\right)\right]
{\cal G}^{\beta'\beta}_{b'b}(k),
\nonumber\\
&=&
g^3
{ {\cal T}^{3(B)} }^{a}_{ij\,[i' j' \, b c]}
{ {\cal V}^{3(B)} }^{\mu}_{[i' j' \, b c]}(P,Q;R), 
\end{eqnarray}
where the fundamental and adjoint color decomposition 
for the two quarks with fundamental color indexes $i\,j$ 
and one gluon with an adjoint color index $a$, 
respectively, is given by,
\begin{eqnarray}
{ {\cal T}^{3(B)} }^{a}_{i j \, [i' j' \, b c]}&=&
{\bf t}^{b}_{ii'} {\bf t}^{c}_{j'j} {\bf T}^{a}_{bc},
\end{eqnarray}
and
\begin{eqnarray}
{ {\cal V}^{3(B)} }^{\mu}_{[i' j' \, b c]}(P,Q;R)&=&
\int \frac{d^4 k}{(2\pi)^4}
\gamma^{\beta}\,
i\,
{\cal S}_{i'}(k+P)\,\delta_{i'j'}\,
\gamma^{\gamma}
{\cal G}^{\gamma\gamma'}_{c}(k+P-Q)
\nonumber\\
&~&
\times
\Gamma^{\gamma'\mu\beta'}\left(k+P-Q,-R,-k\right)
{\cal G}^{\beta'\beta}_{b}(k).
\end{eqnarray}
The effective quark-quark-gluon vertex with the corrections 
up to the order $g^{3}$ is obtained by adding 
the first and second term corrections 
which are of order $g^{3}$
to the bar vertex which is of order $g$. 
Hence, the effective quark-quark-gluon vertex becomes
\begin{eqnarray}
^{*}{\Gamma^{3}}^{a\mu}_{ij}(P,Q;R)&=&
-g \gamma^{\mu}{\bf t}^{a}_{ij}
\,+\,g^{3} {{\bf V}^{3}}^{a\mu}_{ij}(P,Q;R),
\end{eqnarray}
where the correction up to the order of $g^{3}$ 
is given by
\begin{eqnarray}
{{\bf V}^{3}}^{a\mu}_{ij}(P,Q;R)&=&
-{ {\cal T}^{3(A)} }^{a}_{ij\,[i'j'\,b]}\, 
{ {\cal V}^{3(A)} }^{\mu}_{[i'j'\,b]}(P,Q;R)
\nonumber\\
&~&
+
{ {\cal T}^{3(B)} }^{a}_{ij\,[i' j' \, b c]}
{ {\cal V}^{3(B)} }^{\mu}_{[i' j' \, b c]}(P,Q;R). 
\end{eqnarray}

\subsection{2-quarks and 2-gluons vertex}
\begin{figure} 
\includegraphics{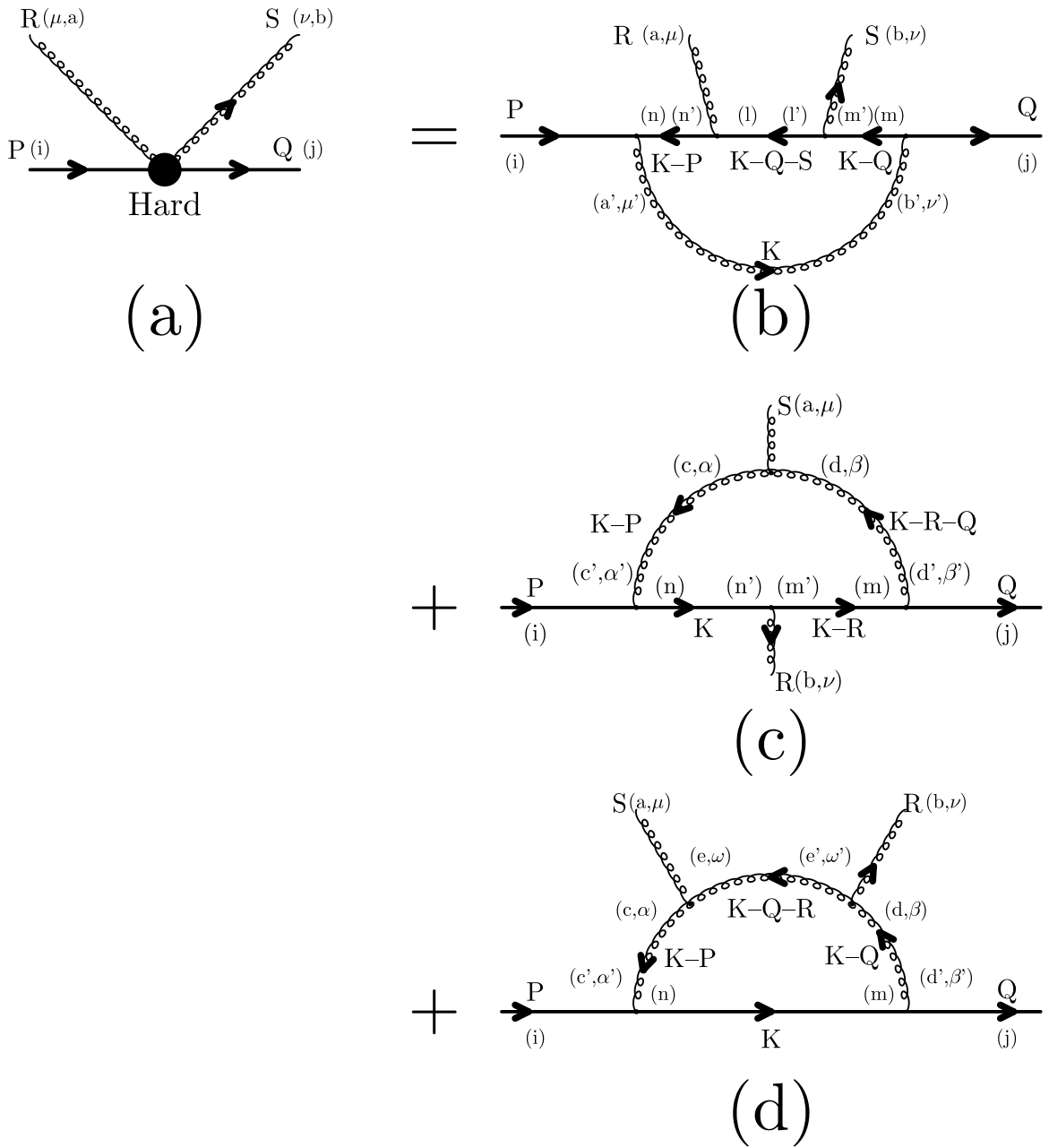}
\caption{\label{effective4pointvertex}
The effective hard 2-quarks and 2-gluons vertex.} 
\end{figure}

When the correction is considered up to the order $g^{4}$, 
the the effective 2-quarks and 2-gluons vertex becomes essential 
in the self-energy corrections (i.e. the radiative corrections). 
When the radiative correction is considered 
up to the order less than 
$g^{4}$, the effective 2-quarks and 2-gluons vertex 
becomes redundant.
The Feynman diagram for the effective 
2-quarks and 2-gluons vertex is displayed 
in Fig.~(\ref{effective4pointvertex} a).  
The effective 2-quarks and 2-gluons vertex is calculated 
from three Feynman diagrams that are displayed in 
Figs.~(\ref{effective4pointvertex} b),  
(\ref{effective4pointvertex} c) 
and  
(\ref{effective4pointvertex} d).
The first correction to the 2-quarks and 2-gluons vertex
is displayed in Fig.~(\ref{effective4pointvertex} b). 
It consists of four quark-quark-gluon vertexes, 
three internal quark lines and one internal gluon line. 
Every internal quark line is identified 
by the fundamental color indexes 
while the gluon line has adjoint color indexes.
The contribution of the first correction is of order $g^{4}$. 
The interaction kernel is constructed as follows
\begin{eqnarray}
{\Gamma^{4(A)}}^{ab\,\mu\nu}_{ij}(P,Q;S,R)
&=&
\int \frac{d^4 k}{(2\pi)^4}
\left[- g \gamma^{\mu'} {\bf t}^{a'}_{n i}\right]
{\cal G}^{\mu'\nu'}_{a'b'}(k)
\left[- g \gamma^{\nu'} {\bf t}^{b'}_{jm}\right]\,
i\,
{\cal S}_{m m'}(k-Q)
\nonumber\\
&~&\times
\left[- g \gamma^{\nu} {\bf t}^{b}_{l'm'}\right]\,
i\,
{\cal S}_{l' l}(k-Q-S)
\left[- g \gamma^{\mu} {\bf t}^{a}_{n'l}\right]\,
i\,{\cal S}_{n' n}(k-P),
\nonumber\\
&=&
g^4\,
{ {\cal T}^{4(A)} }^{ab}_{ij\,[nml\,a']}\,
{ {\cal V}^{4(A)} }^{\mu\nu}_{[nml\,a']}(P,Q;S,R).
\label{2q-2g-correc1}
\end{eqnarray}
The fundamental and adjoint color indexes, namely, 
$i\,j$ and $a\,b$, respectively,
are decomposed by the following projector 
\begin{eqnarray}
{ {\cal T}^{4(A)} }^{ab}_{ij\,[nml\,a']}&=&
{\bf t}^{a'}_{ni}
{\bf t}^{b'}_{jm}
{\bf t}^{b}_{lm}
{\bf t}^{a}_{nl}\,\delta_{a'b'}.
\end{eqnarray}
This projector consists of four fundamental color generators 
due to the four quark-quark-gluon vertexes.
The color projector is defined by
\begin{eqnarray}
{ {\cal V}^{4(A)} }^{\mu\nu}_{[nml\,a']}(P,Q;S,R)
&=&
\int \frac{d^4 k}{(2\pi)^4}
\gamma^{\mu'}
{\cal G}^{\mu'\nu'}_{a'b'}(k)\delta_{a'b'}
\gamma^{\nu'}\,
i\,
{\cal S}_{m}(k-Q)
\gamma^{\nu}\,
i\,
{\cal S}_{l}(k-Q-S)
\nonumber\\
&~&
~~~\times
\gamma^{\mu}\,
i\,
{\cal S}_{n}(k-P).
\end{eqnarray}
The Feynman diagram of the second correction 
is depicted in Fig.~(\ref{effective4pointvertex} c). 
It consists three quark-quark-gluon vertexes 
and one 3-gluons vertex besides two internal gluon lines 
and two internal quark lines.
The interaction kernel for the second term correction for 
the 2-quarks and 2-gluons vertex is constructed as follows
\begin{eqnarray}
{ \Gamma^{4(B)} }^{ab\,\mu\nu}_{ij}(P,Q;S,R)&=&
\int \frac{d^4 k}{(2\pi)^4}
\left[- g \gamma^{\alpha'} {\bf t}^{c'}_{in}\right]\,
i\,
{\cal S}_{n n'}(k)
\left[- g \gamma^{\nu} {\bf t}^{b}_{m' n'}\right]\,
i\,
{\cal S}_{m m'}(k-R)
\nonumber\\
&~&
~~~\times
\left[- g \gamma^{\beta'} {\bf t}^{d'}_{mj}\right]
{\cal G}^{\beta'\beta}_{d'd}(k-R-Q)
\nonumber\\
&~&
~~~\times
\left[
g \left({\bf T}^a\right)_{cd} 
\Gamma^{\mu\alpha\beta}\left(-S,-k+P,k-R-Q\right)
\right]
\nonumber\\
&~&
~~~\times
{\cal G}^{\alpha\alpha'}_{cc'}(k-P).
\end{eqnarray}
From the preceding equation, 
the second term correction is written as follows
\begin{eqnarray}
{ \Gamma^{4(B)} }^{ab\,\mu\nu}_{ij}(P,Q;S,R)&=&
-g^4\,
{ {\cal T}^{4(B)} }^{ab}_{ij\,[nm\,cd]}
\,
{ {\cal V}^{4(B)} }^{\mu\nu}_{[nm\,cd]}(P,Q;S,R),
\label{2q-2g-correc2}
\end{eqnarray}
where the decomposition of the 
fundamental and adjoint color indexes 
for quarks and gluons, respectively, 
is given by
\begin{eqnarray}
{ {\cal T}^{4(B)} }^{ab}_{ij\,[n m \, c d]}
&=&
{\bf t}^{c}_{in}
{\bf t}^{b}_{mn}
{\bf t}^{d}_{mj}
\left({\bf T}^{a}\right)_{cd},
\end{eqnarray}
and
\begin{eqnarray}
{ {\cal V}^{4(B)} }^{\mu\nu}_{[n m \, c d]}(P,Q;S,R)
&=&
\int \frac{d^4 k}{(2\pi)^4}
\gamma^{\alpha'}\,
i\, 
{\cal S}_{n}(k)
\gamma^{\nu}\,
i\, 
{\cal S}_{m}(k-R)
\gamma^{\beta'} 
{\cal G}^{\beta'\beta}_{d'd}(k-R-Q)\delta_{d'd}
\nonumber\\
&~&
~~~\times
\Gamma^{\mu\alpha\beta}\left(-S,-k+P,k-R-Q\right)
{\cal G}^{\alpha\alpha'}_{cc'}(k-P)\delta_{c'c}.
\end{eqnarray}
%
Finally, the third correction for 
the 2-quarks and 2-gluons vertex 
is given by the Feynman diagram that is depicted 
in Fig.~(\ref{effective4pointvertex} d).
It consists two quark-quark-gluon vertexes 
and two 3-gluons vertexes besides 
three internal gluon segments (i.e. lines) 
and one internal quark segment (i.e. line). 
The interaction kernel for the third correction 
is furnished by 
\begin{eqnarray}
{ \Gamma^{4(C)} }^{a b \, \mu \nu}_{ij}(P,Q;S,R)&=&
\int \frac{d^4 k}{(2\pi)^4}
\left[- g \gamma^{\alpha'} {\bf t}^{c'}_{in}\right]\,
i\,
{\cal S}_{n m}(k)
\left[- g \gamma^{\beta'} {\bf t}^{d'}_{m j}\right]
{\cal G}^{\beta'\beta}_{d'd}(k-Q)
\nonumber
\\
&~&
~~~\times\left[
g \left({\bf T}^{b}\right)_{e'd} 
\Gamma^{\nu\omega'\beta}\left(-R,-(k-Q-R),k-Q\right)
\right]
{\cal G}^{\omega'\omega}_{e'e}(k-Q-R)
\nonumber
\\
&~&
~~~\times\left[
g \left({\bf T}^{a}\right)_{ce} 
\Gamma^{\mu\alpha\omega}\left(-S,-(k-P),k-Q-R\right)
\right]
{\cal G}^{\alpha\alpha'}_{c c'}(k-P).
\label{vertex-2q-2g-a1}
\end{eqnarray}
There are two fundamental color generators 
and two adjoint color generators 
to represent the quark and gluon couplings.
Eq.~(\ref{vertex-2q-2g-a1}) is reduced to
\begin{eqnarray}
{ \Gamma^{4(C)} }^{a b \, \mu \nu}_{ij}(P,Q;S,R)&=&
g^4\,
{ {\cal T}^{4(C)} }^{ab}_{i j \, [n \, c d e]}
\,
{ {\cal V}^{4(C)} }^{ \mu \nu }_{[ n \, c d e ]}(P,Q;S,R),
\label{2q-2g-correc3}
\end{eqnarray}
where 
\begin{eqnarray}
{ {\cal T}^{4(C)} }^{ab}_{i j \, [n \, c d e]}
&=&
{\bf t}^{c}_{in}
{\bf t}^{d}_{nj}
\left({\bf T}^{b}\right)_{ed}
\left({\bf T}^{a}\right)_{ce},
\end{eqnarray}
and
\begin{eqnarray}
{ {\cal V}^{4(C)} }^{ \mu \nu }_{[ n \, c d e ]}(P,Q;S,R)
&=&
\int \frac{d^4 k}{(2\pi)^4}
\gamma^{\alpha'}\,
i\, 
{\cal S}_{n m}(k)\delta_{nm}
\gamma^{\beta'}
{\cal G}^{\beta'\beta}_{d'd}(k-Q)\delta_{d'd}
\nonumber\\
&~&
~~~\times
\Gamma^{\nu\omega'\beta}\left(-R,-(K-Q-R),K-Q\right)
{\cal G}^{\omega'\omega}_{e'e}(K-Q-R)\delta_{e'e}
\nonumber\\
&~&
~~~\times
\Gamma^{\mu\alpha\omega}\left(-S,-(K-P),K-Q-R\right)
{\cal G}^{\alpha\alpha'}_{c c'}(K-P)\delta_{c'c}.
\end{eqnarray}
%
Therefore, the effective 2-quarks and 2-gluons vertex up 
to the order of $g^{4}$
is given by adding the first, second and third correction terms 
those are given by 
Eqns.~(\ref{2q-2g-correc1}), (\ref{2q-2g-correc2}), 
(\ref{2q-2g-correc3}), respectively.
The result for the effective 2-quarks and 2-gluons vertex 
becomes
\begin{eqnarray}
{\,^{*}\Gamma^{4}}^{a b \, \mu \nu}_{ij}(P,Q;S,R)&=&
{\Gamma^{4(A)}}^{a b \, \mu \nu}_{ij}(P,Q;S,R)
+
{\Gamma^{4(B)}}^{a b \, \mu \nu}_{ij}(P,Q;S,R)
\nonumber\\
&~&
+{\Gamma^{4(C)}}^{a b \, \mu \nu}_{ij}(P,Q;S,R),
\nonumber\\
&=& 
g^4\,
\left[\,
{ {\cal T}^{4(A)} }^{ab}_{ij\,[nml\,a']}
\,
{ {\cal V}^{4(A)} }^{\mu\nu}_{[nml\,a']}(P,Q;S,R) 
\right.
\nonumber\\
&~&~~~
-
{ {\cal T}^{4(B)} }^{ab}_{ij\,[nm\,cd]}
\,
{ {\cal V}^{4(B)} }^{\mu\nu}_{[nm\,cd]}(P,Q;S,R)
\nonumber\\
&~&~~~\left.
+
{ {\cal T}^{4(C)} }^{ab}_{i j \, [n \, c d e]}
\,
{ {\cal V}^{4(C)} }^{ \mu \nu }_{[ n \, c d e ]}(P,Q;S,R)
\,
\right].
\end{eqnarray}

\section{\label{section7} Effective quark self-energy 
up to the order $O(g^2)$}

\begin{figure} 
\includegraphics{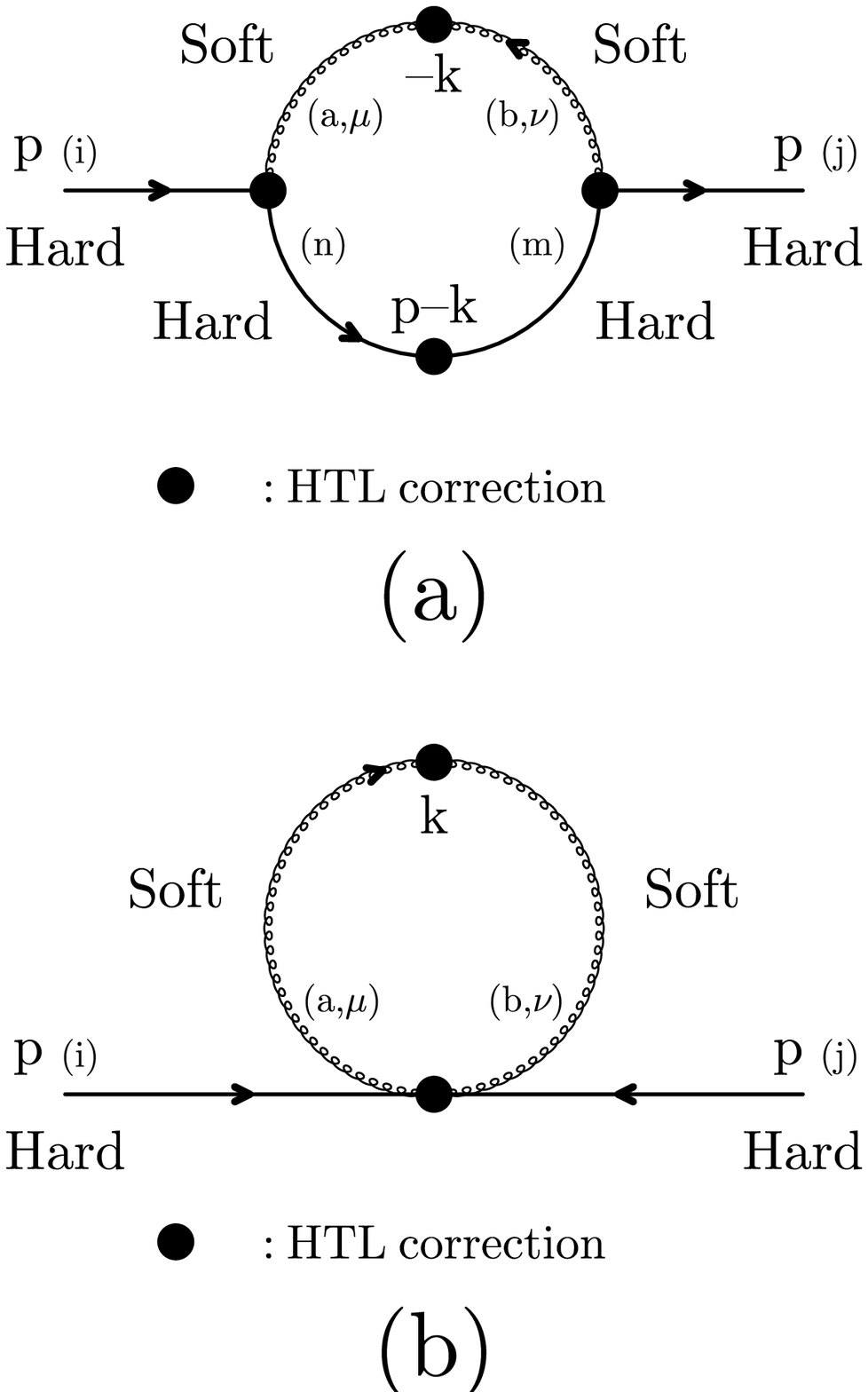}
\caption{\label{effectivequark}
Effective self-energy for the hard thermal quark.} 
\end{figure}

In order to calculate the colored quark decay rate in
the ultra-relativistic heavy ion collisions, the effective 
quark self-energy must be calculated with the HTL approximation.
In this case the momentum of the external quark line 
is hard (i.e. of order $\sim\,T$) while the momentum 
of the internal loop is soft (i.e. of order $\sim\,g\,T$). 
The vertexes that appear in the correction loop are hard ones. 
The effective quark self-energy is calculated 
up to the order of $g^{2}$. 
The relevant Feynman diagrams  are 
depicted in Fig.~(\ref{effectivequark}). 

In Fig.~(\ref{effectivequark} a), 
the self-energy interaction for 
the hard quark line is represented by 
an internal quark-gluon loop where 
the internal loop is composed 
of a quark segment 
with an effective hard thermal quark self-energy correction
and a gluon segment with 
an effective hard thermal 
gluon self-energy correction.
Since the quark is considered hard, 
the bar quark propagator 
is sufficient in the calculation.
The quark segment is shown by the internal lower semi-circle 
and it is identified by fundamental color indexes 
while the gluon segment is identified by adjoint
color indexes and it appears as an internal 
upper semi-circle that complements 
the quark lower semi-circle in order 
to form the internal quark-gluon loop.
The internal hard thermal quark-gluon loop 
consists of two hard quark-quark-gluon vertexes 
and one internal hard quark line (i.e. the HTL is not necessary)
and one internal soft gluon line segment 
with the HTL correction.
Furthermore, the external quark momentum 
$(p_{0},\vec{p})$ is taken to be hard one 
with respect to the internal momentum 
$(k_{0},\vec{k})$ with the assumption that
$|\vec{k}|\,/\,|\vec{p}| \le g \ll 1$. 
Since the external momentum $|\vec{p}|$ is assumed to be 
of the order $|\vec{p}| \,\sim\, T$, 
the internal momentum becomes of order 
$|\vec{k}| \,\sim\, g\,p \,\sim\, g\,T$.
The lowest order correction to be considered 
for the effective quark self-energy
is of order of $g^2$ in the present work.
The other contribution to the hard quark line's 
self-energy is given by the hard quark line 
with a gluon tadpole which is depicted 
in Fig.~(\ref{effectivequark} b). 
The tadpole consists one 
effective hard thermal 2-gluons and 2-quarks 
vertex and an internal gluon loop 
with an effective HTL gluon correction 
and two external quark legs. 
The lowest order correction of the tadpole 
for the hard quark line is of order $O(g^{4})$.
When the effective quark self-energy is calculated 
up to the order of $g^{2}$, 
the quark-gluon loop interaction which is displayed 
in Fig.~(\ref{effectivequark} a) is sufficient 
while the tadpole interaction which 
is displayed in Fig.~(\ref{effectivequark} b) 
can be neglected. 
The effective quark self-energy is calculated 
by integrating the internal soft momentum, 
namely, $(k_{0},\vec{k})$ as follows, 
\begin{eqnarray}
{{\,^{*}\Sigma}_{Q}}_{ij}(p_0,\vec{p})&=&
\int \frac{d^{4}k}{(2\pi)^{3}}\,
{{\,^{*}\Sigma}_{Q}}_{ij}(p_0,\vec{p},k_{0},\vec{k}).
\label{effec-quark-eng1}
\end{eqnarray}
The internal momentum $k$ appears in the interaction 
kernels that are displayed in 
Fig.~(\ref{effectivequark}).
The effective interaction kernel with 
the internal thermal quark-gluon loop 
which is displayed in Fig.~(\ref{effectivequark} b) 
is written as follows:
\begin{eqnarray}
{{\,^{*}\Sigma}_{Q}}_{ij}
\left(p_0,\vec{p},k_{0},\vec{k}\right)
&=&
{ {\,^{*}\Gamma}^{3} }^{a\,\mu}_{ii'}\,
i\,
{ {\,^*{\cal S}}_{Q} }_{i'j'}(p-k)
\,
{ {\,^{*}\Gamma}^{3} }^{b\,\nu}_{j'j}
\,
{\,^{*}{\cal G}}^{\mu \nu}_{a b}(k),
\nonumber\\
&\approx&
{ { \Gamma}^{3} }^{a\,\mu}_{ii'}\,
i\,
{ { {\cal S}}_{Q} }_{i'j'}(p-k)
\,
{ {\Gamma}^{3} }^{b\,\nu}_{j'j}
\,
{\,^{*}{\cal G}}^{\mu \nu}_{a b}(k),
\label{effec-quark-eng1a}
\end{eqnarray}
where
\begin{eqnarray}
{ {\,^{*}\Gamma}^{3} }^{a\,\mu}_{ii'}&=&
{ {\,^{*}\Gamma}^{3} }^{a\,\mu}_{ii'}(p,-p+k,-k),
\nonumber\\
&\approx& -g\,\gamma^{\mu}\, {\bf t}^{a}_{ii'}
+ O(g^{3}),
\end{eqnarray}
and
\begin{eqnarray}
{ {\,^{*}\Gamma}^{3} }^{b\,\nu}_{j'j}
&=&
{ {\,^{*}\Gamma}^{3} }^{b\,\nu}_{j'j}(p-k,-p,k)
\nonumber\\
&\approx&
-g\,\gamma^{\nu}\, {\bf t}^{b}_{j'j}
+ O(g^{3}).
\end{eqnarray}
Since the external quark line is assumed to be hard,
the internal quark loop segment is assumed to carry 
the hard part of the external momentum. 
The effective internal quark propagator 
is approximated to the bar one.  
Eq.~(\ref{effec-quark-eng1a}) 
is approximated up to the order of $g^{2}$ 
to 
\begin{eqnarray}
{{\,^{*}\Sigma}_{Q}}_{ij}(p_0,\vec{p},k_{0},\vec{k})
&\approx& 
({\bf t}^a)_{in} ({\bf t}^a)_{nj}\,
{{\,^{*}\Sigma}_{Q}}^{a}_{n}(p_0,\vec{p},k_{0},\vec{k}),
\label{effec-quark-kernel1}
\end{eqnarray}
where
\begin{eqnarray}
{{\,^{*}\Sigma}_{Q}}^{a}_{n}(p_0,\vec{p},k_{0},\vec{k})
&\approx& 
g^2\,
\left[
\gamma_0\, 
i\,
{\,^*{\cal S}}_{n}(p-k)\, \gamma_0\, 
{\,^*{\cal G}}^{a}_{L}(k)
\right]
\nonumber\\
&~&~~~
+
g^2\, 
\sum^3_{l,k=1}\left[
\gamma_{l}\, 
i\,
{\,^*{\cal S}}_{n}(p-k)\, \gamma_{k}\,
\left(\delta_{lk}-\hat{k}_{l}\hat{k}_{k}\right)\,
{\,^*{\cal G}}^{a}_{T}(k)
\right],
\nonumber\\
&\approx& 
g^2\,
\left[
\gamma_0\, 
i\,
{{\cal S}}_{n}(p-k)\, \gamma_0\, 
{\,^*{\cal G}}^{a}_{L}(k)
\right]
\nonumber\\
&~&~~~
+
g^2\, 
\sum^3_{l,k=1}\left[
\gamma_{l}\, 
i\,
{{\cal S}}_{n}(p-k)\, \gamma_{k}\,
\left(\delta_{lk}-\hat{k}_{l}\hat{k}_{k}\right)\,
{\,^*{\cal G}}^{a}_{T}(k)
\right].
\end{eqnarray}
Furthermore, by considering 
Eq.~(\ref{effec-quark-kernel1}),
Eq.~(\ref{effec-quark-eng1a}) 
is reduced to
\begin{eqnarray}
{{\,^{*}\Sigma}_{Q}}_{ij}(p_0,\vec{p})&=&
({\bf t}^a)_{in} ({\bf t}^a)_{nj}\,
{{\,^{*}\Sigma}_{Q}}^{a}_{n}(p_0,\vec{p}),
\label{effec-quark-eng-red2}
\end{eqnarray}
where
\begin{eqnarray}
{{\,^{*}\Sigma}_{Q}}^{a}_{n}(p_0,\vec{p})
&=&
\int \frac{d^{4}k}{(2\pi)^{3}}
{{\,^{*}\Sigma}_{Q}}^{a}_{n}(p_0,\vec{p},k_{0},\vec{k}).
\label{effec-quark-eng-red2b}
\end{eqnarray}
The Foldy-Wouthuysen energy transformation 
decomposes the interaction kernel 
${{\,^*{\Sigma}}_{Q}}^{a}_{n}(p_0,\vec{p},k_0,\vec{k})$ 
to positive and negative energy 
components as follows
\begin{eqnarray}
{{\,^*{\Sigma}}_{Q}}^{a}_{n}(p_0,\vec{p},k_0,\vec{k})
&\approx& 
g^2\,
\left[
\sum_{r=\pm}
\gamma_0\, h^{(r)}_{Q}\left(\vec{p}-\vec{k}\right)\, 
\gamma_0\,
{ {\Delta} }^{(r)}_{Q\,n}(p-k)
\, 
{\,^*{\cal G}_{L}}^{a}(k)
\right.
\nonumber\\
&~&\left.
+
\sum_{r=\pm}
\sum^3_{l,k=1}
\gamma_{l}\,  h^{(r)}_{Q}\left(\vec{p}-\vec{k}\right)
\, \gamma_{k}\,
\left(\delta_{lk}-\hat{k}_{l}\hat{k}_{k}\right)
\,
{ {\Delta} }^{(r)}_{Q\,n}(p-k)
\, 
{\,^*{\cal G}_{T}}^{a}(k)
\right].
\nonumber\\
\label{effec-quark-FW-1}
\end{eqnarray}
Performing the integration over 
the time-component variable $k_{0}$ reduces 
Eq.~(\ref{effec-quark-FW-1}) 
to the following result
\begin{eqnarray}
{{\,^*{\Sigma}}_{Q}}^{a}_{n}(p_0,\vec{p},\vec{k})
&=&
g^2\,
\left[
\sum_{r=\pm}
\gamma_0\, h^{(r)}_{Q}\left(\vec{p}-\vec{k}\right)\, 
\gamma_0\,
{{\,^{*}{\cal K}}^{(r)}_{L}}^{a}_{Q\,n}
(p_{0},\vec{p},\vec{k})
\right.
\nonumber\\
&~&\left.
+\sum_{r=\pm}
\sum^3_{i,j=1}
\gamma_{i}\,  h^{(r)}_{Q}\left(\vec{p}-\vec{k}\right)
\, \gamma_{j}\,
\left(\delta_{ij}-\hat{k}_{i}\hat{k}_j\right)
{{\,^{*}{\cal K}}^{(r)}_{T}}^{a}_{Q\,n}
(p_{0},\vec{p},\vec{k})
\right],
\label{effec-quark-p-k-new1}
\end{eqnarray}
where
\begin{eqnarray}
{{\,^{*}{\cal K}}^{(r)}_{S}}^{a}_{Q\,n}
(p_{0},\vec{p},\vec{k})
&=&
\int \frac{d k_0}{2\pi}
{\,^*{\Delta}}^{(r)}_{Q\,n}(p-k)
\,{\,^*{\cal G}_{S}}^{a}(k), 
\nonumber\\
&=&
\int \frac{d k_0}{2\pi} 
\int^{\beta}_{0} d\tau
\int^{\beta}_{0} d\tau'
e^{\left[(p_0-k_0)-\mu_{Q}-\frac{i\theta_{n}}{\beta}\right]\tau'}\,
e^{\left[k_0-\frac{i\phi^{a}}{\beta}\right]\tau}\,
\nonumber\\
&~& ~~~ ~~~ ~~~
\times
{ {\Delta} }^{(r)}_{Q\,n}(\tau',\vec{p}-\vec{k})
\, 
{\,^*{\cal G}_{S}}^{a}(\tau,\vec{k}),
\nonumber\\
&=&
\int^{\beta}_{0} d\tau e^{\left[p_0-\mu_{Q}
-\frac{i\theta_{n}}{\beta}
-\frac{i\phi^{a}}{\beta}\right]\tau}
{ {\Delta} }^{(r)}_{Q\,n}(\tau,\vec{p}-\vec{k})
\, 
{\,^*{\cal G}_{S}}^{a}(\tau,\vec{k}).
\label{effec-quark-p-k-new2}
\end{eqnarray}
The subscript $S=L,T$ is referred to the longitudinal 
and transverse component, respectively.
The integrations over the variables $k_{0}$ and $\tau$ 
are evaluated explicitly in 
Eqns.~(\ref{effec-quark-p-k-new1}) 
and 
~(\ref{effec-quark-p-k-new2}). 
The results for the positive and negative Foldy-Wouthuysen 
energy components of the interaction kernel are reduced to
\begin{eqnarray}
{{\,^{*}{\cal K}^{(+)}_{S}}}^{a}_{n}
(p_{0},\vec{p},\vec{k})
&=&
\int^{\infty}_{-\infty}\frac{d\xi_0}{2\pi}
\left[
\frac{1-n_{F}\left(
\epsilon_{Q}(\vec{p}-\vec{k})-\mu_Q-i\frac{\theta_n}{\beta}
\right)
+N_{G}\left({\xi_0}-i\frac{\phi^{a}}{\beta}\right)}
{
p_0-\epsilon_{Q}(\vec{p}-\vec{k})
-\xi_0
}\right]
\nonumber\\
&~&~~~\times
{{\,^{*}\rho}_{S}}^{a}
\left(\xi_0-i\frac{\phi^{a}}{\beta},\vec{k}\right),
\label{effective-kernel-positive1}
\end{eqnarray}
and
\begin{eqnarray}
{{\,^{*}{\cal K}}^{(-)}_{S}}^{a}_{n}
(p_{0},\vec{p},\vec{k})
&=&
\int^{\infty}_{-\infty}\frac{d\xi_0}{2\pi}
\left[
\frac{n_{F}\left(
\epsilon_{Q}(\vec{p}-\vec{k})+\mu_Q+i\frac{\theta_n}{\beta}
\right)
+N_{G}\left({\xi_0}-i\frac{\phi^{a}}{\beta}\right)}
{p_0+\epsilon_{Q}(\vec{p}-\vec{k})-\xi_0}\right]
\nonumber\\
&~&~~~\times
{{\,^{*}\rho}_{S}}^{a}
\left(\xi_0-i\frac{\phi^{a}}{\beta},\vec{k}\right),
\label{effective-kernel-negative1}
\end{eqnarray}
respectively.
By adopting the approximation of 
$|\vec{k}|/|\vec{p}|\approx g\ll 1$,
the pole 
of Eq.~(\ref{effective-kernel-positive1}) 
is allocated in the following position,
\begin{equation}
\begin{array}{l}
|\vec{p}| \approx p_0, \\
\epsilon_{Q}(\vec{p}-\vec{k}) \approx
|\vec{p}|-|\vec{k}|\hat{k}\cdot\hat{p},
\\
p_{0}-\epsilon_{Q}(\vec{p}-\vec{k})-\xi_{0} 
\approx
|\vec{k}|\hat{k}\cdot\hat{p}-\xi_{0}.
\end{array}
\end{equation} 
The conservation of color charges connects both 
the fundamental and the adjoint color chemical 
potentials for the quark-quark-gluon vertex 
in the following way
\begin{eqnarray}
\phi^{a}&=& \phi^{\underbrace{(ii')}}, 
\nonumber\\
&=&\theta_{i}-\theta_{i'}.
\end{eqnarray}
This leads to 
$\theta_{i}=\theta_{n}+\phi^{a}$ 
in the quark-gluon loop interaction 
that is depicted in Fig.~(\ref{effectivequark} a).
Therefore, by using the preceding results 
and preforming the analytic continuation 
over the external momentum's time-component  
${p}_{0}\rightarrow {p}_{0}+i\,\eta$,
Eq.~(\ref{effective-kernel-positive1}) 
is approximated to the following result:
\begin{eqnarray}
{{\,^{*}{\cal K}^{(+)}_{S}}}^{a}_{n}
(p_{0}+i\eta,\vec{p},\vec{k})
&=&
\int^{\infty}_{-\infty}\frac{d\xi_0}{2\pi}
\,
\left[
P\left(
\frac{1}{|\vec{k}|\,\hat{k}\cdot\hat{p}-\xi_{0}} 
\right)+i\,\pi\,\delta\left(
\xi_{0}-|\vec{k}|\,\hat{k}\cdot\hat{p}
\right)
\right]
\nonumber\\
&~&~~~~~~\times
\left[
1-
n_{F}\left(
\epsilon_{Q}(\vec{p}-\vec{k})-\mu_Q-i\frac{\theta_n}{\beta}
\right)
+
N_{G}\left({\xi_0}-i\frac{\phi^{a}}{\beta}\right)
\right]
\nonumber\\
&~&~~~~~~\times
{{\,^{*}\rho}_{S}}^{a}
\left(\xi_0-i\frac{\phi^{a}}{\beta},\vec{k}\right).
\label{K-positive-S-1}
\end{eqnarray}
In this class of function (i.e. Eq.~(\ref{K-positive-S-1})), 
the real and imaginary parts do not overlap 
with each other and they do not enhance simultaneously 
in the momentum complex plane.
The real part exists only in the time-like energy domain 
$\vec{p}^{2}\,<\,p_{0}^{2}$. 
When the momentum turns to run
over the space-like energy domain 
$\vec{p}^{2} \,\ge\, {p}_{0}^{2}$,
the imaginary part is developed 
while the real part is strongly suppressed.
In the limit
$\vec{p}^{2} \,\rightarrow\, {p}_{0}^{2}$ 
that when $\vec{p}^{2}$ 
approaches ${p}_{0}^{2}$ 
from below,
the time-like energy domain switches 
to the space-like energy domain 
and consequently the imaginary part 
is developed while the real part disappears
under the present HTL approximation. 
Therefore, the resultant imaginary part 
is reduced to
\begin{eqnarray}
{{\,^{*}{\cal K}^{(+)}_{S}}}^{a}_{n}
(p_{0}+i\eta,\vec{p},\vec{k})
&=&
\int^{\infty}_{-\infty}\frac{d\xi_0}{2\pi}
\,
\left[
1-
n_{F}\left(
\epsilon_{Q}(\vec{p}-\vec{k})-\mu_Q-i\frac{\theta_n}{\beta}
\right)
+
N_{G}\left({\xi_0}-i\frac{\phi^{a}}{\beta}\right)
\right]
\nonumber\\
&~&~~~\times\,
i\,\pi\,\delta\left(
\xi_{0}-|\vec{k}|\,\hat{k}\cdot\hat{p}
\right)\,
{{\,^{*}\rho}_{S}}^{a}
\left(\xi_0-i\frac{\phi^{a}}{\beta},\vec{k}\right),
\nonumber\\
&~&~~~~~~~~~~~~~~(\mbox{in space-like energy domain}),
\label{result-kernel-im-1}
\end{eqnarray}
for the external hard $p$-momentum  
(i.e. the external hard quark line) 
and the internal soft $k$-momentum 
(i.e. the internal soft gluon line). 
It should be noted that the spectral density 
${{\,^{*}\rho}_{S}}^{a}\left(\xi_0,\vec{k}\right)$  
is given in the space-like energy domain.
On the other hand, 
the negative Foldy-Wouthuysen energy component,
that is given 
by Eq.~(\ref{effective-kernel-negative1}), 
is approximated to
\begin{eqnarray}
{{\,^{*}{\cal K}}^{(-)}_{S}}^{a}_{n}
({p}_{0},\vec{p},\vec{k})
&\approx&
\frac{1}{2|\vec{p}|}\int^{\infty}_{-\infty}\frac{d\xi_0}{2\pi}
\left[
n_{F}\left(
\epsilon_{Q}(\vec{p}-\vec{k})+\mu_Q+i\frac{\theta_n}{\beta}
\right)
+N_{G}\left({\xi_0}-i\frac{\phi^{a}}{\beta}\right)
\right]
\nonumber\\
&~&~~~\times
{{\,^{*}\rho}_{S}}^{a}
\left(\xi_0-i\frac{\phi^{a}}{\beta},\vec{k}\right),
\nonumber\\
{{\,^{*}{\cal K}}^{(-)}_{S}}^{a}_{n}
({p}_{0}+i\eta,\vec{p},\vec{k})
&\approx& \mbox{dropped from the calculation}.
\label{effective-kernel-negative-approx1}
\end{eqnarray}
It is evident that 
the negative Foldy-Wouthuysen component 
has no pole and 
it does not develop any imaginary part for 
the external $p$-momentum even 
when the momentum turns 
to run over the space-like energy domain.
This means that 
Eq.~(\ref{effective-kernel-negative-approx1})
can be dropped from the calculation since 
it does not develop an imaginary part 
in the space-like energy domain.

The relevant quantity for the quark decay rate is
\begin{eqnarray}
\frac{1}{4|\vec{p}|}
\mbox{tr}\left[ p\cdot\gamma\, {\,^{*}\Sigma}
\left(p_{0},\vec{p}\right)
\right],
\end{eqnarray}
in the limit 
$|\vec{p}|\rightarrow {p}_{0}+i\,\eta$ 
from below (i.e. the brink of space-like energy domain).
It is useful to introduce the following 
approximations
\begin{eqnarray}
\frac{1}{4|\vec{p}|}\mbox{tr}
\left[
(p\cdot\gamma) \gamma_0 h_{+}
\left(\vec{p}-\vec{k}\right)\gamma_0 \right]
&\approx& 1,
\end{eqnarray}
and
\begin{eqnarray}
\frac{1}{4|\vec{p}|}\mbox{tr}
\left[
(p\cdot\gamma) \gamma_i h_{+}
\left(\vec{p}-\vec{k}\right)\gamma_j \right]\approx 
\hat{p}_i\hat{p}_j,
\end{eqnarray}
under the assumption of the hard external 
$p$-momentum 
and 
the soft internal loop $k$-momentum  
(i.e. $p\,\gg\,k$ and $k/p\sim g$).
The decay rate for the quark line with 
external fundamental color 
indexes $\delta_{ij}$ 
and the hard momentum is given by
\begin{eqnarray}
{\gamma_{Q}}_{ij}&=& 
\lim_{|\vec{p}|\rightarrow p_{0}}
\frac{1}{4|\vec{p}|}\,
\mbox{tr}
\left[
\left(p\cdot\gamma\right)
\,
{\,^{*}\Sigma_{Q}}_{ij}(p_{0}+i\eta,\vec{p})
\right],
\nonumber\\
&=& 
\sum_{a}\sum_{n}\,{\bf t}^{a}_{in}\,{\bf t}^{a}_{nj}\,  
{\gamma_{Q}}^{a}_{n}.
\end{eqnarray}
In order to be specific, 
the decay rate is given by taking the 
imaginary part of 
${\gamma_{Q}}_{ij}$ and ${\gamma_{Q}}^{a}_{n}$.
The decay rate part of the internal quark-gluon loop, 
namely, ${\gamma_{Q}}^{a}_{n}$
with an internal quark segment 
with a fundamental color index $n$
and a gluon segment with 
an adjoint color index $a$ 
is given by
\begin{eqnarray}
{\gamma_{Q}}^{a}_{n}
&=&
\lim_{ |\vec{p}|\rightarrow p_{0} }\,
\frac{1}{4|\vec{p}|}\,
\mbox{tr}
\left[
\left(p\cdot\gamma\right)
\,
{\,^{*}\Sigma_{Q}}^{a}_{n}(p_{0}+i\eta,\vec{p})
\right],
\nonumber\\
&=&
\lim_{|\vec{p}|\rightarrow p_{0}}\,
\frac{1}{4|\vec{p}|}\,
\mbox{tr}
\left[
\left(p\cdot\gamma\right)
\,\int \frac{d^{3}\vec{k}}{(2\pi)^{3}}
{\,^{*}\Sigma_{Q}}^{a}_{n}
(p_{0}+i\eta,\vec{p},\vec{k})
\right].
\end{eqnarray}
%
%
In the limit $|\vec{p}|\rightarrow {p}_{0}$,
the quark's decay rate of the effective 
internal thermal quark-gluon loop is reduced to
\begin{eqnarray}
{\gamma_{Q}}^{a}_{n}
&=&
\frac{1}{4|\vec{p}|}\,\mbox{tr}\left[
\left(p\cdot\gamma\right)
\,
{\,^{*}\Sigma_{Q}}^{a}_{n}
(p_{0}+i\,\eta,\vec{p},\vec{k})
\right],
\nonumber\\
&=& 
g^{2} 
\left[
\int \frac{d^{3}\vec{k}}{(2\pi)^{3}}
{{\,^{*}{\cal K}}^{(+)}_{L}}^{a}_{n}
(p_{0}+i\,\eta,\vec{p},\vec{k})
\right.
\nonumber\\
&~& ~~~~~~\left.
+ 
\int \frac{d^{3}\vec{k}}{(2\pi)^{3}}
\left(1-(\hat{p}\cdot\hat{k})^2\right)
{{\,^{*}{\cal K}}^{(+)}_{T}}^{a}_{n}
( p_{0}+i\,\eta,\vec{p},\vec{k} )
\right].
\label{decay-internal-1}
\end{eqnarray}
By using Eq.~(\ref{result-kernel-im-1}), 
the quark's decay rate which 
is given by 
Eq.~(\ref{decay-internal-1}) 
is reduced to
\begin{eqnarray}
{\gamma_{Q}}^{a}_{n}&=&
\lim_{|\vec{p}|\rightarrow p_{0}}\,
\frac{1}{4|\vec{p}|}\,
\mbox{tr}
\left[
\left(p\cdot\gamma\right)
\,
{\,^{*}\Sigma_{Q}}^{a}_{n}(p_{0}+i\,\eta,\vec{p})
\right],
\nonumber\\
&=& 
g^{2}\,(i\,\pi)\,
\int^{\infty}_{-\infty}\frac{d\xi_0}{2\pi}\,
\int \frac{d^{3}\vec{k}}{(2\pi)^{3}}\,
\delta\left(
\xi_{0}-|\vec{k}|\,\hat{k}\cdot\hat{p}
\right)\,
 \left[
{{\,^{*}\rho}_{L}}^{a}
\left(\xi_0-i\frac{\phi^{a}}{\beta},\vec{k}\right)
\right.
\nonumber\\
&~&~~~\left.
+ 
\left(1-(\hat{p}\cdot\hat{k})^2\right)
{{\,^{*}\rho}_{T}}^{a}
\left(\xi_0-i\frac{\phi^{a}}{\beta},\vec{k}\right)
\right]\,
\nonumber\\
&~&~~~\times
\left[
1-
n_{F}\left(
\epsilon_{Q}(\vec{p}-\vec{k})-\mu_Q-i\frac{\theta_n}{\beta}
\right)
+
N_{G}\left({\xi_0}-i\frac{\phi^{a}}{\beta}\right)
\right].
\label{internal-decay-n-ax1}
\end{eqnarray}
%
%
%
The Dirac $\delta$-function is transformed to
\begin{eqnarray}
\delta\left(
\xi_{0}-|\vec{k}|\,\hat{k}\cdot\hat{p}
\right)
&=&
\frac{1}{|\vec{k}|}\,
\delta\left(
\cos\theta-\frac{\xi_{0}}{|\vec{k}|}
\right).
\end{eqnarray}
Furthermore, by transforming the $\delta$-function using
$\delta\left(
\xi_{0}-|\vec{k}|\,\hat{k}\cdot\hat{p}
\right)\rightarrow \left[1/|\vec{k}|\right]
\delta\left(
\cos\theta-\xi_{0}/|\vec{k}|\right)$
and evaluating the integral that is given 
in Eq.~(\ref{internal-decay-n-ax1})
over 
$\hat{k}\cdot\hat{p}=\cos\theta$ 
where 
$\cos\theta=\xi_{0}/|\vec{k}|$, 
the calculation leads to the interval constraint
$|\vec{k}|\,\ge\,\xi_{0}\,\ge\,-|\vec{k}|$.
The decay rate of the internal quark-gluon 
loop is reduced to
\begin{eqnarray}
{\gamma_{Q}}^{a}_{n}
&=& 
g^{2}\,(i\,\pi)\,
\int^{\infty}_{0} \frac{d|\vec{k}|}{(2\pi)^{2}}\,
|\vec{k}|\,
\int^{|\vec{k}|}_{-|\vec{k}|}\frac{d\xi_0}{2\pi}\,
\left[
{{\,^{*}\rho}_{L}}^{a}
\left(\xi_0-i\frac{\phi^{a}}{\beta},\vec{k}\right)
\right.
\nonumber\\
&~&~~~
\left.
+ 
\left(1-\frac{\xi^2_{0}}{\vec{k}^2}\right)
{{\,^{*}\rho}_{T}}^{a}
\left(\xi_0-i\frac{\phi^{a}}{\beta},\vec{k}\right)
\right]\,
\nonumber\\
&~&~~~\times\left.
\left[
1-
n_{F}\left(
\epsilon_{Q}(\vec{p}-\vec{k})-\mu_Q-i\frac{\theta_n}{\beta}
\right)
+
N_{G}\left({\xi_0}-i\frac{\phi^{a}}{\beta}\right)
\right]
\right|_{\hat{k}\cdot\hat{p}=\xi_{0}/|\vec{k}|}.
\label{xxx_rate1}
\end{eqnarray}
The spectral density is given by
\begin{eqnarray}
{{\,^{*}\rho}_{S}}^{a}\left(\xi_0,\vec{k}\right)=
{\beta_{S}}^{a}\left(\xi_0,\vec{k}\right)
\theta\left(
\vec{k}^2-\xi^2_{0}
\right), ~~~(\mbox{where}~~~ S=T,L),
\label{spectral_xxx_rate1}
\end{eqnarray}
in the space-like energy domain.
Therefore, the decay rate is re-written as follows
\begin{eqnarray}
{\gamma_{Q}}^{a}_{n}
&=& 
g^{2}\,(i\,\pi)\,
\int^{\infty}_{0} \frac{d|\vec{k}|}{(2\pi)^{2}}\,
|\vec{k}|\,
\int^{|\vec{k}|}_{-|\vec{k}|}\frac{d\xi_0}{2\pi}\,
\left[
{\beta_{L}}^{a}\left(\xi_0-i\frac{\phi^{a}}{\beta},\vec{k}\right)
\right.
\nonumber\\
&~&~~~
\left.
+ 
\left(1-\frac{\xi^2_{0}}{\vec{k}^2}\right)
{\beta_{T}}^{a}\left(\xi_0-i\frac{\phi^{a}}{\beta},\vec{k}\right)
\right]\,
\nonumber\\
&~&~~~\times\left.
\left[
1-
n_{F}\left(
\epsilon_{Q}(\vec{p}-\vec{k})-\mu_Q-i\frac{\theta_n}{\beta}
\right)
+
N_{G}\left({\xi_0}-i\frac{\phi^{a}}{\beta}\right)
\right]
\right|_{\hat{k}\cdot\hat{p}=\xi_{0}/|\vec{k}|}.
\label{decay_rate_2}
\end{eqnarray}
The explicit expressions for 
the longitudinal and transverse spectral 
densities are given by
\begin{eqnarray}
{\beta_{L}}^{a}\left(\xi_0,\vec{p}\right)
&=&
\frac{ {\Pi_{L}}^{a}(\xi_{0},\vec{p}) }
{\left(\vec{p}^{2}\right)^{2}+
\left[{\Pi_{L}}^{a}(\xi_{0},\vec{p})\right]^{2}},
\nonumber\\
{\beta_{T}}^{a}\left(\xi_0,\vec{p}\right)
&=&
-\frac{ {\Pi_{T}}^{a}(\xi_{0},\vec{p}) }
{\left(\xi_{0}^{2}-\vec{p}^{2}\right)^{2}+
\left[{\Pi_{T}}^{a}(\xi_{0},\vec{p})\right]^{2}},
\end{eqnarray}
respectively, where
\begin{eqnarray}
{\Pi_{L}}^{a}(\xi_{0},\vec{p})&=&
({\bf m}^{2}_{D})^{a}
\left(
\frac{\pi\,\xi_{0}}{2|\vec{p}|}
\right),
\nonumber\\
{\Pi_{T}}^{a}(\xi_{0},\vec{p})&=&
-\frac{1}{2}
({\bf m}^{2}_{D})^{a}
\left(1-\frac{\xi_{0}^2}{|\vec{p}|^{2}}\right)
\left(\frac{\pi\,\xi_{0}}{2|\vec{p}|}\right).
\end{eqnarray}
Since it is supposed that 
$\xi_{0}-i\phi^{a}/\beta\le g\,T\ll 1$, 
the gluon's partition becomes the dominant term.
Hence, it is possible to assume the following 
approximation
\begin{eqnarray}
\left.\left[
1-
n_{F}\left(
\epsilon_{Q}(\vec{p}-\vec{k})-\mu_Q-i\frac{\theta_n}{\beta}
\right)
+
N_{G}\left({\xi_0}-i\frac{\phi^{a}}{\beta}\right)
\right]
\right|_{\hat{k}\cdot\hat{p}=\xi_{0}/|\vec{k}|}
&\approx&
N_{G}\left({\xi_0}-i\frac{\phi^{a}}{\beta}\right),
\nonumber\\
&\approx& \frac{T}{\xi_{0}-i\frac{\phi^{a}}{\beta}},
\nonumber\\
&\rightarrow&
\frac{T}{\xi_{0}-{\mu_{C}}^{a}}.
\label{approx-zero-like-potential}
\end{eqnarray}
This approximation simplifies the calculation 
of the hard quark decay rate drastically.
For instance, the gluonic part of 
the colored fermion Landau frequency 
and the colored gluon's Debye mass
develop nontrivial imaginary terms for the (real-) 
flavor and (real/imaginary-) 
color chemical potentials. 
These imaginary parts vanish when 
the adjoint color potentials vanish.
The condition 
${\mu_{C}}^{a}=\left({\mu_{C}}_{A}-{\mu_{C}}_{B}\right)=0$ 
(where $a=1, \cdots, N^{2}_c-1$), 
means that fundamental chemical potentials 
become equal to each other for different color species, 
namely, blue, green and red colors. 
The equality of the fundamental color chemical potentials
is possible in the $U(N_c)$ symmetry group.
In the $SU(N_c)$ symmetry group, 
the equality constraint implies that the color chemical 
potentials vanish because of the unimodular condition 
$\sum^{N_c}_{i=1}\,\theta_{i}=0$. 
Therefore, in order to set $\mu^{a}=0$, either 
the fundamental color chemical potentials vanish 
or the quark lines carry all the color information
in the sense that the quark segment of internal loop 
carries the same color charge of the external quark line.
The later possibility imposes an additional color 
contraction $\delta_{i\,n}$ 
where 
$i$ and $n$ are the external and internal quark lines indexes, 
respectively.
In the case that all the fundamental color chemical potentials 
vanish the system turns to be color neutral and 
this is equivalent to neglect the explicit 
color degrees of freedom for quarks and gluons. 
Under the extreme conditions when the
the free mobile of colored quarks 
and gluons matter is reached, 
it is possible that 
the adjoint color chemical potentials of 
the internal gluon's segment
become finite and subsequently 
the fundamental color chemical potentials 
turn to be finite.
In this case, the Debye mass squared becomes 
an imaginary one.
The external and internal quark lines are presumed hard 
while the internal gluon segment is considered to be soft 
then it is reasonable to imagine that 
in the deconfinement matter most 
of the energy information is carried by 
the hard colored quark lines.
Under this assumption,  the internal adjoint color 
potential ${\mu_{C}}^{a}$ vanishes 
(i.e. ${\mu_{C}}^{a}\rightarrow 0$) 
and 
Eq.~(\ref{approx-zero-like-potential})
is reduced to
\begin{eqnarray}
\left.\left[
1-
n_{F}\left(
\epsilon_{Q}(\vec{p}-\vec{k})-\mu_Q-i\frac{\theta_n}{\beta}
\right)
+
N_{G}\left({\xi_0}-i\frac{\phi^{a}}{\beta}\right)
\right]
\right|_{\hat{k}\cdot\hat{p}=\xi_{0}/|\vec{k}|}
&\rightarrow&
\frac{T}{\xi_{0}}.
\label{approx-zero-like-potential-2}
\end{eqnarray} 
The resultant decay rate with the approximation
those are given by 
Eqns.~(\ref{approx-zero-like-potential}) and 
 ~(\ref{approx-zero-like-potential-2})
is reduced to
\begin{eqnarray}
{\gamma_{Q}}^{a}_{n}
&=& 
i\,\frac{g^{2}\,T}{4\pi^2}\,
\int^{\infty}_{0} d|\vec{k}|\,
|\vec{k}|\,
\int^{|\vec{k}|}_{-|\vec{k}|}\frac{d\xi_0}{2\pi}\,
\frac{1}{\left(\xi_{0}-i\frac{\phi^{a}}{\beta}\right)}
\left[
{\beta_{L}}^{a}\left(\xi_0-i\frac{\phi^{a}}{\beta},
\vec{k}\right)
\right.
\nonumber\\
&~&
\left.
~~~ ~~~ ~~~
+ 
\left(1-\frac{\xi^2_{0}}{\vec{k}^2}\right)
{\beta_{T}}^{a}\left(\xi_0-i\frac{\phi^{a}}{\beta},
\vec{k}\right)
\right],
\nonumber\\
&\approx&
i\,\frac{g^{2}\,T}{4\pi^2}\,
\int^{\infty}_{0} d|\vec{k}|\,
|\vec{k}|\,
\int^{|\vec{k}|}_{-|\vec{k}|}\frac{d\xi_0}{2\pi}\,
\frac{1}{\xi_{0}}
\left[
{\beta_{L}}^{a}\left(\xi_0,\vec{k}\right)
+ 
\left(1-\frac{\xi^2_{0}}{\vec{k}^2}\right)
{\beta_{T}}^{a}\left(\xi_0,\vec{k}\right)
\right].
\label{decay_rate_simple}
\end{eqnarray}
Nonetheless, the validity of 
these kind of approximations need 
to be scrutinized numerically 
and to be tested experimentally at LHC. 
The decay rate for the quark with 
the external fundamental color indexes, namely, 
$i\,j$ 
is determined by
\begin{eqnarray}
{\gamma_{Q}}_{i\,j}&=&
\sum^{N_c}_{n}
\sum^{N^2_c-1}_{a}\,
\left({\bf t}^{a}\right)_{in}
\,
\left({\bf t}^{a}\right)_{nj}
\,
{\gamma_{Q}}^{a}_{n},
\nonumber\\
&=&
i\,\sum^{N_c}_{n}
\sum^{N^2_c-1}_{a}\,
\left({\bf t}^{a}\right)_{in}
\left({\bf t}^{a}\right)_{nj}
\nonumber\\
&~&~\times\,
\frac{g^{2}\,T}{4\pi^2}\,
\int^{\infty}_{0} d|\vec{k}|\,
|\vec{k}|\,
\int^{|\vec{k}|}_{-|\vec{k}|}\frac{d\xi_0}{2\pi}\,
\frac{1}{ \xi_{0} }
\left[
{\beta_{L}}^{a}\left(\xi_0,\vec{k}\right)
+ 
\left(1-\frac{\xi^2_{0}}{\vec{k}^2}\right)
{\beta_{T}}^{a}\left(\xi_0,\vec{k}\right)
\right].
\end{eqnarray} 
The decay rate is calculated by taking 
$\Im{m}\, 
\left({\gamma_{Q}}_{i\,j}\right)$
where 
\begin{eqnarray}
{\gamma_{Q}}_{i\,j}
&=&
i\,\delta_{ij}\,
\frac{1}{2N_c}
\left[
N_{c}\sum^{N_c}_{n}{\gamma_{Q}}^{\underbrace{(in)}}
-
{\gamma_{Q}}^{\underbrace{(00)}}
\right],
\nonumber\\
&\approx&
i\,\delta_{ij}\,
\frac{N^2_c-1}{2N_c}
\,{\gamma_{Q}}^{\underbrace{(00)}},
\end{eqnarray}
and
\begin{eqnarray}
{\gamma_{Q}}^{\underbrace{(AB)}}
&=&
\frac{g^{2}\,T}{4\pi^2}\,
\int^{\infty}_{0} d|\vec{k}|\,
|\vec{k}|\,
\int^{|\vec{k}|}_{-|\vec{k}|}\frac{d\xi_0}{2\pi}\,
\frac{1}{ \xi_{0} }
\left[
{\beta_{L}}^{\underbrace{(AB)}}
\left(\xi_0,\vec{k}\right)
\right.
\nonumber\\
&~&~~~~~~
\left.
+ 
\left(1-\frac{\xi^2_{0}}{\vec{k}^2}\right)
{\beta_{T}}^{\underbrace{(AB)}}
\left(\xi_0,\vec{k}\right)
\right].
\end{eqnarray}
The observable quantities such as 
the decay rate of colored quark turn to be dependent 
on the fundamental color chemical potentials
and the interaction details.
The fluid characteristic of the quark-gluon plasma in RHIC energy
support the idea of the role of the color degrees of freedom
and the weakly color coupling in the quark-gluon plasma.
%

\section{\label{section8} Discussion and conclusion}

It is argued that the quark-gluon plasma above the tri-critical 
point of the phase transition from the low-lying hadronic 
phase to the quark-gluon plasma is not true deconfined matter
but rather a weakly interacting quarks and gluons. 
There are strong indications from RHIC that quark-gluon plasma 
is a perfect fluid with a low shear viscosity and 
not a true deconfined matter.
Hence, it is natural to assume that the explicit role 
of the color degrees of freedom becomes important
above the deconfinement 
phase transition line in the phase transition diagram 
from the hadronic matter to the quark-gluon plasma.  
Furthermore, it is naive to believe that the quarks and gluons 
carry color charges and the color fugacities 
tend to vary in the medium. 
The color chemical potentials appear explicitly 
in the quark and gluon partition functions.
In this case, the color degrees of freedom couple 
with the other degrees of freedom such as 
the kinematic degree of freedom because 
the color chemical potentials appear explicitly 
in the quark and gluon partition functions.
The color chemical potentials for the colored quarks
are represented by the fundamental color chemical potentials 
while color chemical potentials for the colored gluons 
are given by the adjoint ones. 
The imaginary chemical potentials for quarks and gluons 
are given by the Fourier variables $i\,\theta_{i}$ 
with a fundamental index, namely, $i$ 
that runs over $i=1,\cdots, N_c$ 
and $i\,\phi^{a}\equiv\,i\,(\theta_{A}-\theta_{B})$
with an adjoint index, namely, $a=\underbrace{(AB)}$
that runs over $a=1,\cdots,N^2_{c}-1$, respectively. 
The imaginary fundamental and adjoint color chemical potentials, 
namely,  $i\,\theta_{i}/\beta$ and $i\,\phi^{a}/\beta$, 
respectively, maintain the conservation of color charges
and/or project a specific internal color symmetry when 
they are integrated over the invariance Haar measure 
and an appropriate color wave-function. 
It is very relevant to calculate the equation of state 
with an expansion of the weak coupling corrections
for the quark-gluon bag with a specific 
internal color structure in order to understand 
the mechanism of the deconfinement phase transition.
The real fundamental and adjoint color chemical potentials, 
namely, 
${\mu_{C}}_{i}$ and 
${\mu_{C}}^{a}={\mu_{C}}_{A}-{\mu_{C}}_{B}$
for quarks and gluons, respectively, adjust 
the color fugacities and determine the color densities.
Despite of the apparent complexity of the color construction, 
the calculation is found simple and straightforward. 
The hard thermal loops with soft and hard momenta are 
extended in a straightforward manner
to include the color degrees of freedom for quarks and gluons. 
In general, there is always a way to decompose 
the external fundamental and adjoint color indexes 
with respect to the internal loop color indexes.
The Fermion plasma frequency is decomposed as follows
\begin{eqnarray}
\left(
\omega^{2}_{0\,Q}
\right)_{ij}&=&
\left(
\omega^{2}_{0\,Q}
\right)_{i}\,\delta_{ij},
\nonumber\\
&=&
\sum^{N^{2}_{c}-1}_{a}
\,
\sum^{N_{c}}_{n=1}
\,
{\bf t}^{a}_{in}\, {\bf t}^{a}_{nj}\, 
\left[
{\left(
\omega^{2}_{0\,Q (Q)}
\right)}_{n}
+
{\left(
\omega^{2}_{0\,Q (G)}
\right)}^{a}
\right],
\nonumber\\
&~&
~~~~(\mbox{there is no sum over the fundamental color indexes}~i\,,\,j) ,
\end{eqnarray}
where
\begin{eqnarray}
\left({\omega}^2_{0\,Q (Q)}\right)_n
&=&
\frac{g^2}{4\pi^2}\,
\left[
\frac{\pi^2}{6}\, T^2
+\frac{1}{2}
\left(
\mu_{Q}+{\mu_{C}}_{n}
\right)^2
\right],
\end{eqnarray}
and
\begin{eqnarray}
\left({\omega}^2_{0\,Q (G)}\right)^{a}
&=&
\frac{g^2}{4\pi^2}\,
\left[
\frac{\pi^2}{3}\, T^{2}
-\frac{1}{2}\left(
{\mu_{C}}_{A}-{\mu_{C}}_{B}
\right)^2
+i\,\pi\,T\,\left(
{\mu_{C}}_{A}-{\mu_{C}}_{B}
\right)
\right], 
\nonumber\\
&~&
~~~\left(\mbox{given that}~ a=\underbrace{(A\,B)}\right).
\end{eqnarray}
The external fundamental color indexes $i\,j$ 
are parametrized in terms of 
the internal loop's fundamental color index $n$
and
the internal loop's adjoint color index $a$
using the matrix elements of fundamental 
and/or adjoint group generators. 
The colored gluon's Debye mass is decomposed as follows 
\begin{eqnarray}
\left({\bf m}^2_{D}\right)^{a'a}
&=&
\left({\bf m}^2_{D}\right)^{a}\,\delta^{aa'},
\nonumber\\
&=&
\sum^{N^2_{c}-1}_{b,c=1}
\,
\sum^{N_c}_{i,j=1}
\,
\left[
({\bf T}^{a'})_{cb}\,({\bf T}^{a})_{bc}^{\dagger}\,
\left({\bf m}^2_{D\,(G)}\right)^{b} 
+
{\bf t}^{a'}_{ij}\, ({\bf t}^{a\dagger})_{ji}\,
\left({\bf m}^2_{D\,(Q)}\right)_{i}
\right],
\nonumber\\
&~&
~~~~(\mbox{there is no sum over the adjoint color indexes}~ a\,,\,a'),
\end{eqnarray}
where 
\begin{eqnarray}
\left({\bf m}^2_{D\,(G)}\right)^{b}
&=&
\frac{g^2}{\pi^2}\,
\left[
\frac{\pi^2}{3}\,T^{2}
-\frac{1}{2}\left(
{\mu_{C}}_{A}-{\mu_{C}}_{B}\right)^2
+i\,\pi\,T\,\left(
{\mu_{C}}_{A}-{\mu_{C}}_{B}\right)
\right],
\nonumber\\
&~&
~~~\left(\mbox{given that}~ b=\underbrace{(A\,B)}\right),
\end{eqnarray}
and
\begin{eqnarray}
\left({{\bf m}^{2}_{D\,(Q)}}\right)_{i}
&=&
\frac{2g^2}{\pi^2}\,\sum^{N_{F}}_{Q=1}
\left[
\frac{\pi^2}{6}\,T^{2}
+\frac{1}{2}
\left(\mu_{Q}+{\mu_{C}}_{i}\right)^2
\right].
\end{eqnarray}
The external adjoint color indexes $a'\,a$ 
for gluons are parameterized in terms 
of the internal adjoint color indexes $b\,c$
for the gluon and ghost loops and tadpole 
as well through the elements
of adjoint generators (i.e. adjoint matrices) 
while they are parameterized in terms 
of the internal fundamental color indexes $i\,j$ 
for the quark-loop.
The astonished result is that the gluon part of both 
the fermion Landau frequency and the Debye mass develop
an imaginary part for the real flavor 
and color chemical potentials. 
This imaginary part is canceled when 
the color adjoint chemical potentials vanish. 
This nontrivial solution leads to the conclusion
that the fundamental color potential of various color species 
(i.e. red, green and blue) tend to be equal to each other. 
However, the equality of the fundamental color potentials 
is possible in the $U(N_c)$ symmetry group 
while this equality in the $SU(N_c)$ symmetry group
means that the fundamental color potentials are 
all identical to zero.
The possible alternate solution 
for vanishing the adjoint color chemical 
potentials is that the quarks could carry all the color information
and they do not violate the color species in the interaction.
These phenomena enrich the physics 
of the quark-gluon plasma above the deconfinement phase transition.

On the other hand, the imaginary part 
disappears when both the imaginary flavor 
and the imaginary color chemical potentials are adopted 
in the calculation. 
The imaginary chemical potentials correspond Fourier variables
for the grand canonical ensemble where the flavor and color charges
are conserved 
(i.e. the charge densities are calculated by the Fourier 
transformation of the Fourier variables).
In this case both the flavor and the color fugacities are eliminated. 

The gluon polarization tensor usually generates
quadratic temperature terms 
(i.e $\left(\Delta\,{\bf m}^2_{D\,(G)}\right)^{b}\,
\propto\, T^{3}$ where $b$ 
is any internal adjoint color index) 
in the virtual internal loop interaction level
~\cite{Hidaka:2008dr,Hidaka:2009hs,Hidaka:2009ma}.
Those quadratic temperature terms are fortunately found 
to cancel each other in the final result 
of the polarization tensor for the real colored gluon 
in the medium. 
This cancellation takes place when the sum 
of all the internal adjoint color indexes is considered.

The decay rate for the colored hard quark is written 
as follows
\begin{eqnarray}
{\gamma_{Q}}_{i\,j}&=&
{\gamma_{Q}}_{i}\,\delta_{ij},
\nonumber\\
&=&
\sum^{N_c}_{n}
\sum^{N^2_c-1}_{a}\,
\left({\bf t}^{a}\right)_{in}
\,
\left({\bf t}^{a}\right)_{nj}
\,
{\gamma_{Q}}^{a}_{n},
\nonumber\\
&~&
~~~~(\mbox{there is no sum over the fundamental indexes}~ i\,,\,j),
\end{eqnarray}
where the external fundamental color indexes $i\,j$ are coupled
to the internal fundamental and adjoint color indexes 
of the internal loop
through the matrix elements of the group generators.
The color fugacities enter explicitly the quarks' and gluons'
partition functions and moreover they appear explicitly 
in the fundamental quark and adjoint gluon propagators. 
They also appears in the physical quantities such as
the quark's self-energy and gluon's polarization tensor.
The finite color chemical potentials seem to modify 
the decay rate for the colored hard quarks.
Therefore, it is possible at the extreme conditions 
to see the decay rates of the colored quarks to depend 
explicitly on the (fundamental-) color chemical potentials. 
General speaking, despite of the apparent complexity 
of the internal color structure, 
it is possible to extend the soft and hard thermal loop 
calculations to include the color degrees of freedom 
for quarks and gluons explicitly.
The internal color structure of quarks and gluons 
above the deconfinement phase transition 
is rich and non-trivial one. 
\begin{acknowledgments} 
The early stages of this work has been supported 
by Alexander von Humboldt foundation.
The author thanks
Carsten Greiner and Robert Pisarski for the discussion.
\end{acknowledgments}

\bibliography{zak10_bib_rob1}

\begin{thebibliography}{27}
\expandafter\ifx\csname natexlab\endcsname\relax\def\natexlab#1{#1}\fi
\expandafter\ifx\csname bibnamefont\endcsname\relax
  \def\bibnamefont#1{#1}\fi
\expandafter\ifx\csname bibfnamefont\endcsname\relax
  \def\bibfnamefont#1{#1}\fi
\expandafter\ifx\csname citenamefont\endcsname\relax
  \def\citenamefont#1{#1}\fi
\expandafter\ifx\csname url\endcsname\relax
  \def\url#1{\texttt{#1}}\fi
\expandafter\ifx\csname urlprefix\endcsname\relax\def\urlprefix{URL }\fi
\providecommand{\bibinfo}[2]{#2}
\providecommand{\eprint}[2][]{\url{#2}}

\bibitem[{\citenamefont{Zakout and Greiner}(2010)}]{Zakout:2010ep}
\bibinfo{author}{\bibfnamefont{I.}~\bibnamefont{Zakout}} \bibnamefont{and}
  \bibinfo{author}{\bibfnamefont{C.}~\bibnamefont{Greiner}}
  (\bibinfo{year}{2010}), \eprint{1002.3119}.

\bibitem[{\citenamefont{Zakout and Greiner}(2008)}]{Zakout:2007nb}
\bibinfo{author}{\bibfnamefont{I.}~\bibnamefont{Zakout}} \bibnamefont{and}
  \bibinfo{author}{\bibfnamefont{C.}~\bibnamefont{Greiner}},
  \bibinfo{journal}{Phys. Rev.} \textbf{\bibinfo{volume}{C78}},
  \bibinfo{pages}{034916} (\bibinfo{year}{2008}), \eprint{0709.0144}.

\bibitem[{\citenamefont{Zakout et~al.}(2007)\citenamefont{Zakout, Greiner, and
  Schaffner-Bielich}}]{Zakout:2006zj}
\bibinfo{author}{\bibfnamefont{I.}~\bibnamefont{Zakout}},
  \bibinfo{author}{\bibfnamefont{C.}~\bibnamefont{Greiner}}, \bibnamefont{and}
  \bibinfo{author}{\bibfnamefont{J.}~\bibnamefont{Schaffner-Bielich}},
  \bibinfo{journal}{Nucl. Phys.} \textbf{\bibinfo{volume}{A781}},
  \bibinfo{pages}{150} (\bibinfo{year}{2007}), \eprint{nucl-th/0605052}.

\bibitem[{\citenamefont{Begun et~al.}(2009)\citenamefont{Begun, Gorenstein, and
  Greiner}}]{Begun:2009an}
\bibinfo{author}{\bibfnamefont{V.~V.} \bibnamefont{Begun}},
  \bibinfo{author}{\bibfnamefont{M.~I.} \bibnamefont{Gorenstein}},
  \bibnamefont{and} \bibinfo{author}{\bibfnamefont{W.}~\bibnamefont{Greiner}},
  \bibinfo{journal}{J. Phys.} \textbf{\bibinfo{volume}{G36}},
  \bibinfo{pages}{095005} (\bibinfo{year}{2009}), \eprint{0906.3205}.

\bibitem[{\citenamefont{Abir and Mustafa}(2009)}]{Abir:2009sh}
\bibinfo{author}{\bibfnamefont{R.}~\bibnamefont{Abir}} \bibnamefont{and}
  \bibinfo{author}{\bibfnamefont{M.~G.} \bibnamefont{Mustafa}},
  \bibinfo{journal}{Phys. Rev.} \textbf{\bibinfo{volume}{C80}},
  \bibinfo{pages}{051903} (\bibinfo{year}{2009}), \eprint{0905.4140}.

\bibitem[{\citenamefont{Ferroni and Koch}(2009)}]{Ferroni:2008ej}
\bibinfo{author}{\bibfnamefont{L.}~\bibnamefont{Ferroni}} \bibnamefont{and}
  \bibinfo{author}{\bibfnamefont{V.}~\bibnamefont{Koch}},
  \bibinfo{journal}{Phys. Rev.} \textbf{\bibinfo{volume}{C79}},
  \bibinfo{pages}{034905} (\bibinfo{year}{2009}), \eprint{0812.1044}.

\bibitem[{\citenamefont{Noronha-Hostler
  et~al.}(2009{\natexlab{a}})\citenamefont{Noronha-Hostler, Beitel, Greiner,
  and Shovkovy}}]{NoronhaHostler:2009cf}
\bibinfo{author}{\bibfnamefont{J.}~\bibnamefont{Noronha-Hostler}},
  \bibinfo{author}{\bibfnamefont{M.}~\bibnamefont{Beitel}},
  \bibinfo{author}{\bibfnamefont{C.}~\bibnamefont{Greiner}}, \bibnamefont{and}
  \bibinfo{author}{\bibfnamefont{I.}~\bibnamefont{Shovkovy}}
  (\bibinfo{year}{2009}{\natexlab{a}}), \eprint{0909.2908}.

\bibitem[{\citenamefont{Noronha-Hostler
  et~al.}(2008)\citenamefont{Noronha-Hostler, Greiner, and
  Shovkovy}}]{NoronhaHostler:2007jf}
\bibinfo{author}{\bibfnamefont{J.}~\bibnamefont{Noronha-Hostler}},
  \bibinfo{author}{\bibfnamefont{C.}~\bibnamefont{Greiner}}, \bibnamefont{and}
  \bibinfo{author}{\bibfnamefont{I.~A.} \bibnamefont{Shovkovy}},
  \bibinfo{journal}{Phys. Rev. Lett.} \textbf{\bibinfo{volume}{100}},
  \bibinfo{pages}{252301} (\bibinfo{year}{2008}), \eprint{0711.0930}.

\bibitem[{\citenamefont{Noronha-Hostler
  et~al.}(2009{\natexlab{b}})\citenamefont{Noronha-Hostler, Noronha, and
  Greiner}}]{NoronhaHostler:2008ju}
\bibinfo{author}{\bibfnamefont{J.}~\bibnamefont{Noronha-Hostler}},
  \bibinfo{author}{\bibfnamefont{J.}~\bibnamefont{Noronha}}, \bibnamefont{and}
  \bibinfo{author}{\bibfnamefont{C.}~\bibnamefont{Greiner}},
  \bibinfo{journal}{Phys. Rev. Lett.} \textbf{\bibinfo{volume}{103}},
  \bibinfo{pages}{172302} (\bibinfo{year}{2009}{\natexlab{b}}),
  \eprint{0811.1571}.

\bibitem[{\citenamefont{Noronha-Hostler
  et~al.}(2009{\natexlab{c}})\citenamefont{Noronha-Hostler, Ahmad, Noronha, and
  Greiner}}]{NoronhaHostler:2009tz}
\bibinfo{author}{\bibfnamefont{J.}~\bibnamefont{Noronha-Hostler}},
  \bibinfo{author}{\bibfnamefont{H.}~\bibnamefont{Ahmad}},
  \bibinfo{author}{\bibfnamefont{J.}~\bibnamefont{Noronha}}, \bibnamefont{and}
  \bibinfo{author}{\bibfnamefont{C.}~\bibnamefont{Greiner}}
  (\bibinfo{year}{2009}{\natexlab{c}}), \eprint{0906.3960}.

\bibitem[{\citenamefont{Landsman and van Weert}(1987)}]{Landsman:1986uw}
\bibinfo{author}{\bibfnamefont{N.~P.} \bibnamefont{Landsman}} \bibnamefont{and}
  \bibinfo{author}{\bibfnamefont{C.~G.} \bibnamefont{van Weert}},
  \bibinfo{journal}{Phys. Rept.} \textbf{\bibinfo{volume}{145}},
  \bibinfo{pages}{141} (\bibinfo{year}{1987}).

\bibitem[{\citenamefont{Braaten and
  Pisarski}(1990{\natexlab{a}})}]{Braaten:1989mz}
\bibinfo{author}{\bibfnamefont{E.}~\bibnamefont{Braaten}} \bibnamefont{and}
  \bibinfo{author}{\bibfnamefont{R.~D.} \bibnamefont{Pisarski}},
  \bibinfo{journal}{Nucl. Phys.} \textbf{\bibinfo{volume}{B337}},
  \bibinfo{pages}{569} (\bibinfo{year}{1990}{\natexlab{a}}).

\bibitem[{\citenamefont{Braaten and
  Pisarski}(1990{\natexlab{b}})}]{Braaten:1989kk}
\bibinfo{author}{\bibfnamefont{E.}~\bibnamefont{Braaten}} \bibnamefont{and}
  \bibinfo{author}{\bibfnamefont{R.~D.} \bibnamefont{Pisarski}},
  \bibinfo{journal}{Phys. Rev. Lett.} \textbf{\bibinfo{volume}{64}},
  \bibinfo{pages}{1338} (\bibinfo{year}{1990}{\natexlab{b}}).

\bibitem[{\citenamefont{Frenkel and Taylor}(1990)}]{Frenkel:1989br}
\bibinfo{author}{\bibfnamefont{J.}~\bibnamefont{Frenkel}} \bibnamefont{and}
  \bibinfo{author}{\bibfnamefont{J.~C.} \bibnamefont{Taylor}},
  \bibinfo{journal}{Nucl. Phys.} \textbf{\bibinfo{volume}{B334}},
  \bibinfo{pages}{199} (\bibinfo{year}{1990}).

\bibitem[{\citenamefont{Braaten and
  Pisarski}(1990{\natexlab{c}})}]{Braaten:1990it}
\bibinfo{author}{\bibfnamefont{E.}~\bibnamefont{Braaten}} \bibnamefont{and}
  \bibinfo{author}{\bibfnamefont{R.~D.} \bibnamefont{Pisarski}},
  \bibinfo{journal}{Phys. Rev.} \textbf{\bibinfo{volume}{D42}},
  \bibinfo{pages}{2156} (\bibinfo{year}{1990}{\natexlab{c}}).

\bibitem[{\citenamefont{Braaten and
  Pisarski}(1992{\natexlab{a}})}]{Braaten:1991gm}
\bibinfo{author}{\bibfnamefont{E.}~\bibnamefont{Braaten}} \bibnamefont{and}
  \bibinfo{author}{\bibfnamefont{R.~D.} \bibnamefont{Pisarski}},
  \bibinfo{journal}{Phys. Rev.} \textbf{\bibinfo{volume}{D45}},
  \bibinfo{pages}{1827} (\bibinfo{year}{1992}{\natexlab{a}}).

\bibitem[{\citenamefont{Braaten and
  Pisarski}(1992{\natexlab{b}})}]{Braaten:1992gd}
\bibinfo{author}{\bibfnamefont{E.}~\bibnamefont{Braaten}} \bibnamefont{and}
  \bibinfo{author}{\bibfnamefont{R.~D.} \bibnamefont{Pisarski}},
  \bibinfo{journal}{Phys. Rev.} \textbf{\bibinfo{volume}{D46}},
  \bibinfo{pages}{1829} (\bibinfo{year}{1992}{\natexlab{b}}).

\bibitem[{\citenamefont{Smilga}(1997)}]{Smilga:1996cm}
\bibinfo{author}{\bibfnamefont{A.~V.} \bibnamefont{Smilga}},
  \bibinfo{journal}{Phys. Rept.} \textbf{\bibinfo{volume}{291}},
  \bibinfo{pages}{1} (\bibinfo{year}{1997}), \eprint{hep-ph/9612347}.

\bibitem[{\citenamefont{Blaizot and Iancu}(2002)}]{Blaizot:2001nr}
\bibinfo{author}{\bibfnamefont{J.-P.} \bibnamefont{Blaizot}} \bibnamefont{and}
  \bibinfo{author}{\bibfnamefont{E.}~\bibnamefont{Iancu}},
  \bibinfo{journal}{Phys. Rept.} \textbf{\bibinfo{volume}{359}},
  \bibinfo{pages}{355} (\bibinfo{year}{2002}), \eprint{hep-ph/0101103}.

\bibitem[{\citenamefont{Blaizot et~al.}(1999)\citenamefont{Blaizot, Iancu, and
  Rebhan}}]{Blaizot:1999ap}
\bibinfo{author}{\bibfnamefont{J.~P.} \bibnamefont{Blaizot}},
  \bibinfo{author}{\bibfnamefont{E.}~\bibnamefont{Iancu}}, \bibnamefont{and}
  \bibinfo{author}{\bibfnamefont{A.}~\bibnamefont{Rebhan}},
  \bibinfo{journal}{Phys. Lett.} \textbf{\bibinfo{volume}{B470}},
  \bibinfo{pages}{181} (\bibinfo{year}{1999}), \eprint{hep-ph/9910309}.

\bibitem[{\citenamefont{Blaizot et~al.}(2003)\citenamefont{Blaizot, Iancu, and
  Rebhan}}]{Blaizot:2003tw}
\bibinfo{author}{\bibfnamefont{J.-P.} \bibnamefont{Blaizot}},
  \bibinfo{author}{\bibfnamefont{E.}~\bibnamefont{Iancu}}, \bibnamefont{and}
  \bibinfo{author}{\bibfnamefont{A.}~\bibnamefont{Rebhan}}
  (\bibinfo{year}{2003}), \eprint{hep-ph/0303185}.

\bibitem[{\citenamefont{Vija and Thoma}(1995)}]{Vija:1994is}
\bibinfo{author}{\bibfnamefont{H.}~\bibnamefont{Vija}} \bibnamefont{and}
  \bibinfo{author}{\bibfnamefont{M.~H.} \bibnamefont{Thoma}},
  \bibinfo{journal}{Phys. Lett.} \textbf{\bibinfo{volume}{B342}},
  \bibinfo{pages}{212} (\bibinfo{year}{1995}), \eprint{hep-ph/9409246}.

\bibitem[{\citenamefont{Schafer}(2003)}]{Schafer:2003jn}
\bibinfo{author}{\bibfnamefont{T.}~\bibnamefont{Schafer}},
  \bibinfo{journal}{Nucl. Phys.} \textbf{\bibinfo{volume}{A728}},
  \bibinfo{pages}{251} (\bibinfo{year}{2003}), \eprint{hep-ph/0307074}.

\bibitem[{\citenamefont{Levai and Heinz}(1998)}]{Levai:1997yx}
\bibinfo{author}{\bibfnamefont{P.}~\bibnamefont{Levai}} \bibnamefont{and}
  \bibinfo{author}{\bibfnamefont{U.~W.} \bibnamefont{Heinz}},
  \bibinfo{journal}{Phys. Rev.} \textbf{\bibinfo{volume}{C57}},
  \bibinfo{pages}{1879} (\bibinfo{year}{1998}), \eprint{hep-ph/9710463}.

\bibitem[{\citenamefont{Hidaka and Pisarski}(2008)}]{Hidaka:2008dr}
\bibinfo{author}{\bibfnamefont{Y.}~\bibnamefont{Hidaka}} \bibnamefont{and}
  \bibinfo{author}{\bibfnamefont{R.~D.} \bibnamefont{Pisarski}},
  \bibinfo{journal}{Phys. Rev.} \textbf{\bibinfo{volume}{D78}},
  \bibinfo{pages}{071501} (\bibinfo{year}{2008}), \eprint{0803.0453}.

\bibitem[{\citenamefont{Hidaka and
  Pisarski}(2009{\natexlab{a}})}]{Hidaka:2009hs}
\bibinfo{author}{\bibfnamefont{Y.}~\bibnamefont{Hidaka}} \bibnamefont{and}
  \bibinfo{author}{\bibfnamefont{R.~D.} \bibnamefont{Pisarski}},
  \bibinfo{journal}{Phys. Rev.} \textbf{\bibinfo{volume}{D80}},
  \bibinfo{pages}{036004} (\bibinfo{year}{2009}{\natexlab{a}}),
  \eprint{0906.1751}.

\bibitem[{\citenamefont{Hidaka and
  Pisarski}(2009{\natexlab{b}})}]{Hidaka:2009ma}
\bibinfo{author}{\bibfnamefont{Y.}~\bibnamefont{Hidaka}} \bibnamefont{and}
  \bibinfo{author}{\bibfnamefont{R.~D.} \bibnamefont{Pisarski}}
  (\bibinfo{year}{2009}{\natexlab{b}}), \eprint{0912.0940}.

\end{thebibliography}

\newpage 

\newpage 

\newpage 
\newpage 

\newpage 

\end{document}